\documentclass [letter, 12pt]{article}
\usepackage{a4wide, amsmath, amsfonts, amssymb, geometry, shuffle}

\usepackage{bm}
\usepackage{fancyhdr, dsfont}
\usepackage[latin1]{inputenc}
\usepackage[pdftex]{graphicx} 
\usepackage{graphicx}
\usepackage{float}
\usepackage{array}
\usepackage[sc,small]{caption}
\usepackage{hyperref}
\hypersetup{
    colorlinks=true,
    linkcolor=black,
    citecolor=black,
    filecolor=black,
    urlcolor=black,
}
\usepackage{color}
\definecolor{dgreen}{rgb}{0,0.70,0.30}
\definecolor{gold}{rgb}{0.85,.66,0}
\definecolor{purple}{rgb}{1.0,0.3,0.6}

\usepackage{tikz}
\usetikzlibrary{calc} \usetikzlibrary{patterns} \usetikzlibrary{decorations.pathreplacing} \usetikzlibrary{decorations.markings} \usetikzlibrary{decorations.pathmorphing} \usetikzlibrary{positioning}

\restylefloat{figure}
\usepackage{epstopdf}

\def\beq{\begin{equation}}
\def\eeq{\end{equation}}
\def\Re{{\rm Re\,}}
\def\Im{{\rm Im\,}}

\newcommand{\co}{\ , \ \ \ \ \ \ }
\newcommand{\dd}{\mathrm{d}}
\newcommand{\te}{\textrm}
\newcommand{\ap}{\alpha'}

\newcommand{\ZZ}{\mathbb Z}

\newcommand{\efrak}{\mathfrak e}
\newcommand{\ffrak}{\mathfrak f}

\renewcommand{\theta}{\vartheta}
                  
\newcommand{\be}{\begin{equation}}
\newcommand{\ee}{\end{equation}}
\newcommand{\bea}{\begin{eqnarray}}
\newcommand{\eea}{\end{eqnarray}}

\title{\vspace{1cm}\textbf{From maximal to minimal supersymmetry\\ in string loop amplitudes} \\ \ \\} 
\author{\hspace{-0cm}Marcus Berg$^{\dag}$, Igor Buchberger$^{\dag}$, Oliver Schlotterer$^{\star}$\\[5mm]
\hspace{-0cm} $^{\dag}$
{\it Department of Physics, Karlstad University,
 651 88 Karlstad, Sweden} \\
\hspace{-0cm} $^{\star}$ 
 {\it Max-Planck-Institut f\"ur Gravitationsphysik, Albert-Einstein-Institut},\\[-1mm] {\it 14476 Potsdam, Germany}
}

\begin{document}

\maketitle{}
\vspace{1cm}
\begin{abstract}
We calculate one-loop string amplitudes of open and closed strings
with $N=1,2,4$ supersymmetry in four and six dimensions, by compactification
on  Calabi--Yau and K3 orbifolds.
In particular, we develop a method to combine contributions
from all spin structures
 for arbitrary 
number of legs  at minimal supersymmetry. Each amplitude is cast into a compact form
by reorganizing the kinematic building blocks and casting the worldsheet integrals in a basis. 
Infrared regularization plays an important
role to exhibit the expected factorization limits. We comment on 
implications for the one-loop string effective action. 
\end{abstract}

\newpage

\setcounter{tocdepth}{2}
\tableofcontents

\numberwithin{equation}{section}

%
%
%
%
%
%
%

\section{Introduction}

A significant part of the work on string loop amplitudes has been performed in ten dimensions and for
maximal
supersymmetry. A classic example is the 1982 Brink-Green-Schwarz calculation \cite{Green:1982sw}
of a 4-point 1-loop amplitude of gravitons or gauge fields. The state
of the art in maximal supersymmetry has reached 
the first non-vanishing results at 3-loop order \cite{Gomez:2013sla}, made tractable by
the manifestly supersymmetric pure spinor formalism \cite{Berkovits:2000fe}.

We will not focus on phenomenology in this paper, but clearly it is of great interest to develop the state
of the art of string effective actions with {\it minimal} supersymmetry, as opposed to maximal. 
We will argue that even at 1-loop order in minimal supersymmetry, there is much left to be understood
about string amplitudes.
For fundamental problems like moduli stabilization, without which there can be no reliable phenomenology\footnote{See for example 
 the review \cite{Blumenhagen:2006ci}. From the vast literature,
 let us highlight \cite{Conlon:2006wz} from the string side and more recently \cite{Reece:2015qbf} 
 from the phenomenology side as
 two illustrative examples of the crucial role of moduli stabilization in phenomenology.}, the string effective action should be calculated at least to 1-loop order, as some stabilization effects are quantum-mechanical. Half-maximal supersymmetry provides a useful step on the way to minimal supersymmetry.

In this paper we study type II string compactifications on K3 and Calabi--Yau (toroidal) orbifolds
that break supersymmetry down to half-maximal or quarter-maximal. For closed strings, quarter-maximal amounts to 8 supercharges which is $N=2$ supersymmetry in $D=4$ terminology. The basic technology to compute all 1-loop amplitudes in type IIB Calabi--Yau orbifolds (and orientifolds) has in principle been available for decades, but various technical obstacles have prevented progress.

Impressive progress on the gauge boson 1-loop 4-point amplitude in quarter- and half-maximal supersymmetry
was made in 2006 \cite{Bianchi:2006nf}, but in a form that was  difficult
 to process further, for example to check supersymmetry Ward identities.
Last year, this calculation was simplified \cite{Bianchi:2015vsa} by specializing the external polarizations 
to spinor-helicity variables at an early stage of the calculation. 

Recently, work on the graviton 1-loop 4-point amplitude for half-maximal K3$\times T^2$
was presented in \cite{Tourkine:2012vx,Ochirov:2013xba}. In 
contrast to those papers,
we first perform the sum over the spin structures of the Ramond-Neveu-Schwarz (RNS) formalism
and will later perform the field-theory limit.
Also, in addition to K3 compactification to half-maximal supersymmetry, 
we also consider closed strings in Calabi--Yau compactification to 
quarter-maximal supersymmetry. 

In this paper, we will approach the problems in the aforementioned papers from a new angle and present substantial
generalizations of the 1-loop amplitudes discussed in the literature so far. Our main 
results are the methods for all-multiplicity spin sums in section \ref{sec:reduced},
the open-string 3-point and 4-point 1-loop amplitudes \eqref{new3pt} and \eqref{new4pt},
and the closed-string 3-point and 4-point 1-loop amplitudes \eqref{cl3d} and \eqref{nred102}.
The precise improvements on previous work will be clarified in those sections. 
The closed-string expressions are valid for generic massless NSNS external states 
(graviton, dilaton, and  antisymmetric tensor).

One key aspect of these results is that connections between 1-loop amplitudes with different amounts of
supersymmetry are revealed. First, the parity-even kinematic factors of open-string $n$-point amplitudes for half-maximal and quarter-maximal supersymmetry are identical. The amplitudes are only distinguished
by the explicit functions of worldsheet moduli, see (\ref{from2to4}). Second, as will be detailed in section \ref{sec:six}, the structure of half-maximal open-string amplitudes at multiplicity $n$  
is very similar to that of their maximally supersymmetric counterparts at multiplicity $n{+}2$. Finally, the
progress we made on open-string amplitudes reverberates in our closed-string amplitudes in section \ref{sec:closedstring}, where the simplified expression \eqref{nred102} for the 4-point function closely resembles the maximally supersymmetric 6-point amplitude of \cite{maxsusy}.

\section{Superstring effective action}
\label{sec:eff}

We will not extract details of the effective action in this paper,
but here is a short review of expectations and motivations. 

The closed-string
sector is somewhat more universal
than the open-string sector,
so let us begin there, but most of our comments extend
to open strings. 
 In $D=10$ the {\it leading} correction to the type IIB string effective action appears at order $\alpha'^3$. 
 Even this leading correction is not completely known (especially for the RR sector), but many pieces are well understood. The gravitational part of the type IIB action  in Einstein frame is (see e.g.\ 
 \cite{Green:1981ya,Grisaru:1986kw,Gross:1986iv,Green:1997tv}
for some original references, or
 \cite{Policastro:2006vt,Liu:2013dna,
Minasian:2015bxa} for contemporary work)
\be \label{R4}
S_{\te{IIB}}=\int \dd^{10}x  \sqrt{-g} \left(R +  f_{3/2}^{(0,0)}(\tau,\bar{\tau})\, \alpha'^3 R^4+\ldots \right) \ ,
\ee
where $R^4$ is schematic 
for index contractions with the well-known tensor structure  $t_8 t_8+ \epsilon_{10}\epsilon_{10}$ (see e.g.\ \cite{Antoniadis:1997eg}). There is a simple way to include also the other massless NSNS fields: the Kalb-Ramond B-field and dilaton. As 
discussed e.g.\  by \cite{Gross:1986mw} in 1986, and more recently in
e.g.\ \cite{Policastro:2006vt,Liu:2013dna,
Minasian:2015bxa}, the idea is to shift the Riemann tensor by $\nabla H$ and $\nabla \nabla \phi$ as
\be  \label{Rshift}
{\widehat{R}_{mn}}{}^{pq} = {R_{mn}}^{pq} + 2\kappa e^{-\kappa \phi /\sqrt{2}} \nabla_{[m} {H_{n]}}^{pq} -
\sqrt{2}\kappa {\delta_{[m}}^{[p} \nabla_{n]} \nabla^{q]} \phi \ ,
\ee
where $\kappa$ is the gravitational coupling,
$H$ is the NSNS 3-form field strength and $\phi$ is the dilaton. The geometric
interpretation of this shift as torsion is discussed for example in \cite{Liu:2013dna}.
(In eq.\ (\ref{Rshift}) and throughout this work, vector indices $m = 0,1,\ldots,D{-}1$ are taken as $m,n,p,q,\ldots$ from the middle of the latin alphabet, where the number $D$ of dimensions will be clear from context.)
We have not discussed terms that depend on RR fields, that we comment on in the outlook.

The best-understood type IIB
coefficient in $D=10$ is the one above, of the $R^4$ term \cite{Green:1997tv},
\be
f_{3/2}^{(0,0)}(\tau,\bar{\tau})=
E_{3/2}(\tau,\bar{\tau}) \ ,
\ee
where $\Im(\tau)=g_{\rm s}$ and $E_s$ is the nonholomorphic Eisenstein series
with series expansion\footnote{For a recent review of the systematics
of such expansions, also with toroidal compactification, consult \cite{Fleig:2015vky}.}
\be  \label{f32}
f_{3/2}^{(0,0)}(\tau,\bar{\tau}) = 2\zeta(3)g_{\rm s}^{-3/2}+{2\pi^2 \over 3}g_{\rm s}^{1/2} +
\mbox{instanton corrections } (e^{-1/g_s}) \ .
\ee
After compactification to e.g.~$D=4$ on some nontrivial (non-toroidal) space, much  less is known than in $D=10$, since as discussed above, the requisite amplitudes 
for the less-than-maximal type II superstring have not been studied systematically until recently. (There is
substantial literature on related issues in the heterotic string, some of which we review below.) As an illustration of the great simplifications of maximal supersymmetry, in
\eqref{f32} we see
that there are no perturbative corrections
beyond one loop. This non-renormalization theorem does not extend to minimal supersymmetry. 
More relevant for our purposes is that in maximal supersymmetry, the $\alpha'^3 R^4$ correction 
is the leading-order $\alpha'$ correction in a flat background. With minimal supersymmetry
one generically expects all the lower-order terms to appear: $R$, $\alpha'R^2$ and possibly (see below) $\alpha'^2R^3$,
both at tree level and at loop orders.
So with less supersymmetry, the $R^4$ correction above 
that was leading in maximal supersymmetry becomes sub-sub-sub-leading in the $\alpha'$-expansion:
\be
S_{1/4}=\int \dd^{4}x  \sqrt{-g} \left(\Delta_1 R +\Delta_2 \alpha' R^2 + \Delta_3 \alpha'^2 R^3+ \Delta_4 \alpha'^3 R^4+\ldots \right) \ ,
\ee
where the subscript $1/4$ means ``quarter-supersymmetric''. 
The  loop-corrected coefficients $\Delta_i$ in general depend on the moduli,
and one  can   extract aspects of this dependence from the 1-loop string amplitudes in this paper.
Until recently 
this would not have been feasible.

Again, similar comments hold for open strings and string corrections to gauge-field effective actions.

\subsection{Tree-level: review}

As summarized above, in the type II string effective action in $D=10$ there is the famous  $\alpha'^3 R^4$ term that first appears at  string tree level (sphere diagram).
The lower powers $\alpha' R^2$ and $\alpha'^2 R^3$ are  forbidden by 32 supercharges.
By contrast, the heterotic string in $D=10$ with 16 supercharges is known to have a tree-level $R^2$
term \cite{Gross:1986mw}, unlike in type II. 
This can be explained by ``double copy'' (see e.g.\ \cite{Broedel:2012rc}): there is a tree-level $\alpha'F^3$ term on the purely bosonic side
of the heterotic string,
and no $\alpha'F^3$ term on the supersymmetric side. ``Multiplying out'' 
two vectors to give a graviton in the sense of double-copy, one obtains a tree-level subleading term $\alpha' R^2$ in the heterotic string in $D=10$. 
So an $\alpha' R^2$ term is allowed by 16 supercharges,
but it is not required. For type II compactified on K3, Antoniadis et al.~\cite{Antoniadis:1997eg} explain  (p.4)
that there is no tree-level $R^2$  term.

The cubic curvature $R^3$ terms are absent in all these theories.
The first evidence for this was by explicit calculation, but now it is understood more generally, see the next section.

Tree-level interactions of open and closed superstrings involve multiple zeta values (MZVs) upon $\alpha'$-expansion, a first hint being the above single-zeta value in $\alpha'^3 \zeta(3)R^4$. In the closed-string sector, $D^m  R^n$ couplings in \cite{Stieberger:2009rr, Schlotterer:2012ny} can be traced back to the all-multiplicity results for open-string trees in \cite{Mafra:2011nv, Mafra:2011nw} through the KLT relations \cite{Kawai:1985xq} which imply identical graviton interactions in type IIB and type IIA theory. The patterns of MZVs and covariant derivatives can then be generated from the Drinfeld associator \cite{Broedel:2013aza}.
 The study of $D^m  R^n$ interactions is important to assess the UV behavior\footnote{Based on a symmetry analysis of $D^mR^n$ matrix elements initiated in \cite{Brodel:2009hu}, any counterterm with the mass dimension of $D^6 R^4$ and below was ruled out, guaranteeing UV-finiteness of four-dimensional $N=8$ supergravity up to and including 6 loops \cite{Beisert:2010jx}.} of $N=8$ supergravity in four dimensions by testing their compatibility with its $E_{7(7)}$ duality symmetry \cite{Brodel:2009hu, Elvang:2010jv, Elvang:2010kc, Beisert:2010jx}. 

As initially observed in \cite{Stieberger:2009rr}, systematic cancellations obscured by the KLT relations occur when assembling $D^m  R^n$ interactions from open-string amplitudes \cite{Schlotterer:2012ny}, leaving for instance only one tree-level interaction of type $D^m  R^n$ at the mass dimensions of 
$D^{2m}R^4$ with $m=0,2,3,4,5$. The selection rules for the accompanying MZVs were identified with the {\it single-valued projection} \cite{Stieberger:2013wea}. Also beyond tree level, 
there is evidence  that the single-valued MZVs and polylogarithms govern the closed-string $\alpha'$-expansion \cite{Zerbini:2015rss, D'Hoker:2015qmf}. 

In the heterotic theory, an interesting tree-level connection between single-trace interactions in the gauge-sector and the type I superstring was found in \cite{Stieberger:2014hba}, again based on the single-valued projection of MZVs. In the gravitational sector of the heterotic string, half-maximal supersymmetry allows for additional $D^m  R^n$ interactions absent for the superstring whose implications for counterterms of $N=4$ supergravity were studied in \cite{Huang:2015sla}. At a given mass dimension, the $D^mR^n$ interactions accompanied by MZVs of highest transcendental weight are universal to the heterotic and type II theories \cite{inprogress}, and this universality of the leading-transcendental part in fact carries over to the bosonic open and closed string. Accordingly, the MZVs along with non-universal $D^mR^n$ interactions of the heterotic string starting with $R^2$ have lower transcendental weight as compared to their universal counterparts \cite{inprogress}, suggesting a classification by weight at each mass dimension.

\subsection{Factorization and ambiguities: tree level and 1 loop}
\label{sec:ambi}

\begin{figure}[h]
\begin{center}
\includegraphics[width=0.4\textwidth]{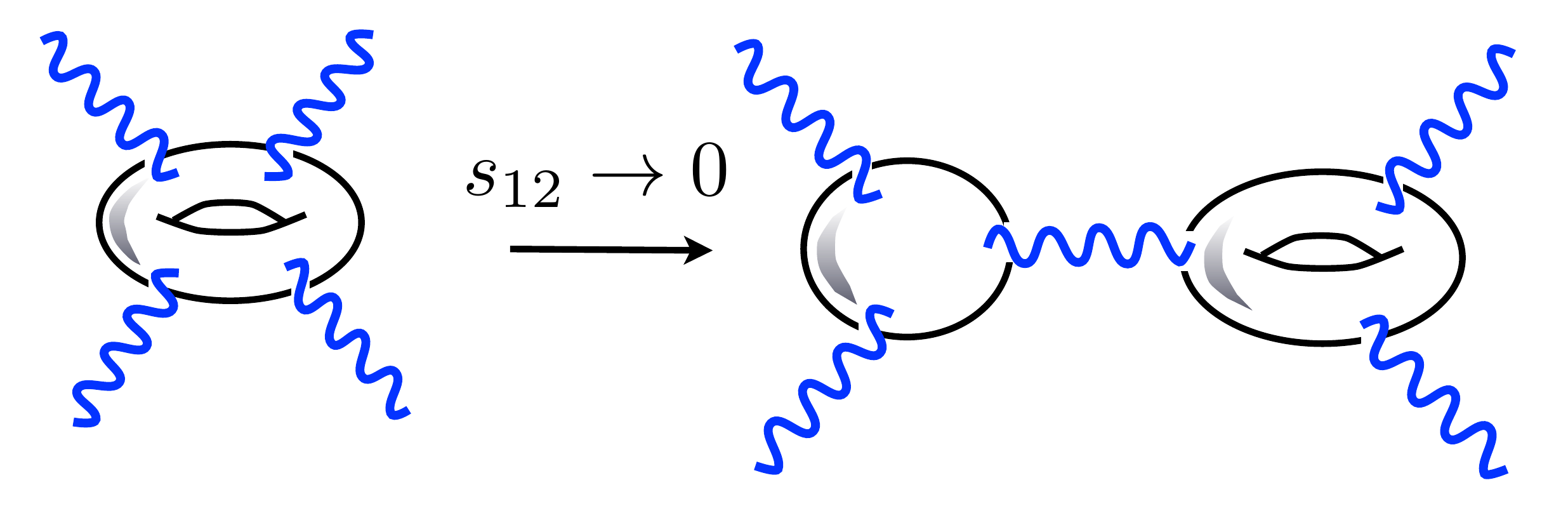}
\vspace{-5mm}
\caption{Factorization onto (tree-level 3-point) $\times$ (propagator) $\times$ (1-loop 3-point).}
\label{fig:fact1}
\end{center}
\end{figure}

The Gross-Sloan paper from 1986  mentioned above \cite{Gross:1986mw}  
contains a  detailed discussion of {\it field-theory pole subtractions}, 
a key piece in the machinery of extracting 
an effective field theory from string amplitudes. 
For less than maximal supersymmetry,
the 4-point string amplitudes at 1-loop order factorizes onto 3-point vertices,
as drawn  in fig.\ \ref{fig:fact1}.
Reducible field-theory diagrams with the gravitational 3-point vertices need to be subtracted from the low-energy
limit of the string amplitude to isolate the irreducible field-theory 4-point coupling corresponding to $D^{2m}R^4$ terms in the effective action.
This is a laborious procedure. As emphasized in \cite{Kiritsis:1997em}, if we are interested
in fewer powers of the Riemann tensor like $R^2$ and $R^3$,
we could in principle extract them from 2-point or 3-point functions, where one could expect
there to be no reducible contributions at all. 
This can be taken as a general argument
that for efficient computation one should strive to compute the lowest number of external legs
that can probe the term of interest in the effective action. 
 
However, 2-point and 3-point functions of massless states vanish on-shell unless they
 are infrared-regularized, as we review in appendix \ref{kinematics}. This regularization is a key point in this paper 
 and we will discuss it in more detail in section \ref{sec:minahan}. Somewhat surprisingly, we will see that
  the same regularization procedure should also be applied to
 $n$-point functions for any $n$, to exhibit the expected factorizations 
 in the spirit of this section.

A related issue is that since our string amplitudes are on-shell,
there are ambiguities coming from field redefinitions,
a typical example being a shift of the graviton $h_{mn}\rightarrow h_{mn}+R_{mn}$
that can shuffle coefficients between the three terms $R_{mnpq}R^{mnpq}$,
$R_{mn}R^{mn}$ and $R^2$ in the string effective action, as explained for example in  \cite{Forger:1996vj}.
The coefficients are moduli- and background-dependent,
so a more general background could lift some of these degeneracies.
We discuss this issue a little further in section \ref{sec:het} and appendix \ref{kinematics},
but the main focus of this paper is  the underlying string amplitudes
and not the explicit construction of a string effective action.

\begin{figure}[h]
\begin{center}
\includegraphics[width=0.4\textwidth]{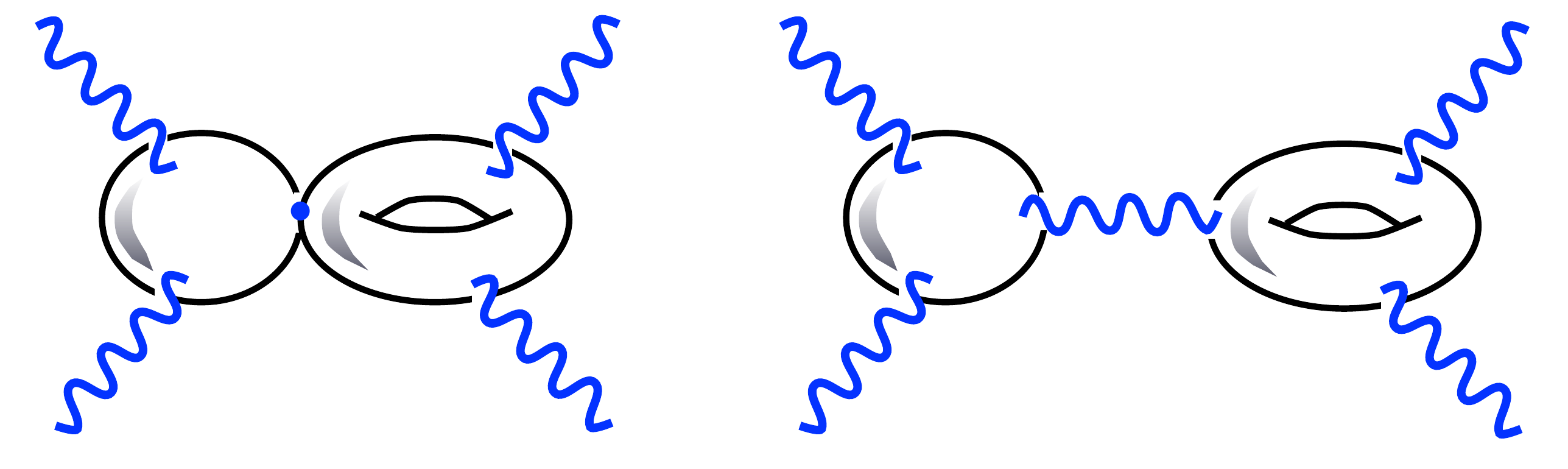}
\vspace{-5mm}
\caption{The distinction between ``delta function'' and ``delta function and propagator''.}
\label{fig:fact2}
\end{center}
\end{figure}

So let us return to the factorization of the closed-string 4-point 1-loop amplitude as illustrated in fig.\ \ref{fig:fact1}. 
(The following discussion will be general, but
a few explicit expressions corresponding to the figures
drawn here are given in appendix \ref{sec:factorization}.)
The moduli space of 
string loop amplitudes is interesting already in this fairly simple example. 
As a first question, in the factorization limit in fig.\ \ref{fig:fact1}, 
which of the two more specific diagrams in fig.\ \ref{fig:fact2} is actually realized?
By conformal invariance of the worldsheet theory,
we can always factor off a sphere from a bulk point on the worldsheet. 
When we draw this sphere explicitly as in fig.\ \ref{fig:fact2},
we mean that the two external states on the left part of each diagram
are closer to each other  than to any other vertex operator\footnote{The moduli space of superstring
amplitudes in figures like in this section is discussed more systematically in for example \cite{Minahan:1987ha},
for open strings in for example \cite{Hashimoto:1996bf}
and more recent discussions include Witten's
extensive notes \cite{Witten:2012bh}. We admit that the qualification ``closer to each other''
restricts us to some class of worldsheet metrics.}.
In the worldsheet computation, this arises from a delta function of two punctures along with a propagator, caused by the collision of 
vertex operators. Now, we 
will encounter  situations where an inverse propagator is generated from the contractions among vertex operators. 
If this cancels the propagator that arose with the delta function, we draw the diagram
in the left panel of  fig.\  \ref{fig:fact2}.
If the propagator is uncancelled, we draw it explicitly as in the right panel of 
fig.\ \ref{fig:fact2}.
In both diagrams in fig.\ \ref{fig:fact2}, the delta function has reduced the number of integrations over punctures
by one, so the moduli space of the remaining 1-loop integral
is that of a 3-point 1-loop torus diagram, but where one of the external momenta 
is the sum $k_1+k_2$ of momenta of the two states on the left side of the 4-point diagram.
Unlike individual momenta of massless states, this sum is not
constrained to be lightlike: $(k_1+k_2)^2\neq 0$. In field theory, the right side
is then called a {\it 1-mass triangle}, but we emphasize
that we have not taken a field-theory limit yet. 

\begin{figure}[h]
\begin{center}
\includegraphics[width=0.4\textwidth]{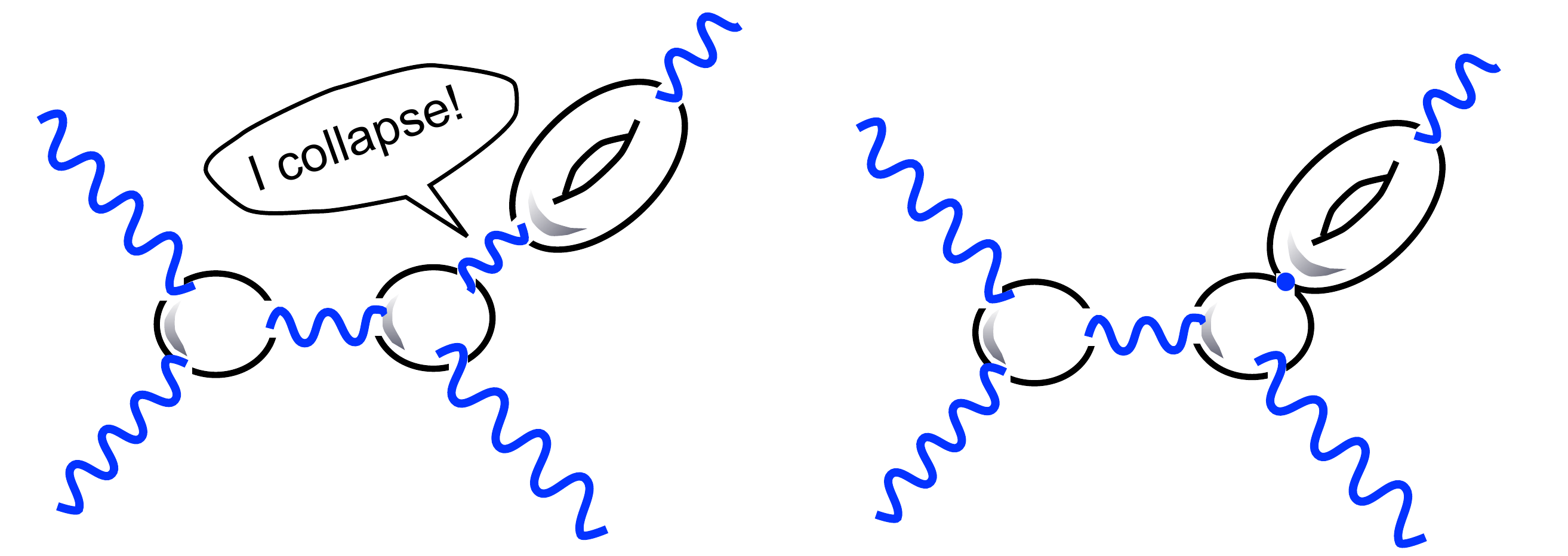}
\vspace{-5mm}
\caption{Collapse of specific propagator avoids double factorization limit.}
\label{fig:fact3}
\end{center}
\end{figure}

Analogously, we can ask whether
there are any {\it 1-mass bubbles},
i.e.\ whether there is further factorization
of the subdiagram on the right of fig.\ \ref{fig:fact2} (3-point 1-loop torus). As indicated in fig.\ \ref{fig:fact3}, 
we find that there is always a Mandelstam variable in the numerator 
that offsets the propagator closest to the torus, so 
this propagator always collapses to a point. 
This is important since an actual double factorization limit
would have generated a 3-particle propagator $(k_1+k_2+k_3)^{-2}=(-k_4)^{-2}$ which is
in fact infrared divergent in the 4-particle 
momentum phase space.
The two spheres in the diagram
on the right in fig.\ \ref{fig:fact3}
each represent a delta function from a particular region in moduli space, so this
 leaves one integration over a puncture, as appropriate
for a torus 2-point function with one fixed vertex operator.
In the field-theory limit, this string amplitude
will indeed generate a 1-mass bubble. 

Factorization in the field-theory limit is an interesting
topic in its own right,
and will be discussed in a companion paper \cite{FT}.

The above discussion was quite detailed, so let us
make one broader statement, that we will explain in more detail in later sections.
In maximal supersymmetry, it is well-known that the 
factorization of the 4-point function as in fig.\ \ref{fig:fact1} does
not occur (i.e.\ has zero residue). In fact, the number of
successive factorizations of an $n$-point function in maximal supersymmetry
is $n{-}4$. We will find that for half-maximal supersymmetry
as well as parity-even contributions to 1-loop amplitudes in quarter-maximal supersymmetry, 
this number is $n{-}2$, as we have illustrated
in the figures in this section. Parity-odd terms in quarter-maximal require a refined analysis,
and preliminary arguments in later sections suggest $n{-}3$ successive factorizations for
open strings and $n{-}2$ for closed strings.

\subsection{One-loop: review}
\label{sec:oneloop}

Now we turn to the 1-loop effective action. First let us note the obvious point that if a coupling is prevented by 
supersymmetry in the 
sense that a superspace lift of the coupling
does not exist, it will be prevented equally well at tree level and  loop level. 
\begin{table}[h]
\begin{center}
\begin{tabular}{|c||c|c|c||c|c|c||c|c|c||}\hline
1-loop term & IIA & IIB & Het & IIA/K3 & IIB/K3  & Het/K3 & IIA/CY & IIB/CY & Het/CY  \\ \hline \hline
$R$\,  &  $\times$ &  $\times$ &  $\times$ & $\times$&  $\times$ &  $\times$&  $\checkmark$&  $\checkmark$&  $\times$  \\ \hline
$R^2$ &  $\times$ &  $\times$ &  $\times$ & $\checkmark$ &  $\times$ &  $\checkmark$ &  $\checkmark$   &  $\checkmark$&  $\checkmark$ \\ \hline
$R^3$ &  $\times$ &  $\times$  &  $\times$& $\times$&  $\times$ &  $\times$&  $\times$ &  $\times$&  $\times$ \\ \hline
$R^4$ & $\checkmark$  & $\checkmark$   & $\checkmark$  & $\checkmark$  & $\checkmark$   & $\checkmark$
& $\checkmark$ &  $\checkmark$&  $\checkmark$ \\ \hline
\end{tabular} \hspace{9mm}
\end{center}
\vspace{-4mm}
\caption{One-loop curvature corrections. The double vertical lines
delineate $D=10,6,4$.} 
\vspace{-2mm}
\end{table}
For IIB on K3, which has 16 supercharges like in the heterotic string in $D=10$
 (or on $T^4$),
 one would expect that
 supersymmetry would allow $R^2$.
The details are  interesting:
it turns that the 1-loop correction to $R^2$ vanishes in IIB on K3 but does not vanish in IIA on K3.
See for example \cite{Kiritsis:1997em,Antoniadis:1997eg} and
especially \cite{Gregori:1997hi} as well as section \ref{sec:seventwo} for a review 
of this string amplitude computation.
From the supergravity point of view, \cite{Antoniadis:1997eg} explains the vanishing of 1-loop $R^2$ corrections in 
the $D=6$ IIB string theory on K3 
from reduction of the ten-dimensional 1-loop term $
(t_8 t_8 \pm \epsilon_{10} \epsilon_{10}) R^4$, where the relative
sign gives cancellation in IIB but not IIA.
There is also a duality argument: for IIA on K3 there should be a 1-loop $R^2$ correction but no tree-level $R^2$, 
because in heterotic on $T^4$ there is a tree-level $R^2$ (as discussed above) and no 1-loop $R^2$,
and  they should be exchanged by heterotic-IIA duality \cite{Antoniadis:1997eg}.\footnote{For an impressive example
of this type of argument in the heterotic-type I duality, see \cite{Stieberger:2002fh,Stieberger:2002wk}.}
 These three arguments
illustrate the variety of techniques that have been developed for half-maximal supersymmetry.

The previous discussion concerned $D=6$. 
Compactification of type II on K3$\times T^2$ to $D=4$ is discussed in  \cite{Harvey:1996ir,Gregori:1997hi,Antoniadis:1997eg},
where the authors calculate moduli-dependent couplings  like
\be  \label{moduli}
\int \! \dd^4 x \,  \sqrt{-g} \,\Delta(U) R^2 \quad \mbox{(IIB) \ ,} \qquad \int \! \dd^4 x  \,\sqrt{-g} \, \Delta(T) R^2 \quad \mbox{(IIA)} \ ,
\ee
 where  $U$ is 
the complex structure and $T$ is
the K\"ahler modulus of the 2-torus, and they are exchanged by T-duality.
Note that despite having the same amount of supersymmetry as in IIB on K3 above, compactification
to $D=4$ on this 2-torus allows an $R^2$ term in IIB. 
 The authors of \cite{Gregori:1997hi}
argue that in the decompactification limit of the 2-torus, the coefficient would need to contain some power
of the K\"ahler modulus $T$ of the 2-torus to survive the large-torus limit, and $\Delta(U)$ does not. This recovers
the vanishing of the $R^2$ term in $D=6$ for IIB and the non-vanishing for IIA. 

Finally,
there is a fairly detailed discussion of the heterotic 1-loop $R^2$ correction in 
\cite{Forger:1996vj}, where the non-renormalization
of the Einstein--Hilbert action is also discussed. See e.g.\ \cite{Soda:1987in} for previous work
and \cite{Peeters:2000qj} for a useful summary of some of the older literature.

Let us move on to $R^3$ corrections. 
Reduction of $R^4$ from $D=10$ on K3 or Calabi--Yau produces
contractions of the schematic form $(R_{\rm external})^3 R_{\rm internal}$, where
$R_{\rm internal}$ leaves no room for anything else than the  Ricci scalar of the compactification manifold,
which vanishes by Ricci-flatness. 
In effective field theory, there is a general superspace argument that no superinvariant containing
$R^3$ as bosonic component can be constructed 
(see for example
 \cite{Tseytlin:1995bi}). Original explicit calculations showing the absence of $R^3$ terms 
 go back to the 1970s, see for example \cite{Grisaru:1976nn,Metsaev:1986yb,
Nilsson:1986rh,Kallosh:1987mb,Bergshoeff:1989de}. Some of these explicit calculations
are being revisited using modern techniques, see e.g.\  \cite{Bern:2017puu}.

However, eq.\ \eqref{R4} together with eq.\ \eqref{Rshift} indicate that
there should be $R^3$ terms in nontrivial backgrounds, like flux backgrounds
or internal dilaton gradients. Constructing such terms from string amplitudes in nontrivial backgrounds
is challenging, see the conclusions for  comments on this.

Finally, string loop corrections to the Einstein--Hilbert action in quarter-maximal supersymmetry
were studied using the background field method in 
\cite{Kiritsis:1994ta,Kohlprath:2002fe}. 
Amplitude calculations of this correction was discussed recently in \cite{Haack:2015pbv}\footnote{This 
paper is mostly about orientifolds, but the torus amplitude
only differs by a factor of $1/2$ from the parent theory.}
which builds on, corrects and extends
results from \cite{Kohlprath:2002fe,Kohlprath:2003pu,Epple:2004ra}.
These results are all extracted from infrared-regularized low-point functions,
in the strong sense that we discuss in detail in section \ref{sec:minahan}. It would be desirable
to compare with our  results, though we do not do so in detail in this paper.
The general conclusion from these papers is 
as expected from the effective supergravity discussion in 
\cite{Antoniadis:1997eg}, section 5:  there is a 1-loop correction to $R$
in type II on Calabi--Yau. It descends from the $\epsilon_{10}\epsilon_{10}$
term in \eqref{R4}, and the relative
sign of the tree-level and 1-loop correction
to $R$  in type IIA is the opposite to that of type IIB. 
For completeness, let us also mention that there is a correction to $R$
in type IIB orientifolds on K3 \cite{Antoniadis:1996vw}, which is also quarter-maximal due to the orientifolding.

Covariant derivatives $D^mR^n$ of Riemann tensors have also been studied at loop level. In ten-dimensional type IIB theory, S-duality was exploited to determine the full moduli-dependent coefficients of the $D^4R^4$ \cite{Green:1999pu} and $D^6R^4$ \cite{Green:2014yxa} interactions, including their non-perturbative completions. S-duality based predictions for  the 2-loop and 3-loop coefficient of $D^6R^4$ \cite{Green:2005ba} were confirmed by the amplitude computations of \cite{D'Hoker:2014gfa} and \cite{Gomez:2013sla}, also see \cite{Gomez:2015uha} for $D^2R^5$ at two loops. 

The amplitude calculations of this paper culminate in the compact expression (\ref{nred102}) for the half-maximal 1-loop amplitude involving four NSNS sector states in type IIB and type IIA. We lay some foundations for a systematic investigation of 1-loop $D^{2n} R^{m\leq 4}$ couplings in half-maximal type II compactifications by identifying the gauge-invariant ``seeds'' in their matrix elements.

Here we have focused mostly on gravitational corrections. Other NSNS corrections involving
$B$ fields and dilatons have been studied somewhat less,
but were discussed for example in \cite{Liu:2013dna,Minasian:2015bxa},
and our results here are equally relevant for those loop corrections, see e.g.\ section \ref{efftwoB}. We comment on RR fields in the conclusions.

\section{Amplitude prescriptions and spin sums}
\label{sec:prescr}

In this section we define the computations that will occupy us for the remainder of the paper,
including efficient  techniques to sum over spin structures of the worldsheet spinors. We study compactifications of type I and type II superstrings on certain $\mathbb{Z}_N$ orbifolds illustrated in fig.\ \ref{fact2} that yield half-maximal and quarter-maximal supersymmetry.
This is textbook material (see e.g.\  \cite{Polchinski:1998rr,Kiritsis:2007zza,Ibanez:2012zz,Blumenhagen:2013fgp}), but
before launching into the detailed prescriptions, we give a quick review.

\subsection{Orbifolds}

\begin{figure}[h]
\begin{center}
\includegraphics[width=0.55\textwidth]{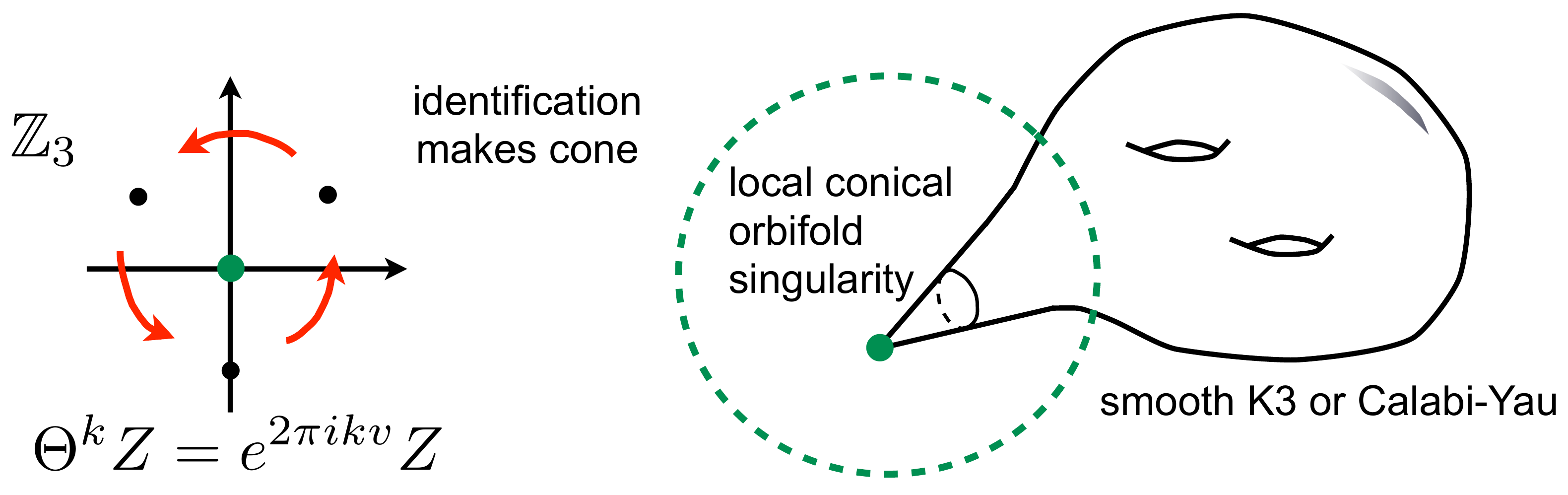}
\vspace{-5mm}
\caption{Orbifold compactification: identification
under ${\mathbb Z}_N$ creates a conical singularity.
The orbifold twist $kv$ (that we will call $\gamma$, see eq.\ \eqref{orbtw}) will occur
in all our amplitudes.}
\label{fact2}
\end{center}
\end{figure}

We will consider 
supersymmetric orbifolds of the form $T^4/\mathbb{Z}_N$,
$T^4/\mathbb{Z}_N\times T^2$ or $T^6/\mathbb{Z}_N$\footnote{We assume
 factorizable tori, i.e. $T^4=(T^2)^2$ and $T^6=(T^2)^3$.
 }.
The ``orbifold group'' $\mathbb{Z}_N$ is a discrete subset of the rotation group and
one identifies points in spacetime 
that are related by the $\mathbb{Z}_N$  action.
With complexified string coordinates $Z_j=X^{2j+2}+U_j X^{2j+3}$, where $j=1,2,3$ and $U_j$ is the complex structure
of the $j^{\te{th}}$ 2-torus, the discrete orbifold rotation is diagonal:
\be
\Theta^k Z_j=e^{2\pi i k v_j}Z_j \ .
\ee
The rational numbers $v_j$ are such that $\Theta^N=1$ (or occasionally
one allows $-1$), and
they  satisfy $v_1+v_2+v_3=0$ to preserve some supersymmetry\footnote{A simple
way to see this is to use the oscillator notation for gamma matrices,
see e.g.\ Appendix B of \cite{Polchinski:1998rr}.}, see table 2 below for examples. 
An orbifold theory is obtained from a ``parent'' theory in $D=10$ by inserting the projector 
$\sum_{k=0}^{N-1} \Theta^k/N$ in amplitude trace computations. 
The power $k$ in the trace is called the {\it sector} of the orbifold. The identification of points in spacetime that are related by the
$\mathbb{Z}_N$  action can create conical singularities at fixed points of the orbifold action, as in fig.\ \ref{fact2}. 
One can also mix in the worldsheet parity operation
in the above orbifold group to make an orientifold, 
in the same sense that type I (open+closed strings) in $D=10$
is an orientifold of type IIB. The  D-branes
on which open strings can end are added to cancel the negative D-brane charge of the orientifold plane.
In noncompact models, one can get away
without orientifolding, but in compact models, there is some additional work
to compute the M\"obius strip and Klein bottle amplitudes that might be needed for specific consistent string models.
We will only consider annulus and torus amplitudes in this paper, but the key simplifications of the integrands 
should carry over straightforwardly  to the remaining topologies. (We note
that for closed strings, our torus amplitudes will be consistent by themselves, but 
 for model-building 
one might want to orientifold also for closed strings, to  allow moduli stabilization in minimal supergravity.)
\begin{table}[h]
\vspace{0mm}
\begin{center}
\begin{tabular}{|c|c|} \hline
\multicolumn{2}{|c|}{$T^4/{\mathbb Z}_N\times T^2$} \\  \hline
${\mathbb Z}_2$ & ${1 \over 2}(1,-1,0)$ \\   \hline
${\mathbb Z}_3$ & ${1 \over 3}(1,-1,0)$ \\   \hline
${\mathbb Z}_6$ & ${1 \over 6}(1,-1,0)$ \\   \hline
\end{tabular}\hspace{2cm}
\begin{tabular}{|c|c|} \hline
\multicolumn{2}{|c|}{$T^6/{\mathbb Z}_N$} \\  \hline
${\mathbb Z}_3$ & ${1 \over 3}(1,1,-2)$ \\   \hline
${\mathbb Z}_4$ & ${1 \over 4}(1,1,-2)$ \\   \hline
${\mathbb Z}_6'$ & ${1 \over 6}(1,2,-3)$ \\   \hline
\end{tabular}
\end{center}
\vspace{-5mm}
\caption{Examples of $(v_1,v_2,v_3)$ for supersymmetric orbifolds/orientifolds, see e.g.\ \cite{Blumenhagen:2013fgp}.}
\label{orbiexample}
\end{table}
\vspace{-5mm}

\subsection{Open-string prescriptions}
\label{sec:fourone}

One-loop scattering amplitudes among unoriented open-string states receive contributions from cylinder and M\"obius-strip diagrams. In this work, we will discuss the planar cylinder (annulus)
with modular parameter $\tau_2$ as a representative diagram where all external states are inserted
 on the same boundary component, and the corresponding color factor is a single-trace of gauge-group generators. In a  parametrization of the non-empty cylinder boundary via purely imaginary coordinates $z_i$ with $0 \leq \Im (z_i) \leq \tau_2$, the universal $n$-point open-string integration measure will be denoted by 
 \beq
\int \dd \mu_{12\ldots n}^D \equiv {V_D \over 8N} \int^{\infty}_0 \frac{ \dd \tau_2 }{(8 \pi^2 \alpha' \tau_2)^{D/2}} \! \! \! \! \!  \int \limits_{0 \leq \Im(z_1) \leq \Im(z_2) \leq \ldots \leq \Im(z_n) \leq \tau_2} \! \! \! \! \! \! \! \! \! \! \! \! \! \! \!  \dd z_1 \, \dd z_2 \, \ldots \, \dd z_n \, \delta(z_1) \Pi_n \ .
\label{openmeas}
\eeq 
We have incorporated the regularized external volume $V_D$, the order $N$ of the orbifold ${\mathbb Z}_N$ as well as the ubiquitous Koba--Nielsen factor $\Pi_n$ of eq.\ \eqref{red23} below, which arises from the plane-wave factors
of the vertex operators, see section \ref{sec:fiveone}.
The measure (\ref{openmeas}) with modular parameter $\tau_2$ can straightforwardly be  adjusted to the remaining worldsheet topologies, and the delta-function $\delta(z_1)$ fixes the translation invariance of genus-one surfaces, by fixing one puncture to the origin. The number $D$ of uncompactified spacetime dimensions is denoted as a superscript of $\dd \mu_{12\ldots n}^D$, and the subscript $12\ldots n$ refers to the cyclic ordering of the open-string states along the boundary as well as the trace-ordering of the accompanying color factor.

\subsubsection{Half-maximal supersymmetry}
\label{sec:halfmaximalsusy}

If one of the twist vector entries vanishes but the other
two are nonzero, say $v_3=0$ and therefore $v_1 = - v_2$ as in table \ref{orbiexample}, the orbifold only breaks half of the supersymmetries.
These orbifolds can be characterized by a single rational real number $v$ that enters the 
partition functions through the vector $\vec{v}_k \equiv k(v,-v)$.
For brevity we will mostly discuss half-maximally supersymmetric
orbifolds in their maximal spacetime dimension $D=6$,
i.e.\  arising from compactification from $D=10$ on $T^4/{\mathbb Z}_N$, which 
are special points in the moduli space of K3 manifolds. The 1-loop amplitude of $n$ gauge bosons in this setting is given by 
(for textbook examples, see e.g.\ \cite{Blumenhagen:2013fgp})
\be   
\label{AN2}
{\cal A}_{1/2}(1,2,\ldots,n) = \int \dd \mu_{12\ldots n}^{D=6} \, \left\{ \Gamma_{\cal C}^{(4)} c_0 \,{\cal I}_{n,\te{max}}\, + \, 
\sum_{k=1}^{N-1} c_k\, \hat \chi_k  \, {\cal I}_{n,1/2}(\vec{v}_k) 
\right\} \ ,
\ee 
where the subscript ``1/2'' means ``half-maximal'', $\Gamma_{\cal C}^{(n)}$ denotes lattice sums over $n$-dimensional internal momenta, $c_0$ and $c_k$ are model-dependent constants determined by the action of the orbifold group on the Chan--Paton factors\!
\footnote{In toy models with just one gauge group, $c_k=(\!\text{ tr}\, \gamma_k)^2$, cf. appendix \ref{opf}. In models with more than one gauge group the traces are over sub-blocks of the matrices $\gamma_k$. Explicit expressions are given in the companion paper \cite{FT}.},
and the generalities of the constants $\hat \chi_k=-[\sin(\pi k v)/\pi]^2$ are explained in appendix \ref{opf}. The external-state information is encoded in the integrands ${\cal I}_{\ldots}$ whose dependence on the integration variables $\tau_2$ and $z_i$ of the measure (\ref{openmeas}) will usually be suppressed. The subscripts ``$\te{max}$'' or ``$1/2$''  distinguish orbifold sectors that preserve all or half the supersymmetries, respectively. 
While the maximally supersymmetric integrand is parity-even\footnote{In general, for amplitudes of solely external states, like amplitudes of gauge bosons in $D=6$, there is never a parity-odd contribution to the maximally supersymmetric integrand. With only external excitation it is impossible to saturate the fermionic zero modes along the internal directions.}, the half-maximal integrand receives both parity-even and parity-odd contributions labelled by superscripts $e$ and $o$. We write 
\beq
 {\cal I}_{n,1/2} (\vec{v}_k)  \equiv{\cal I}^e_{n,1/2}(\vec{v}_k) \, + \, {\cal I}^{o}_{n,D=6} \,,
 \label{combine}
\eeq
where $\vec{v}_k$ highlights the dependence of the parity-even contribution on non-trivial orbifold sectors, i.e.\ on the internal partition function. The dependence of parity-odd integrands on orbifold twists $\vec{v}_k$ cancels between the contributions  to the partition function due to worldsheet bosons and worldsheet fermions in the odd spin structure. Explicitly, we have
\begin{align}
{\cal I}_{n,\te{max}} &\equiv  \frac{1}{\Pi_n} \sum_{\nu=2}^{4} (-1)^{\nu-1} \left[ \frac{\theta_{\nu}(0,\tau)  }{\theta'_1(0,\tau)} \right]^4  \, \langle V_1^{(0)}(z_1) V_2^{(0)}(z_2) \, \ldots \, V_n^{(0)}(z_n) \rangle_\nu
\label{IN4} \\
{\cal I}^e_{n,1/2}(\vec{v}_k) &\equiv \frac{1}{\Pi_n}  \sum_{\nu=2}^{4} (-1)^{\nu} \left[ \frac{\theta_{\nu}(0,\tau)  }{\theta'_1(0,\tau)} \right]^2
\left[ \frac{\theta_{\nu}(kv,\tau)  }{\theta_1(kv,\tau)} \right]^2  \, \langle V_1^{(0)}(z_1) V_2^{(0)}(z_2) \, \ldots \, V_n^{(0)}(z_n) \rangle_\nu \ ,
\label{IN2}  
\end{align}
where the second argument of the $\theta$-functions is the purely imaginary $\tau= i\tau_2$ for the planar cylinder under consideration. The inverse of the Koba--Nielsen factor $\Pi_n$ in \eqref{red23} compensates for its inclusion in the measure (\ref{openmeas}) and facilitates  bookkeeping in later sections. Here $\nu=2,3,4$ are the even spin structures of the RNS worldsheet spinors, and the standard explicit
form of the vertex operators $V_j^{(0)}$ for gauge-bosons will be written down in (\ref{red21}). The maximally supersymmetric integrand (\ref{IN4}) has been discussed in many places of the literature including \cite{Green:1982sw, Tsuchiya:1988va, Stieberger:2002wk} and can be obtained from pure spinor computations such as \cite{Berkovits:2004px, Mafra:2012kh, maxsusy} upon dimensional reduction. For the $(n\leq 4)$-point amplitudes under discussions, the result is \cite{Green:1982sw}
\beq
{\cal I}_{n,\te{max}} = 0 \ , \ \te{if} \ n\leq 3 \ , \ \ \ \ \ \ 
{\cal I}_{4,\te{max}} = - 2 t_8(1,2,3,4) \ ,
\label{maxint}
\eeq
see (\ref{red42}) and (\ref{red43}) for the $t_8$ tensor.

The parity-odd part of the integrand
\beq 
{\cal I}^{o}_{n,D} \equiv  \frac{1}{\Pi_n} \langle P^{(+1)}(z_0) V_1^{(-1)}(z_1) V_2^{(0)}(z_2) \, \ldots \, V_n^{(0)}(z_n) \rangle_{\nu=1}^{D} \ ,
\label{IN1O}
\eeq
uses the picture changing operator $P^{(+1)}$ and the vertex operator $V_1^{(-1)}$ in the $-1$ superghost picture, see (\ref{oddstuff}). They
are required for zero-mode saturation in the superghost sector. 
The integrand
only receives contributions from the odd spin structure $\nu=1$, and the path integral over worldsheet spinors requires $D$ zero-mode components to be saturated according to  \cite{Polchinski:1998rr,Green:1987mn}
\beq
\psi^{m_1} \psi^{m_2}\ldots \psi^{m_D} \rightarrow i\epsilon^{m_1 m_2 \ldots m_D} 
\label{zeropsi}
\eeq
with the $D$-dimensional Levi-Civita symbol on the right-hand side. The dependence on the position $z_0$ of $P^{(+1)}$ drops out on kinematic grounds, 
as expected from general arguments (see e.g.\ \cite{Verlinde:1988tx,Polchinski:1998rr}),
which we check in detail in appendix \ref{app:dim}. Note that the expression  \eqref{AN2} for half-maximal amplitudes in $D=6$ straightforwardly generalizes
to $D=4$, i.e.\ compactification on K3$\times T^2$ instead of just K3,
\be
\label{AN3}
{\cal A}_{1/2}^{D=4} (1,2,\ldots,n) = \int \dd \mu_{12\ldots n}^{D=4} \, \left\{ \Gamma_{\cal C}^{(6)} c_0\, {\cal I}_{n,\te{max}}\, + \,\Gamma_{\cal C}^{(2)} \sum_{k=1}^{N-1} c_k \,\hat \chi_k\, {\cal I}^e_{n,1/2}(\vec{v}_k)
\right\} \ .
\ee 
Similarly to the maximally supersymmetric integrand, when there are  internal directions
that are unaffected by the orbifold rotation, the parity-odd contribution vanishes for external excitations.
To save writing, we will mainly give $D=6$ expressions, but the point here was to illustrate that the extrapolation is trivial, before performing $\tau$ integrals.

\subsubsection{Quarter-maximal supersymmetry}

A similar prescription applies to orbifolds with quarter-maximal supersymmetry which we will discuss in their maximal spacetime dimension $D=4$. The quarter-maximal counterpart of (\ref{AN2}),
\be
\label{AN1}
{\cal A}_{1/4}(1,2,\ldots,n) = \int \dd \mu_{12\ldots n}^{D=4} \left\{\! \Gamma_{\cal C}^{(6)} c_0\, {\cal I}_{n,\te{max}}-\Gamma_{\cal C}^{(2)} \!\!\sum_{\exists^1\, kv_j \in {\mathbb Z}}\!\! c_k\,\hat \chi_k\, {\cal I}^e_{n,1/2}(\vec{v}_k)
 +\! \sum_{kv_j \notin {\mathbb Z}}\! c_k\,\hat \chi_k \, {\cal I}_{n,1/4}(\vec{v}_k)\!
\right\} \ ,
\ee
contains two kinds of lattice sums $\Gamma_{\cal C}^{(6)}, \Gamma_{\cal C}^{(2)}$, and the twist vector is $\vec{v}_k=k(v_1,v_2,v_3)$ with $v_1+v_2+v_3=0$.
Half-maximal contributions arise in orbifold models where one of the three internal tori is fixed under the action of some orbifold sectors, e.g.\ if $\Theta^k Z_3= Z_3$ for some $k$ (i.e.\ $kv_3\in \mathbb{Z}$). In this case $kv_1= kv_2$, and ${\cal I}^e_{n,1/2}(\vec{v}_k)$ is determined by (\ref{IN2}) with $v \rightarrow v_1$.

The quarter-maximal integrand contains  parity-even and parity-odd contributions,
\beq 
{\cal I}_{n,1/4}(\vec{v}_k) \equiv{\cal I}^e_{n,1/4}(\vec{v}_k) \, + \, {\cal I}^{o}_{n,D=4}\,,
\eeq
where ${\cal I}^{o}_{n,D=4}$ is a special case of (\ref{IN1O}), and the parity-even part
\beq
{\cal I}^e_{n,1/4}(\vec{v}_k) \equiv  \frac{1}{\Pi_n} \sum_{\nu=2}^{4} (-1)^{\nu-1}  \frac{\theta_{\nu}(0,\tau)  }{\theta'_1(0,\tau)} 
\left[ \prod_{j=1}^3 \frac{\theta_{\nu}(kv_j,\tau)  }{\theta_1(kv_j,\tau)} \right]  \, \langle V_1^{(0)}(z_1) V_2^{(0)}(z_2) \, \ldots \, V_n^{(0)}(z_n) \rangle_\nu 
\label{IN1} 
\eeq
is understood to depend on $kv_j \,\notin \mathbb{Z}$ for all  $j=1,2,3$. 

\subsection{Spin sums}
\label{sec:fourtwo}

A major challenge in the evaluation of string amplitudes with half- and quarter-maximal supersymmetry is to perform the spin sums in the parity-even integrands (\ref{IN2}) and (\ref{IN1}). As elaborated in section \ref{sec:fiveone}, the worldsheet spinors in vertex operators $V_i^{(0)}$ cause the correlators  to depend on the spin structure $\nu$ through their two-point function, the Szeg\"o kernel
\beq
S_\nu(z,\tau)  \equiv \frac{\theta_1'(0,\tau) \theta_\nu(z,\tau)}{\theta_\nu(0,\tau)\theta_1(z,\tau)} \ .
\label{red15}
\eeq 
Individual sectors with $\nu=2,3,4$ contain spurious worldsheet singularities that
cancel upon summation, as a consequence of supersymmetry. 
Such spurious singularities are an inconvenient feature of the RNS formalism, and their cancellation in maximally supersymmetric cases can be manifested through the techniques of \cite{Tsuchiya:1988va, Broedel:2014vla}. In this section, we will demonstrate that the method of the references can be adapted to address situations with reduced supersymmetry as well.

\subsubsection{Worldsheet functions}

We follow the notation of \cite{Broedel:2014vla} where a doubly-periodic function  $f^{(n)}$ for each non-negative integer $n$ is defined by a non-holomorphic Kronecker-Eisenstein series
\beq
\Omega(z,\alpha,\tau) \equiv \exp\Big(2\pi i \alpha \frac{ \Im z}{\Im \tau} \Big) \frac{ \theta_1'(0,\tau) \theta_1(z+\alpha,\tau) }{\theta_1(z,\tau)\theta_1(\alpha,\tau)} 
\equiv
\sum_{n=0}^{\infty} \alpha^{n-1} f^{(n)}(z,\tau) \ ,
\label{red11}
\eeq
starting with
\begin{align}
f^{(0)}(z,\tau) &\equiv 1 \co f^{(1)}(z,\tau) =   \partial \ln \theta_1(z,\tau) +  2\pi i \, \frac{ \Im (z) }{\Im (\tau)}
\label{red12}
\\
f^{(2)}(z,\tau) & \equiv \frac{1}{2} \Big\{  \Big( \partial \ln \theta_1(z,\tau) +  2\pi i \, \frac{ \Im (z) }{\Im (\tau)} \Big)^2 + \partial^2 \ln \theta_1(z,\tau) - \frac{\theta_1'''(0,\tau)}{ 3\theta_1'(0,\tau)} \Big\} \ .
\label{red13}
\end{align}
Note that $f^{(1)}$ is the only singular term of (\ref{red11}) with a simple pole at the origin as well as its translations $z=n+m\tau$ with $m,n\in \ZZ$. For ease of notation, the dependence on the modular parameter $\tau$ will be suppressed in the following. 

We note in passing that $\Omega(z,\alpha,\tau)$ is closely related
to the twisted fermion Green's function, which is in turn
a nonholomorphic Eisenstein-Kronecker function $E_s^{(k)}(w,z,\tau)$, as
discussed for example in \cite{Berg:2014ama}.
We note that there, $w$ has direct interpretation as a twist of external orbifold-charged states,
while here $\alpha$ is a formal expansion parameter. 

\subsubsection{Maximal supersymmetry}

After pairwise contractions of the worldsheet spinors to Szeg\"o kernels (\ref{red15}) via Wick's theorem, RNS amplitudes with maximal supersymmetry give rise to the spin sum
\beq
{\cal G}_n(x_1,x_2,\ldots,x_n) \equiv \sum_{\nu=2,3,4} (-1)^{\nu-1} \left[ \frac{ \theta_\nu(0) }{\theta_1'(0)} \right]^4 S_\nu(x_1) S_\nu(x_2)\ldots S_\nu(x_n) \ , \ \ \ \ \ \sum_{i=1}^{n} x_i = 0 \ .
\label{red14}
\eeq
An efficient method to evaluate (\ref{red14}) and to make its pole structure manifest  was introduced in \cite{Tsuchiya:1988va} (also see \cite{Stieberger:2002wk} for a variation). The functions $f^{(n)}$ in (\ref{red11}) allow to streamline the results as  \cite{Broedel:2014vla}
\begin{align}
{\cal G}_n(x_1,x_2,\ldots,x_n) &= 0 \co n \leq 3 
\label{red16} \\
{\cal G}_4(x_1,x_2,x_3,x_4) &= 1
\label{red17} \\
{\cal G}_5(x_1,x_2,\ldots,x_5) &= \sum_{j=1}^5 f^{(1)}_j 
\label{red18} \\
{\cal G}_6(x_1,x_2,\ldots,x_6) &= \sum_{j=1}^6 f^{(2)}_j +\sum_{1\leq j < k}^6 f^{(1)}_j f^{(1)}_k
\label{red19} \\
{\cal G}_7(x_1,x_2,\ldots,x_7) &=  \sum_{j=1}^7 f^{(3)}_j +\sum_{1\leq j < k}^7 ( f^{(2)}_j f^{(1)}_k+f^{(1)}_j f^{(2)}_k)+ \sum_{1\leq j < k<l}^7 f^{(1)}_j f^{(1)}_k f^{(1)}_l
\label{red110} \\
{\cal G}_8(x_1,x_2,\ldots,x_8) &= \sum_{j=1}^8 f^{(4)}_j  +\sum_{1\leq j < k}^8 ( f^{(3)}_j f^{(1)}_k+f^{(2)}_j f^{(2)}_k+f^{(1)}_j f^{(3)}_k)  +\!\!\!\! \sum_{1\leq j < k<l<m}^8 \!\!\!\!  f^{(1)}_j f^{(1)}_k f^{(1)}_l f^{(1)}_m\notag \\
& \ \ \ \ +\sum_{1\leq j < k<l}^8 (f^{(2)}_j f^{(1)}_k f^{(1)}_l +f^{(1)}_j f^{(2)}_k f^{(1)}_l +f^{(1)}_j f^{(1)}_k f^{(2)}_l )  + 3 G_4 \ ,
\label{red111} 
\end{align}
using the shorthand $f^{(n)}_i \equiv f^{(n)}(x_i)$. The appearance of the holomorphic Eisenstein series $G_4$ as an extra constant in (\ref{red111}) generalizes in a pattern described in \cite{Tsuchiya:1988va,Broedel:2014vla}, and
see also an alternative method in \cite{Dolan:2007eh}. The associated $x_j$-dependence in ${\cal G}_{N\geq 9}$ can be cast into a convenient form through the notation
\begin{align}
V_1(x_1,x_2,\ldots,x_n) &\equiv \sum_{j=1}^n f^{(1)}_j  \co V_2(x_1,x_2,\ldots,x_n) \equiv \sum_{j=1}^n f^{(2)}_j +\sum_{1\leq j < k}^n f^{(1)}_j f^{(1)}_k \label{red112} \\
V_3(x_1,x_2,\ldots,x_n) &\equiv
 \sum_{j=1}^n f^{(3)}_j +\sum_{1\leq j < k}^n ( f^{(2)}_j f^{(1)}_k+f^{(1)}_j f^{(2)}_k)+ \sum_{1\leq j < k<l}^n f^{(1)}_j f^{(1)}_k f^{(1)}_l \label{red113} \\
V_4(x_1,x_2,\ldots,x_n) &\equiv \sum_{j=1}^n f^{(4)}_j  +\sum_{1\leq j < k}^n ( f^{(3)}_j f^{(1)}_k+f^{(2)}_j f^{(2)}_k+f^{(1)}_j f^{(3)}_k)  +\!\!\!\! \sum_{1\leq j < k<l<m}^n \!\!\!\!  f^{(1)}_j f^{(1)}_k f^{(1)}_l f^{(1)}_m\notag \\
& \ \ \ \ +\sum_{1\leq j < k<l}^n (f^{(2)}_j f^{(1)}_k f^{(1)}_l +f^{(1)}_j f^{(2)}_k f^{(1)}_l +f^{(1)}_j f^{(1)}_k f^{(2)}_l ) \ . 
\label{red114} 
\end{align}
A general definition can be compactly given in terms of the generating series $\Omega(z,\alpha)$ in (\ref{red11}),
\begin{align}
V_w(x_1,x_2,\ldots,x_n)\equiv \alpha^n \Omega(x_1,\alpha)\Omega(x_2,\alpha)\ldots \Omega(x_n,\alpha) \big|_{\alpha^w} \ .
\label{red114b}
\end{align}
The virtue of the functions $V_w$ to express ${\cal G}_{n}$ at higher multiplicity is exemplified by \cite{Broedel:2014vla}
\begin{align}
{\cal G}_n(x_1,x_2,\ldots,x_n) &= V_{n-4}(x_1,x_2,\ldots,x_n) \co 4\leq n \leq 7 
\label{red114a} \\
{\cal G}_8(x_1,x_2,\ldots,x_8) &=
V_{4}(x_1,x_2,\ldots,x_8) + 3 G_4
\label{red115} \\
{\cal G}_9(x_1,x_2,\ldots,x_9) &= V_{5}(x_1,x_2,\ldots,x_9) + 3 G_4 V_{1}(x_1,x_2,\ldots,x_9) 
\label{red116} \\
{\cal G}_{10}(x_1,x_2,\ldots,x_{10}) &= V_{6}(x_1,x_2,\ldots,x_{10}) + 3 G_4 V_{2}(x_1,x_2,\ldots,x_{10})  + 10 G_6 \ .
\label{red117} 
\end{align}
We see that without resorting to specific
Riemann identities for large numbers of theta functions, these results
let us write relatively compact expressions
for integrands up to at least 10 external states
without too much effort, incorporating the cancellations mentioned above.

\subsubsection{Reduced supersymmetry}
\label{sec:reduced}

The results in the maximally supersymmetric sector that we  reviewed above will now be extended to  the most general spin sum in half-maximal and quarter-maximal amplitudes (\ref{AN2}) and (\ref{AN1}). The key idea is 
to rewrite the orbifold-twisted partition functions (which reflect reduced supersymmetry)
in terms of fermion Green's functions with the twist as an insertion (which 
``uses up'' additional external states). To this end, we
rewrite (\ref{IN2}) and (\ref{IN1}) 
by pulling out a factor like that of the maximal case \eqref{red14} by hand:
\begin{align}
{\cal I}^e_{n,1/2}(\vec{v}_k) &= \frac{1}{\Pi_n}  \sum_{\nu=2}^{4} (-1)^{\nu-1} \left[ \frac{\theta_{\nu}(0)  }{\theta'_1(0)} \right]^4
S_\nu(kv) S_\nu(-kv) \, \langle V_1^{(0)}(z_1) V_2^{(0)} (z_2)\, \ldots \, V_n^{(0)}(z_n) \rangle_\nu
\label{IN2p} \\
{\cal I}^e_{n,1/4}(\vec{v}_k)&=\frac{1}{\Pi_n} 
\sum_{\nu=2}^{4} (-1)^{\nu-1} \left[ \frac{\theta_{\nu}(0)  }{\theta'_1(0)} \right]^4
\left[ \prod_{j=1}^3 S_\nu(kv_j) \right]\, \langle V_1^{(0)}(z_1) V_2^{(0)}(z_2) \, \ldots \, V_n^{(0)}(z_n) \rangle_\nu \ ,
\label{IN1p}
\end{align}
using the definition (\ref{red15}) of the Szeg\"o kernel. The correlators of $V_i^{(0)}$ yield the same cycles of two-point contractions $S_\nu(x_1)S_\nu(x_2)\ldots S_\nu(x_n)$ with $\sum_{i=1}^{n}x_i=0$ as seen in the maximal case. Hence, the most general spin sum resulting from (\ref{IN2p}) and (\ref{IN1p}), respectively, is given by
\begin{align}
\sum_{\nu=2}^{4} (-1)^{\nu-1} \left[ \frac{\theta_{\nu}(0)  }{\theta'_1(0)} \right]^4
S_\nu(\gamma) S_\nu(-\gamma) \, S_\nu(x_1)S_\nu(x_2) \ldots S_\nu(x_n) &= {\cal G}_{n+2}(x_1,x_2,\ldots,x_n,\gamma,-\gamma)
\label{IN2x} \\
\sum_{\nu=2}^{4} (-1)^{\nu-1} \left[ \frac{\theta_{\nu}(0)  }{\theta'_1(0)} \right]^4
\left[ \prod_{j=1}^3 S_\nu(\gamma_j) \right]S_\nu(x_1) S_\nu(x_2)\ldots S_\nu(x_n) &={\cal G}_{n+3}(x_1,x_2,\ldots,x_n,\gamma_1,\gamma_2,\gamma_3)  \ .
\label{IN1x}
\end{align}
In order to avoid proliferation of factors $k$, we introduce the shorthands
\beq
\gamma \equiv kv \co  \gamma_j \equiv kv_j \co \gamma_1+\gamma_2+\gamma_3=0 \ 
\label{orbtw}
\eeq
for the orbifold twists.
The expressions can be identified with the prototype spin sum (\ref{red14}) from the maximal case by viewing $\gamma,-\gamma$ as $x_{n+1},x_{n+2}$ and $\gamma_1,\gamma_2,\gamma_3$ as $x_{n+1},x_{n+2},x_{n+3}$, respectively. They preserve the requirement on the $x_j$ to sum to zero, and they additionally imply that subsets of the arguments in the enlarged ${\cal G}_{n+2}$ and ${\cal G}_{n+3}$ add up to zero. As a convenient way to explore the resulting cancellations, we rewrite the expressions in (\ref{IN2x}) and (\ref{IN1x}) such as to manifest the symmetries $S_\nu(-x)=-S_\nu(x)$ of Szeg\"o kernels, and exploit $f^{(n)}(-x) = (-1)^n f^{(n)}(x)$:
\begin{align}
{\cal G}_{n+2}&(\gamma,-\gamma,x_1,x_2,\ldots,x_n) = \frac{1}{4}\big[ {\cal G}_{n+2}(\gamma,-\gamma,x_1,x_2,\ldots,x_n) + {\cal G}_{n+2}(-\gamma,\gamma,x_1,x_2,\ldots,x_n)\notag \\
&\!\!\!\!\! \!\!\!\!\! +
(-1)^n{\cal G}_{n+2}(\gamma,-\gamma,-x_1,-x_2,\ldots,-x_n) + (-1)^n {\cal G}_{n+2}(-\gamma,\gamma,-x_1,-x_2,\ldots,-x_n) \big]  \label{red120}\\
{\cal G}_{n+3}&(\gamma_1,\gamma_2,\gamma_3,x_1,\ldots,x_n) = \frac{1}{4} \big[ {\cal G}_{n+3}(\gamma_1,\gamma_2,\gamma_3,x_1,\ldots,x_n) - {\cal G}_{n+3}(-\gamma_1,-\gamma_2,-\gamma_3,x_1,\ldots,x_n)\notag \\
&\!\!\!\!\! \!\!\!\!\! +
(-1)^n{\cal G}_{n+3}(\gamma_1,\gamma_2,\gamma_3,-x_1,\ldots,-x_n) - (-1)^n {\cal G}_{n+3}(-\gamma_1,-\gamma_2,-\gamma_3,-x_1,\ldots,-x_n) \big] \ .
 \label{red121}
\end{align}
As a result, the $\gamma$-dependence in the half-maximal (\ref{red120}) conspires to functions of even modular weight,
\begin{align}
F_{1/2}^{(0)}(\gamma) &\equiv 1 \co F_{1/2}^{(2)}(\gamma) \equiv 2 f^{(2)}(\gamma) -f^{(1)}(\gamma)^2 
\label{red122}
\\
F_{1/2}^{(4)}(\gamma) &\equiv 2 f^{(4)}(\gamma) - 2 f^{(3)}(\gamma) f^{(1)}(\gamma)
+ f^{(2)}(\gamma)^2 \ .  \label{red123}
\end{align}
In fact, all the $F_{1/2}^{(k)}(\gamma)$ past $k=2$ will be identified below as independent of $\gamma$, but we will keep the generic notation $F_{1/2}^{(k)}(\gamma)$  to emphasize similarities to the quarter-maximal case.

The analogous manipulations in the quarter-maximal case (\ref{red121}) only admit odd modular weight for the dependence on $\gamma_j$,
\begin{align}
F_{1/4}^{(1)}(\gamma_j) &\equiv  f^{(1)}(\gamma_1) + f^{(1)}(\gamma_2) + f^{(1)}(\gamma_3) 
\label{red124}
\\
F_{1/4}^{(3)}(\gamma_j) &\equiv f^{(1)}(\gamma_1) f^{(1)}(\gamma_2) f^{(1)}(\gamma_3) + f^{(3)}(\gamma_1) + f^{(3)}(\gamma_2) + f^{(3)}(\gamma_3) \notag \\
& \ \ \ + \sum_{1\leq i<j}^3 (f^{(1)}(\gamma_i)  f^{(2)}(\gamma_j)  + f^{(2)}(\gamma_i)  f^{(1)}(\gamma_j) )
  \ .  \label{red125}
\end{align}
More generally, the $\gamma$-dependence in the results (\ref{red120}) and (\ref{red121}) is organized in terms of $V_n(\ldots)$ from \eqref{red114b} above:
\beq
F_{1/2}^{(n)}(\gamma) \equiv V_n(\gamma,-\gamma) \co F_{1/4}^{(n)}(\gamma_j) \equiv V_n(\gamma_1,\gamma_2,\gamma_3)
\label{red126}
\eeq
with appropriate parity for $n$. With these definitions and the functions $V_n(x_1,\ldots,x_n)$ of worldsheet positions in (\ref{red112}) to (\ref{red114b}), the spin sums for reduced supersymmetry can be evaluated as
\begin{align}
{\cal G}_{2+2}(\gamma,-\gamma,x_1,x_2) &= 1
\label{red127a} \\
{\cal G}_{2+3}(\gamma,-\gamma,x_1,x_2,x_3) &= V_1(x_1,x_2,x_3) = f^{(1)}(x_1)+f^{(1)}(x_2)+f^{(1)}(x_3)
\label{red127b} \\
{\cal G}_{2+4}(\gamma,-\gamma,x_1,\ldots,x_4) &= F_{1/2}^{(2)}(\gamma) +  V_2(x_1,\ldots,x_4) 
\label{red127c} \\
{\cal G}_{2+5}(\gamma,-\gamma,x_1,\ldots,x_5) &= F_{1/2}^{(2)}(\gamma) V_1(x_1,\ldots,x_5)  +  V_3(x_1,\ldots,x_5) 
\label{red127d} \\
{\cal G}_{2+6}(\gamma,-\gamma,x_1,\ldots,x_6) &= F_{1/2}^{(4)}(\gamma)  + 3G_4+ F_{1/2}^{(2)}(\gamma)  V_2(x_1,\ldots,x_6) + V_4(x_1,\ldots,x_6)
\label{red127e} \\
{\cal G}_{2+7}(\gamma,-\gamma,x_1,\ldots,x_7) &=  (F_{1/2}^{(4)}(\gamma)+3G_4)  V_1(x_1,\ldots,x_7) \notag \\
& \ \ + F_{1/2}^{(2)}(\gamma)  V_3(x_1,\ldots,x_7) +  V_5(x_1,\ldots,x_7)
\label{red127f} \\
{\cal G}_{2+8}(\gamma,-\gamma,x_1,\ldots,x_8) &= F_{1/2}^{(6)}(\gamma) + 10 G_6 + F_{1/2}^{(4)}(\gamma) V_2(x_1,\ldots,x_8) +  F_{1/2}^{(2)}(\gamma) V_4(x_1,\ldots,x_8)\notag \\
& \ \   + 3 G_4 ( F_{1/2}^{(2)}(\gamma) + V_2(x_1,\ldots,x_8) )  + V_6(x_1,\ldots,x_8) \ ,
\label{red127g} 
\end{align}
which suffices for eight-point amplitudes in half-maximal compactifications. 
Comparing to results derived by standard methods, the first
three are well-known:  ${\cal G}_{2+2}$ comes
from two fermion bilinears
after so-called ``spin sum collapse'', using
a standard theta function identity whose proof is outlined for example in \cite{Berg:2004ek},
eqs.\ (120) to (132). To obtain ${\cal G}_{2+3}$ from three fermion bilinears,
one adapts a calculation from \cite{Bianchi:2006nf}, in particular their eq.\ (3.37) that reads
$S_{\nu}(x_{13})S_{\nu}(x_{23})=S_{\nu}(x_{12})V_1(x_1,x_2,x_3)+\partial_x S_{\nu}(x_{12})
$. While similar methods were used in \cite{Stieberger:2002wk,Bianchi:2006nf} to determine ${\cal G}_{2+4}$, 
we are not aware of explicit results for ${\cal G}_{2+n}$ with $n\geq 5$ in the literature. 

The quarter-maximal analogues of (\ref{red127a}) to (\ref{red127g}) sufficient for seven-point amplitudes are given by
\begin{align}
%
{\cal G}_{3+2}(\gamma_1,\gamma_2,\gamma_3,x_1,x_2) &= F_{1/4}^{(1)}(\gamma_j)
\label{red128b} \\
{\cal G}_{3+3}(\gamma_1,\gamma_2,\gamma_3,x_1,\ldots,x_3) &= F_{1/4}^{(1)}(\gamma_j)   V_1(x_1,x_2,x_3) 
\label{red128c} \\
{\cal G}_{3+4}(\gamma_1,\gamma_2,\gamma_3,x_1,\ldots,x_4) &= F_{1/4}^{(3)}(\gamma_j)   +  F_{1/4}^{(1)}(\gamma_j) V_2(x_1,\ldots,x_4) 
\label{red128d} \\
{\cal G}_{3+5}(\gamma_1,\gamma_2,\gamma_3,x_1,\ldots,x_5) &=  F_{1/4}^{(3)}(\gamma_j)  V_1(x_1,\ldots,x_5) +  F_{1/4}^{(1)}(\gamma_j) V_3(x_1,\ldots,x_5)
\label{red128e} \\
{\cal G}_{3+6}(\gamma_1,\gamma_2,\gamma_3,x_1,\ldots,x_6) &=  F_{1/4}^{(5)}(\gamma_j) + 3 F_{1/4}^{(1)}(\gamma_j) G_4 + F_{1/4}^{(3)}(\gamma_j)  V_2(x_1,\ldots,x_6) \notag \\
&\ \ +   F_{1/4}^{(1)}(\gamma_j) V_4(x_1,\ldots,x_6) 
\label{red128f} \\
{\cal G}_{3+7}(\gamma_1,\gamma_2,\gamma_3,,x_1,\ldots,x_7) &= (F_{1/4}^{(5)}(\gamma_j)  + 3F_{1/4}^{(1)}(\gamma_j) G_4  )V_1(x_1,\ldots,x_7)    \notag \\
& \ \ + F_{1/4}^{(3)}(\gamma_j)  V_3(x_1,\ldots,x_7) +  F_{1/4}^{(1)}(\gamma_j) V_5(x_1,\ldots,x_7)   \ .
\label{red128g} 
\end{align}
With standard methods, ${\cal G}_{3+2}$ has been computed in the spin sum
for two fermion bilinears, and 
the proof of the required spin sum  identity is outlined for example in \cite{Berg:2004ek},
eq.\ (130).
Note that $F_{1/4}^{(1)}$ in ${\cal G}_{3+2}$ is reminiscient of $V_1$ above
but is independent of $x_i$, just
like for the two-fermion-bilinear piece in the half-maximal case. 
With three and four fermion bilinears, computations in
\cite{Stieberger:2002wk,Bianchi:2006nf} can be adapted to yield
${\cal G}_{3+3}$ and ${\cal G}_{3+4}$ above, but starting from ${\cal G}_{3+5}$ we believe
the results are new.

In addition to new explicit results, we emphasize the general applicability
of this method. As an example, the following observation
would be difficult to make without our strategy. 
For $n\geq 2$, the structure of $V_k(x_1,\ldots,x_n)$ is obviously identical in the above expressions for ${\cal G}_{2+n}$ and ${\cal G}_{3+n}$. If $1=F_{1/2}^{(0)}$ is inserted in each term of ${\cal G}_{2+n}$ without an extra factor of $F_{1/2}^{(k\neq 0)}$, the correspondence between (\ref{red127a}) to (\ref{red127f}) and (\ref{red128b}) to (\ref{red128g}) can be summarized by
\beq
{\cal G}_{3+n}(\gamma_1,\gamma_2,\gamma_3,,x_1,\ldots,x_n) = 
{\cal G}_{2+n}(\gamma,-\gamma,,x_1,\ldots,x_n) 
\Big|_{
F_{1/2}^{(k)}(\gamma)\rightarrow F_{1/4}^{(k+1)}(\gamma_j)} \ .
\label{promote}
\eeq
Hence, the resulting scattering amplitudes in half-maximal and quarter-maximal compactifications have the same structure in the parity-even sector, i.e.\ their integrands can be straightforwardly mapped into each other upon replacing $F_{1/2}^{(k)}(\gamma)\rightarrow F_{1/4}^{(k+1)}(\gamma_j)$. However, the parity-odd contributions to half-maximal and quarter-maximal cases will exhibit differences as we will comment on in sections \ref{sec:fiveseven}, \ref{sec:fiveeight} and \ref{sec:sevensix}.

As noted above, all the $F_{1/2}^{(k)}(\gamma)$ past $k=2$ turn out to be independent of $\gamma$. In 
fact they are given by holomorphic Eisenstein series:
\beq
F_{1/2}^{(k)}(\gamma) = (k-1) G_k \co k=4,6,8,\ldots \ ,
\label{simp127}
\eeq
and then (\ref{red127e}) to (\ref{red127g}) can be further simplified to
\begin{align}
{\cal G}_{2+6}(\gamma,-\gamma,x_1,\ldots,x_6) &= 6G_4+ F_{1/2}^{(2)}(\gamma)  V_2(x_1,\ldots,x_6) + V_4(x_1,\ldots,x_6)
\notag \\
{\cal G}_{2+7}(\gamma,-\gamma,x_1,\ldots,x_7) &= 6G_4  V_1(x_1,\ldots,x_7)  + F_{1/2}^{(2)}(\gamma)  V_3(x_1,\ldots,x_7) +  V_5(x_1,\ldots,x_7)
\label{simpred127f} \\
{\cal G}_{2+8}(\gamma,-\gamma,x_1,\ldots,x_8) &= 15 G_6 + 6 G_4 V_2(x_1,\ldots,x_8) +  F_{1/2}^{(2)}(\gamma) (V_4(x_1,\ldots,x_8) + 3 G_4 )  + V_6(x_1,\ldots,x_8) \ . \notag
\end{align}
It would be interesting to find analogous simplifications in the quarter-maximal case.


\subsection{Closed-string prescriptions}
\label{sec:fourthree}

In this section, we recall the starting point for 1-loop closed-string amplitudes with non-maximal supersymmetry, specifically we consider half-maximal and quarter-maximal compactifications of type IIA and type IIB theories. Similar to the open-string integration measure (\ref{openmeas}), we capture the integration over inequivalent worldsheets of torus topology by the closed-string measure 
\beq
\int \dd \rho_n^D \equiv   {V_D \over 8N}  \int_{{\cal F}} \frac{ \dd^2 \tau }{(4\pi^2\alpha' \tau_2)^{D/2}} \int_{{\cal T}(\tau)^n} \dd^2 z_1 \, \dd^2 z_2 \, \ldots \, \dd^2 z_n \, \delta^2 (z_1,\bar z_1) \Pi_n  \ .
\label{closedmeas}
\eeq
As before, the regularized external volume $V_D$, the order $N$ of the orbifold group ${\mathbb Z}_N$, and the Koba--Nielsen factor $\Pi_n$ in (\ref{red23}) are incorporated for later convenience.
By modular invariance, the torus modulus $\tau$ is integrated over the fundamental domain ${\cal F}$ defined by $|\Re(\tau)| \leq \frac{1}{2}$ and $|\tau|\geq 1$. External-state insertions $z_i$ are integrated over the torus ${\cal T}(\tau)$ parametrized by the parallelogram in $\mathbb C$ that  is bounded by $0,1,\tau{+}1,\tau$. 

\subsubsection{Half-maximal supersymmetry}

The half-maximal 1-loop amplitude for $n$ external states in $D=6$ can be written as 
\beq
{\cal M}_{1/2}(1,2,\ldots, n) = \int \dd \rho^{D=6}_n  \Bigg\{ \Gamma_{\mathcal{T}}^{(4)}\,{\cal J}_{n,{\rm max}}\,
+ \sum_{\underset{(k, k') \neq (0,0)}{k,k'=0}}^{N-1} \hat{\chi}_{k,k'}{\cal J}_{n,1/2}(\vec{v}_{k,k'}) \Bigg\} \ ,
\label{TOR2}
\eeq 
where $\Gamma_{\mathcal{T}}^{(n)}$ denotes $n$-dimensional closed-strings lattice sums, and $\hat{\chi}_{k,k'}$ are constant coefficients that encode the degeneracies of orbifold-charged
(``twisted'') states, see appendix \ref{opf} and e.g.\ \cite{Font:2005td}.
Similarly as for open strings, the maximally supersymmetric integrand ${\cal J}_{n,{\rm max}}$ can only receive contributions from the even-even sector. By contrast, the half-maximal integrand  in general receives non-trivial contributions from all parity sectors, we write
\beq\label{JN2} 
{\cal J}_{n, 1/2}(\vec{v}_{k,k'}) = {\cal J}^{e,\tilde{e}}_{n, 1/2}(\vec{v}_{k,k'}) + {\cal J}^{e,\tilde{o}}_{n, 1/2}(\vec{v}_{k,k'}) + {\cal J}^{o,\tilde{e}}_{n, 1/2}(\vec{v}_{k,k'}) + {\cal J}^{o,\tilde{o}}_{n, 1/2} \,,
\eeq 
where $\vec{v}_{k,k'}=(k+k'\tau)(v,-v)$ gives  the dependence of the corresponding integrands on the internal partition function.
At genus one, the total picture number of the vertex operators in the $(e,\,\tilde{e})$ sector must be $(0,0)$ \cite{Polchinski:1998rr}, and as is customary we choose all of them in the $(0,0)$ picture. In shorthand notation where $\gamma_{k,k'}\equiv (k+k'\tau)v $,
we have
\begin{align} \label{JN4ee} 
 {\cal J}_{n,\te{max}}  &\equiv  \frac{1}{\Pi_n} \sum_{\nu,\tilde{\nu}=2}^{4} \! (-1)^{\nu+ \tilde{\nu}} 
 \left[ \frac{\theta_{\nu}(0,\tau)  }{\theta'_1(0,\tau)}  \frac{\bar{\theta}_{\tilde{\nu}}(0,\bar{\tau})  }{\bar{\theta}'_1(0,\bar{\tau})} \right]^4  
 \, \langle V_1^{(0,0)} V_2^{(0,0)} \!\! \ldots  V_n^{(0,0)}\rangle_{\nu,\tilde{\nu}}
\\
\label{JN2ee}
\!\!\! {\cal J}^{e, \tilde{e}}_{n,1/2}(\vec{v}_{k,k'}) &\equiv
   \frac{1}{\Pi_n} \sum_{\nu,\tilde{\nu}=2}^{4} \! (-1)^{\nu+\tilde{\nu}} \!
 \left[ \frac{\theta_{\nu}(0,\tau)  }{\theta'_1(0,\tau)}  
 \frac{\bar{\theta}_{\tilde{\nu}}(0,\bar{\tau})  }{\bar{\theta}'_1(0,\bar{\tau})} \right]^4  \!\!\!
 \left[S_{\nu}(\gamma_{k,k'} ,\tau) \bar{S}_{\tilde{\nu}}( \bar  \gamma_{k,k'} ,\bar \tau )\right]^2
\, \langle V_1^{(0,0)}  \!\! \ldots  V_n^{(0,0)} \rangle_{\nu,\tilde{\nu}}\,,
\end{align}
where, analogously as for the open-string integrand \eqref{IN2p}, we have expressed parts of the partition function in terms of Szeg\"o kernels (\ref{red15}). Similarly, the inverse of the Koba--Nielsen factor $\Pi_n$ compensates for its inclusion into the measure \eqref{closedmeas}, and the vertex operators $V^{(0,0)}_j$ (whose arguments $z_j$ are suppressed for ease of notation) are defined in \eqref{nred9}.

Note that the contributions $\bar{\theta}_{\tilde{\nu}}(0,\bar{\tau})$ and $\bar{S}_{\tilde{\nu}}( \bar  \gamma_{k,k'},\bar \tau  )$ from the right-moving sector in (\ref{JN4ee}) and (\ref{JN2ee}) are understood to be the complex conjugate of $\theta_{\tilde{\nu}}(0,\tau)$ and $S_{\tilde{\nu}}( \gamma_{k,k'} ,\tau )$, respectively. The same sort of notation will appear in later equations on closed-string amplitudes, and will drop the obvious dependence of the above functions on $\tau$ and $\bar \tau$.

In the $(e,\,\tilde{o})$ sector the super-moduli structure of the torus requires the total picture number of the vertex operators to be $(0,-1)$ and the inclusion of the picture changing operator $P^{(0,+1)}$.  We have
\beq\label{JN2eo} 
 {\cal J}^{e, \tilde{o}}_{n,1/2}(\vec{v}_{k,k'}) \equiv \pm \, \frac{1}{\Pi_n} \sum_{\nu=2}^{4} (-1)^{\nu} \left[ \frac{\theta_{\nu}(0)  }{\theta'_1(0)} \right]^4  \!\!  \left[S_{\nu}(\gamma_{k,k'} )\right]^2\, 
\langle P^{(0,+1)}_0V_1^{(0,-1)} V_2^{(0,0)}  \, \ldots \, V_n^{(0,0)} \rangle_{\nu,\,\tilde{\nu}=1}^{D=6}\,,
\eeq
where the GSO projection of the type IIB and type IIA theories yields a $+$ sign and a $-$  sign, respectively, see appendix \ref{opf}. The expression for ${\cal J}^{o, \tilde{e}}_{n,1/2}(\vec{v}_{k,k'})$ in the $(o,\tilde{e})$ sector obviously follows from (\ref{JN2eo}) upon exchange of left- and right-movers except for a uniform sign $\pm \rightarrow +$ in both type IIA and type IIB. Note that the spin sums in (\ref{JN4ee}), (\ref{JN2ee}) and (\ref{JN2eo}) can be addressed through the methods of section \ref{sec:fourtwo}.

In the $(o,\tilde{o})$ sector we have
\beq\label{JN2oo} 
 {\cal J}^{o, \tilde{o}}_{n,\,1/2} \equiv \pm \, \frac{1}{\Pi_n} 
\langle P^{(+1,+1)}_0V_1^{(-1,-1)} V_2^{(0,0)}  \, \ldots \, V_n^{(0,0)} \rangle_{\nu=1, \tilde{\nu}=1, }^{D=6}\,.
\eeq 
In close analogy with (\ref{AN3}) for open strings, the half-maximal amplitude \eqref{TOR2} in $D=6$ easily generalizes for half-maximal models in $D=4$,
\beq
{\cal M}_{1/2}^{D=4}(1,2,\ldots, n) = \int \dd \rho^{D=4}_n  \Bigg\{ \Gamma_{\mathcal{T}}^{(6)}\,{\cal J}_{n,{\rm max}}\,
+ \Gamma_{\mathcal{T}}^{(2)} \sum_{\underset{(k, k') \neq (0,0)}{k,k'=0}}^{N-1} \hat{\chi}_{k,k'}{\cal J}_{n,1/2}^{e,\tilde{e}} (\vec{v}_{k,k'})\Bigg\}\,,
\label{TOR3}
\eeq 
where now also for the half-maximal integrand the only non-vanishing contribution is from the $(e,\tilde{e})$ sector with 
${\cal J}_{n,1/2}^{e,\tilde{e}}$ given by eq.\ \eqref{JN2ee}.

\subsubsection{Quarter-maximal supersymmetry}
 
For compactifications down to four dimensions leading to quarter-maximal supersymmetry, the amplitude for $n$ external NSNS states reads\footnote{This amplitude prescription is valid only for compactifications  on $T^6/\mathbb{Z}_N$ with $N$ prime. 
Orbifold groups of non-prime rank give rise to sectors with fixed tori, leading to half-maximal contributions to \eqref{TOR1} with two-dimensional lattice sums similar to (\ref{AN1}). For ease of presentation, we do not contemplate this case.} 
\beq
{\cal M}_{1/4}(1,2,\ldots, n)= \int \dd \rho^{D=4}_n  \,  \Bigg\{ \Gamma_{\mathcal{T}}^{(6)}\,{\cal J}_{n,{\rm max}}
\,+\!
 \sum_{\underset{(k, k') \neq (0,0)}{k,k'=0}}^{N-1} \hat{\chi}_{k,k'} \,{\cal J}_{n,1/4}(\vec{v}_{k,k'})
\Bigg\}
\label{TOR1} \,.
\eeq
 The quarter-maximal integrand  in general receives non-vanishing contributions from all parity sectors,
\beq\label{JN1} 
{\cal J}_{n, 1/4}(\vec{v}_{k,k'}) = {\cal J}^{e,\tilde{e}}_{n, 1/4}(\vec{v}_{k,k'}) + {\cal J}^{e,\tilde{o}}_{n, 1/4}(\vec{v}_{k,k'}) + {\cal J}^{o,\tilde{e}}_{n, 1/4}(\vec{v}_{k,k'}) + {\cal J}^{o,\tilde{o}}_{n, 1/4} \,,
\eeq 
where $\vec{v}_{k,k'}\equiv(k + k' \tau)(v_1,\,v_2,\,v_3)$. In the shorthand notation $\gamma_{k,k'}^j \equiv (k+k'\tau)v_j$, we have
\begin{align}\label{JN1ee}
\!\!\! {\cal J}^{e, \tilde{e}}_{n,1/4}(\vec{v}_{k,k'}) &\equiv
  \sum_{\nu,\tilde{\nu}=2}^{4} \!     \frac{(-1)^{\nu+\tilde{\nu}} }{\Pi_n}   \!
 \left[ \frac{\theta_{\nu}(0)  }{\theta'_1(0)}  
 \frac{\bar{\theta}_{\tilde{\nu}}(0)  }{\bar{\theta}'_1(0)} \right]^4  \!
 \left[\prod_{j=1}^3 S_{\nu}(\gamma_{k,k'}^j) \bar{S}_{\tilde{\nu}}(\bar \gamma_{k,k'}^j)\right] 
\, \langle V_1^{(0,0)}   \!\! \ldots  V_n^{(0,0)} \rangle_{\nu,\tilde{\nu}}
\\
\label{JN1eo} 
 {\cal J}^{e, \tilde{o}}_{n,1/4}(\vec{v}_{k,k'}) &\equiv \pm \sum_{\nu=2}^{4} \frac{(-1)^{\nu-1}}{\Pi_n}    \!\!\left[ \frac{\theta_{\nu}(0)  }{\theta'_1(0)} \right]^4  \!  \left[\prod_{j=1}^3 S_{\nu}(\gamma_{k,k'}^j)\right]
\langle P^{(0,+1)}_0V_1^{(0,-1)} V_2^{(0,0)}  \! \!  \ldots  V_n^{(0,0)} \rangle_{\nu,\,\tilde{\nu}=1}^{D=4} \\
\label{JN1oo} 
 {\cal J}^{o, \tilde{o}}_{n,\,1/4} &\equiv \pm \, \frac{1}{\Pi_n} 
\langle P^{(+1,+1)}_0V_1^{(-1,-1)} V_2^{(0,0)}  \, \ldots \, V_n^{(0,0)} \rangle_{\nu=1,\, \tilde{\nu}=1}^{D=4}\,,
\end{align}
and $\pm$ in the last two equations is a $+$ sign for type IIB and $-$ for type IIA .

\section{Open-string scattering amplitudes}
\label{sec:stringopen}

This section is devoted to the polarization-dependent part of open-string amplitudes (\ref{AN2}) and (\ref{AN1}) with $n=3$ and $n=4$ external states and less-than-maximal supersymmetry. We evaluate the correlation functions $\langle V^{(0)}_1 V^{(0)}_2 \ldots V^{(0)}_n \rangle_\nu$ of vertex operators and exploit the simplifications due to the sum over parity-even spin structures along the lines of section \ref{sec:fourtwo}. In contrast to \cite{Tourkine:2012vx,Ochirov:2013xba,Bianchi:2015vsa}, we do not make use of four-dimensional spinor-helicity variables and mostly keep polarizations $e^m$ and momenta $k^m$ dimension-agnostic. This is crucial for the infrared regularization scheme in section \ref{sec:minahan} and to simultaneously address the $D=4$ and $D=6$ realizations of half-maximal supersymmetry. More generally, this approach reveals parallels between various spacetime dimensions and varying amounts of supersymmetry, culminating in the simple dictionary (\ref{from2to4}) between the parity-even contributions to amplitudes with half-maximal and quarter-maximal supersymmetry.

\subsection{Vertex operators and CFT basics}
\label{sec:fiveone}

Gauge bosons as massless excitations of the open superstring are represented by vertex operators\footnote{A note on conventions.
To avoid proliferation of imaginary units, we absorb a factor of $i$ in $ X^m$. In doing so we
 depart from the  standard form $e^{ik\cdot X}$ of the plane-wave part of the vertex operator. 
We also   set $\alpha'=1/2$ for open strings, and for closed strings, $\alpha'=2$. \label{convfootnote}}
\beq
V^{(0)}(e,k) \equiv e_m (\partial X^m + (k\cdot \psi) \psi^m) e^{k\cdot X} 
\label{red21}
\eeq
in the zero superghost picture. 
The BRST invariance of these vertex operators is ensured
by having lightlike momenta and transverse polarization vectors,
\beq
k_m k^m = 0 \co k_m e^m=0 \ .
\label{onshl}
\eeq
For the parity-odd sector (\ref{IN1O}) we also need
 the vertex in the $-1$ superghost picture as well as the picture changing operator,
\beq
V^{(-1)}(e,k) \equiv e_m \psi^m e^{-\phi} e^{k \cdot X} \co
P^{(+1)} \equiv \partial X_m \psi^m e^{\phi} \ ,
\label{oddstuff}
\eeq
where the fields $e^{\pm \phi}$ from bosonizing the $\beta$-$\gamma$ superghost system \cite{Friedan:1985ey, Friedan:1985ge} only enter through their zero modes in this work. 

Correlation functions of the free conformal fields $\partial X^m(z)$ and $\psi^m(z)$ of weight $h=1$ and $h=\frac{1}{2}$ are determined by their two-point contractions on genus-one worldsheets,
\beq
\langle \partial X^m(z) X^n(0) \rangle = \eta^{mn} f^{(1)}(z) \co \langle  \psi^m(z) \psi^n(0) \rangle_\nu = \eta^{mn} \left\{ \begin{array}{cl} S_\nu(z) &: \ \nu=2,3,4 \\ f^{(1)}(z) &: \ \nu = 1 \end{array} \right. \ ,
\label{red22}
\eeq
where $f^{(1)}$ and $S_\nu$ are defined in (\ref{red12}) and (\ref{red15}), respectively, and the modular parameter $\tau$ is suppressed. The plane waves $e^{k\cdot X}$ in the vertex operators yield the ubiquitous Koba--Nielsen factor,
\beq
\Pi_n \equiv  \langle e^{ k_1 \cdot X(z_1)}e^{ k_2 \cdot X(z_2)}\ldots e^{ k_n \cdot X(z_n)} \rangle   =\prod_{1\leq i<j}^n e^{s_{ij} G_{ij}} \ ,
\label{red23}
\eeq
which is absorbed into the integration measure (\ref{openmeas}) by our conventions for the integrands ${\cal I}^{\ldots}_{\ldots}$ in (\ref{IN4}), (\ref{IN2}), (\ref{IN1O}) and (\ref{IN1}). The boson Green's function $G_{ij}$ is
\beq
G_{ij}\equiv G(z_i,z_j,\tau) = \log \left| \frac{ \theta_1(z_i-z_j,\tau)}{\theta_1'(0,\tau)} \right|^2 - \frac{2 \pi}{\Im (\tau)} \big[ \Im(z_i-z_j) \big]^2
\label{bosGF}
\eeq
and satisfies
\beq
\partial_i G_{ij} \equiv \frac{ \partial G_{ij}}{\partial z_i}  = f^{(1)}_{ij}  \co f^{(n)}_{ij} \equiv f^{(n)}(z_i-z_j) \ ,
\label{red24}
\eeq
where $s_{ij}$ are Mandelstam variables
\beq
s_{ij} \equiv k_i \cdot k_j \co s_{i_1i_2\ldots i_p} \equiv \frac{1}{2}(k_{i_1}+k_{i_2}+\ldots+k_{i_p})^2 \ .
\label{red25}
\eeq
Factors of $\partial X^m$ in the vertex operators can  contract 
among themselves via $\partial X^m(z) \partial X^n(0) \rightarrow -\partial f^{(1)}(z)$, see (\ref{red22}), or interact with the exponentials to yield
\beq
Q^m_i \equiv \sum_{j\neq i} k_j^m f^{(1)}_{ij} \ .
\label{red27}
\eeq
Contractions of the fermions lead to the spin sums evaluated in section \ref{sec:fourtwo}. The associated kinematic factors are gauge invariant Lorentz-traces over linearized field strengths $e^{[m} k^{n]}$ which will be denoted by
\begin{align}
t(1,2) &\equiv (e_1\cdot k_2) (e_2 \cdot k_1) - (e_1\cdot e_2) (k_1\cdot k_2)
\label{red28} \\
t(1,2,3) &\equiv (e_1 \cdot k_2) (e_2\cdot k_3)(e_3 \cdot k_1) - (e_1 \cdot k_2) (e_2\cdot e_3)(k_3 \cdot k_1) \notag \\
&- (e_1 \cdot e_2) (k_2\cdot k_3)(e_3 \cdot k_1) + (e_1 \cdot e_2) (k_2\cdot e_3)(k_3 \cdot k_1) \notag \\
&-(k_1 \cdot k_2) (e_2\cdot k_3)(e_3 \cdot e_1) + (k_1 \cdot k_2) (e_2\cdot e_3)(k_3 \cdot e_1) \notag \\
&+ (k_1 \cdot e_2) (k_2\cdot k_3)(e_3 \cdot e_1) - (k_1 \cdot e_2) (k_2\cdot e_3)(k_3 \cdot e_1) 
\label{red29} \\
t(1,2,\ldots,n) &\equiv (e_1 \cdot k_2) (e_2\cdot k_3)(e_3 \cdot k_4)\ldots (e_{n-1}\cdot k_n)(e_n\cdot k_1) \notag \\
& - \te{antisymmetrization in all} \ (k_j\leftrightarrow e_j) \ .
\label{red30}
\end{align}
They are convenient to track intermediate steps of the subsequent computations, but an alternative system of kinematic building blocks will be introduced in section \ref{sec:six} to obtain simpler and more compact representations of the correlators and to highlight parallels with maximally supersymmetric cases.

\subsection{Infrared regularization by minahaning}
\label{sec:minahan}

Any 3-point function  of any massless
external states naively vanishes by ``3-point special kinematics''.
This means that all 3-point would-be Mandelstam invariants (\ref{red25}) vanish identically\footnote{We keep the kinematic identities covariant and dimension-agnostic in this work, i.e.\ factorization of $s_{12}= \frac{1}{2}(k_3^2 -k_2^2 - k_1^2)=0$ into four-dimensional spinor brackets $\langle 12\rangle$ and $[12]$ (one of which is often taken to be non-zero for complex momenta, see \cite{Elvang:2010kc}) will not enter the discussion.}, as implied by momentum conservation and $k_j^2 =0$. This infrared zero can 
lead to $0/0$ issues in presence of certain propagators. 
We will regularize by relaxing momentum conservation in intermediate steps: $k_1^m+k_2^m+k_3^m=p^m$ for
a lightlike ``deformation'' vector $p^2=0$.
The three Mandelstam invariants $s_{12}$, $s_{23}$, $s_{13}$ then become nonzero, but subject to the single condition
\be   \label{Mina1}
{1\over 2}(k_1+k_2+k_3)^2=s_{12}+s_{23}+s_{13} =0 \; . 
\ee
This  is needed to ensure that  exponentials of boson propagators 
in the Koba--Nielsen factor (\ref{red23}) of the string integrand 
are modular invariant. 
Other conditions on the  
deformed Mandelstam variables, for example the more symmetric but stronger $s_{12}=s_{23}=s_{13}$,
would violate modular invariance, as explained by Minahan in 1987 \cite{Minahan:1987ha}.
To see directly how the ``deformation''  momentum $p^m$
allows for nonzero Mandelstam invariants in the 3-point function,
take scalar products with for example $k_1$:
\be
k_1 \cdot p = k_1 \cdot (k_1+k_2+k_3)=k_1\cdot k_2+
k_1\cdot k_3=-s_{23} \; ,
\label{Mina5}
\ee
i.e.\ the $s_{ij}$ in the 3-point function are only nonzero due to the deformation $p^m$. 
We give some more details on this in appendix \ref{kinematics}. 
In general, we will refer to the procedure 
of relaxing momentum conservation 
subject to the constraint $\sum_{i<j}^n s_{ij}=0$ as ``minahaning'' an $n$-point function. 
For the 3-point amplitude the need for some kind of infrared regularization is clear
because of the infrared zero of 3-point special kinematics, but
we will argue that there is a sense in which this should be done for any $n$-point amplitude.

As a first step in the subsequent calculations, we will combine the regularized (i.e.\ nonzero) Mandelstam invariants as in (\ref{Mina5}) with vanishing propagator denominators from string theory such that all indeterminate 0/0 expressions are taken care of. Then, for the purposes of this paper, we can safely set the  deformation $p^m$ to zero in our final expressions for amplitudes.\footnote{We note that the original procedure in \cite{Minahan:1987ha}
was a slightly stronger form of regularization, when terms in the effective action are computed
without setting the deformation  to zero at the end. This allows for the
extraction of effective couplings from two-point functions, more recently used for example in \cite{Berg:2014ama}
and references therein, which will not be discussed in this work. To distinguish the stronger form of regularization from the weaker ``minahaning'' used in this paper, one might be tempted
to call the stronger procedure ``maxahaning''. We will resist this temptation.}

\subsection{Half-maximal parity-even 3-point amplitude}
\label{sec:fivethree}

For three external states, the well-known half-maximal integrand in (\ref{IN2}) is given by
\begin{align}
{\cal I}_{3,1/2}^e   &= {\cal G}_{2}(\gamma,-\gamma) \big[   \big( \partial f^{(1)}_{12}(e_1\cdot e_2) (e_3\cdot Q_3) + (3\leftrightarrow 2,1) \big) - (e_1 \cdot Q_1) (e_2\cdot Q_2) (e_3\cdot Q_3) \big] \label{red32}  \\
&+ \big[ {\cal G}_{4}(\gamma,-\gamma,z_{12},z_{21}) t(1,2) (e_3 \cdot Q_3) + (3\leftrightarrow 2,1) \big] + {\cal G}_{5}(\gamma,-\gamma,z_{12},z_{23},z_{31})t(1,2,3) \ , \notag
\end{align}
recalling that the Koba--Nielsen factor is absorbed into the measure (\ref{openmeas}) and the definition (\ref{red27}) of $Q^m_i$. The spin sum ${\cal G}_2$ in the first line evaluates to zero whereas ${\cal G}_4$ and ${\cal G}_5$ in the second line are given by (\ref{red127a}) and (\ref{red127b}), respectively. Hence, the correlator (\ref{red32}) is homogeneous in $f^{(1)}_{ij}$,
\beq
{\cal I}_{3,1/2}^e = f^{(1)}_{12} K_{12|3} + (12\leftrightarrow 13,23) \ ,
\label{red33}
\eeq
whose antisymmetric kinematic factors $K_{12|3}=-K_{21|3}$ can be simplified using relaxed momentum conservation \eqref{Mina1} and order-$p$ transversality (\ref{onshl}) via $(e_1\cdot k_3) = - (e_1 \cdot k_2)$. We find
\begin{align}
K_{12|3}= t(1,2,3) + (e_1\cdot k_2) t(2,3) -  (e_2\cdot k_1) t(1,3) =  s_{12}(e_1\cdot e_2) (e_3 \cdot k_1) \ .
\label{red34}
\end{align}
The singular function $f^{(1)}_{12} \sim (z_1-z_2)^{-1}$ integrates to a kinematic pole in presence of the Koba--Nielsen factor $\Pi_3$, i.e.~$\int_{z_1} \dd z_2 f^{(1)}_{12} e^{s_{12} G_{12}} \sim {1}/{s_{12}}$, such that 3-particle momentum conservation for massless states would make this $1/0$. However, the minahaning procedure explained in section \ref{sec:minahan} yields a finite integral for the function 
\beq
X_{ij} \equiv s_{ij}f^{(1)}_{ij} \ ,
\label{red35}
\eeq
i.e.~the prefactor $s_{12}$ in the kinematic factors (\ref{red34}) can be used to identify appropriate building blocks (\ref{red35}) which make the finiteness of the $z$-integral manifest:
\beq
{\cal I}_{3,1/2}^e =   X_{12} (e_1\cdot e_2) (e_3 \cdot k_1) + (3\leftrightarrow 2,1)
\label{red36}
\eeq
Again, minahaning means to first use only  $\sum^3_{i<j}s_{ij}=0$ and to only after performing the $z$-integrals at nonvanishing values of $s_{ij}$ finally impose momentum conservation again,
$s_{ij}=0$. A similar procedure will be applied for the 4-point function.

In the low-energy limit $\ap\rightarrow 0$, the analytic part\footnote{
In addition to a power-series expansion in $\alpha'$, loop amplitudes in string theories give rise to logarithmic, non-analytic momentum dependence. As will be elaborated in a companion paper \cite{FT}, the integration region of large $\tau_2$ yields Feynman integrals of Yang--Mills along with their threshold singularities in $s_{ij}$, see also \cite{Bern:1990cu, Bern:1991aq, Strassler:1992zr, BjerrumBohr:2008ji}. Following the discussion of closed-string 1-loop amplitudes in \cite{Green:1999pv, Green:2008uj, Richards:2008jg, Green:2013bza}, the analytic parts of the amplitude can be isolated in a well-defined manner.
} of the integrals over all of $X_{12},X_{23}$ and $X_{31}$ yield a constant, and (\ref{red36}) reduces to the 3-point tree-level amplitude,
\beq
{\cal I}_{3,1/2}^e \rightarrow  (e_1\cdot e_2) (e_3 \cdot k_1) + (3\leftrightarrow 2,1) = A^{\te{tree}}(1,2,3) \ .
\label{red37}
\eeq
After discarding total derivatives $\partial_2 \Pi_3$ and $\partial_3 \Pi_3$ of the Koba--Nielsen factor, we can effectively set\footnote{In slight abuse of notation, we will often write equalities such as the present $X_{12}=X_{23}$ at the level of the integrands ${\cal I}^{\ldots}_{\ldots}$ which only hold upon integration over the $z_i$ in presence of $\Pi_n$.} $X_{12}=X_{23}=-X_{31}$ in (\ref{red36}). The correlator then simplifies to
\beq
{\cal I}_{3,1/2}^e =   X_{23} A^{\te{tree}}(1,2,3) \ ,
\label{red36a}
\eeq
which manifests gauge invariance in any dimension $D$ and reduces to the result in section 3 of \cite{Bianchi:2015vsa} upon dimensional reduction to $D=4$ and conversion to spinor-helicity variables.

\subsection{Half-maximal parity-even 4-point amplitude}
\label{sec:fivefour}

Also the half-maximal 4-point amplitude is well-known not to receive any contributions with less than two fermion bilinears,
\begin{align}
{\cal I}_{4,1/2}^e &= \big[ \big((e_1\cdot Q_1) (e_2\cdot Q_2) -(e_1\cdot e_2) \partial f^{(1)}_{12} \big) t(3,4) {\cal G}_4(\gamma,-\gamma,z_{34},z_{43}) 
+(12\leftrightarrow 13,14,23,24,34)\big] \notag\\
&+ \big[ (e_4\cdot Q_4) t(1,2,3)  {\cal G}_5(\gamma,-\gamma,z_{12},z_{23},z_{31})  + (4\leftrightarrow 3,2,1) \big] \notag \\
&+\big[ t(1,2,3,4) {\cal G}_6(\gamma,-\gamma,z_{12},z_{23},z_{34},z_{41})  \notag \\
& \ \ \ \ \ \ \ - t(1,2)t(3,4) {\cal G}_6(\gamma,-\gamma,z_{12},z_{21},z_{34},z_{43}) + \te{cyc}(2,3,4) \big] \ .
\label{red38}
\end{align}
All the spin sums are readily evaluated using (\ref{red127a}) to (\ref{red127c}) and give rise to functions $f^{(1)} f^{(1)}$ or $f^{(2)}$ with various combinations of arguments.

\subsubsection{Double-pole treatment}
\label{sec:double}

\begin{figure}[h]
\begin{center}
\includegraphics[width=0.4\textwidth]{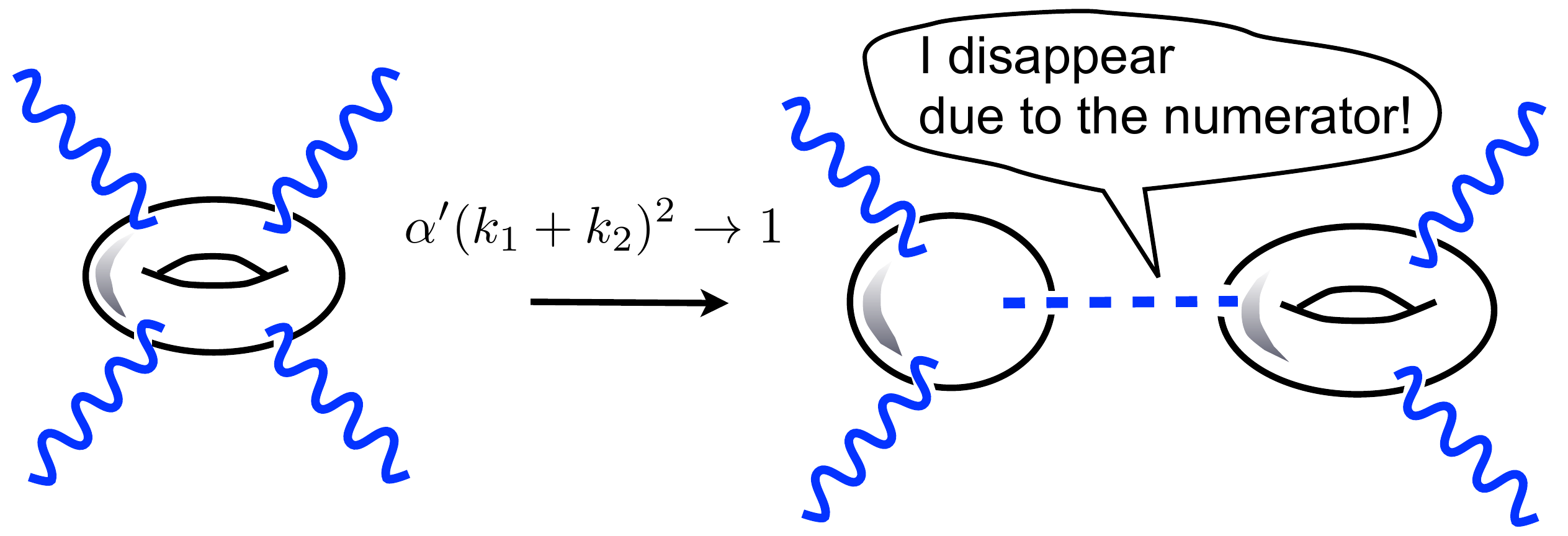}
\vspace{-5mm}
\caption{The double-pole residue
ensures that tachyons do not propagate. This statement holds universally for both the present open-string calculation and its closed-string counterpart in section \ref{sec:closedstring}, and we are drawing its representative for a torus worldsheet.}
\label{fig:tachyon}
\end{center}
\end{figure}

Note that the spin sum ${\cal G}_6$ in the last line of (\ref{red38}) yields double poles in some of the $z_{i}-z_j$,
\beq
{\cal G}_6(\gamma,-\gamma,z_{12},z_{21},z_{34},z_{43}) = F_{1/2}^{(2)}(\gamma) + 2 f^{(2)}(z_{12}) + 2 f^{(2)}(z_{34}) - f^{(1)}(z_{12})^2-f^{(1)}(z_{34})^2 \ ,
\label{red39}
\eeq
see (\ref{red122}) for the definition of $F_{1/2}^{(2)}(\gamma)$. Similar double poles occur in the first line from $ \partial f^{(1)}_{12}(e_1 \cdot e_2)$ and $(e_1\cdot Q_1) (e_2\cdot Q_2) \rightarrow (e_1\cdot k_2)f^{(1)}_{12} (e_2 \cdot k_1)f^{(1)}_{21}$. 
These two sources of double poles conspire such as to cancel the tensor structure $(e_1\cdot k_2) (e_2 \cdot k_1) t(3,4)$ and to yield a double-pole residue $\sim (s_{12}{-}1)$ along with $(e_1 \cdot e_2) t(3,4)$. The prefactor $(s_{12}-1)$ 
does the important job of eliminating tachyon propagation
in that factorization limit where $(k_1+k_2)^2 \rightarrow 2$, see fig.\ \ref{fig:tachyon}.
\begin{align}
{\cal I}_{4,1/2}^e  \big|_{(z_1-z_2)^{-2}} &= - \big[ (f^{(1)}_{12})^2(e_1 \cdot k_2) (e_2 \cdot k_1) + \partial f^{(1)}_{12}(e_1 \cdot e_2) \big] t(3,4) + (f^{(1)}_{12})^2 t(1,2) t(3,4) \notag \\
&= - (\partial f^{(1)}_{12} + s_{12} (f^{(1)}_{12})^2 ) (e_1 \cdot e_2) t(3,4) \ .
\label{red40}
\end{align}
In presence of the Koba--Nielsen factor, total derivatives of $f^{(1)}_{12} \Pi_{4}$ can be added to render the leftover integral in (\ref{red40}) manifestly free of double poles in $z_{i}-z_j$:
\begin{align}
{\cal I}_{4,1/2}^e  \big|_{(z_1-z_2)^{-2}} &= -\frac{1}{2} f^{(1)}_{12} (e_1 \cdot e_2) t(3,4) \big[ s_{23} f^{(1)}_{23}+s_{24} f^{(1)}_{24}-s_{13} f^{(1)}_{13}-s_{14} f^{(1)}_{14}\big] \ .
\label{red41}
\end{align}
In order to maintain manifest permutation invariance, we choose to average over the two possibilities of eliminating the spurious double pole: integration by parts relations involving either $\partial_1 (f^{(1)}_{12} \Pi_4)$ or $-\partial_2 (f^{(1)}_{12} \Pi_4)$. After these manipulations, four classes of worldsheet functions remain:
\begin{itemize}
\item[(i)] $F_{1/2}^{(2)}(\gamma)$ in (\ref{red122}) from the spin sums ${\cal G}_6$ in the last two lines of (\ref{red38})
\item[(ii)] six permutations of $f^{(2)}_{ij}$ from the spin sums ${\cal G}_6$
\item[(iii)] three permutations of $f^{(1)}_{12} f^{(1)}_{34}$ from all the lines of (\ref{red38})
\item[(iv)] twelve permutations of $f^{(1)}_{12} f^{(1)}_{13}$ from all the lines of (\ref{red38}) as well as the integration by parts treatment of double poles given in (\ref{red41}).
\end{itemize}
The first class (i) reproduces the kinematic factor of the maximally supersymmetric case,
\begin{align}
{\cal I}_{4,1/2}^e \big|_{  F_{1/2}^{(2)} } &= t(1,2,3,4) - t(1,2) t(3,4) + \te{cyc}(2,3,4) \label{red42}\\
&= -2 t_8(1,2,3,4) \ ,
\notag
\end{align}
which leads to the famous $t_8$ tensor
\begin{align}
t_8(1,2,3,4) &\equiv s_{12} s_{23}(e_1 \cdot e_3)(e_2\cdot e_4) + (e_1 \cdot e_2) \big[  s_{13} (k_1 \cdot e_4) (k_2 \cdot e_3) + s_{23}(k_1\cdot e_3)(k_2 \cdot e_4)  \big] \notag \\
& \ \ \ \ + (e_3\cdot e_4) \big[s_{13} (k_4\cdot e_1) (k_3\cdot e_2) + s_{23} (k_3\cdot e_1) (k_4\cdot e_2)  \big] + \te{cyc}(2,3,4) \label{red43}
\\
&= s_{12} s_{23} A^{\te{tree}}(1,2,3,4)  \notag
\end{align}
and naturally combines with the maximally supersymmetric orbifold sector in (\ref{maxint}).

\subsubsection{Minahaning the 4-point function}

The second class (ii) of functions $\sim f^{(2)}_{ij}$ is non-singular as $z_i\rightarrow z_j$ and therefore does not contribute to any factorization channel or the low-energy limit. However, the last two classes (iii) and (iv) of functions yield up to two simultaneous kinematic poles from the integration region where $z_i\rightarrow z_j$. Worse, the 3-particle factorization of schematic form $\int_{z_1}\dd z_2 \dd z_3 f^{(1)}_{12} f^{(1)}_{13} \Pi_n \sim (s_{12} s_{123})^{-1}$ due to (iv) involves divergent propagators. This requires minahaning, as discussed in section \ref{sec:minahan}. 

We shall repeat the procedure of the 3-point amplitude and transform the integrals to a basis that is manifestly free of kinematic poles. The replacement $f^{(1)}_{12} f^{(1)}_{34}=\frac{X_{12} X_{34}}{s_{12}s_{34}}$ for the non-overlapping singularities of type (iii) is a straightforward doubling of (\ref{red35}) whereas the class (iv) of functions requires the Fay identity \cite{Broedel:2014vla}
\beq
f_{12}^{(1)} f_{13}^{(1)}+f_{21}^{(1)} f_{23}^{(1)}+f_{31}^{(1)} f_{32}^{(1)} = f^{(2)}_{12}+f^{(2)}_{13}+f^{(2)}_{23} \ ,
\label{red44}
\eeq
which generalizes partial fraction relations and leads to the rearrangement
\beq
f^{(1)}_{12} f^{(1)}_{13} = \frac{ s_{23} }{s_{123}}(f^{(2)}_{12}+f^{(2)}_{13}+f^{(2)}_{23}
) + \frac{ X_{12,3} }{s_{12} s_{123}} + \frac{ X_{13,2} }{s_{13} s_{123}}
\ .
\label{red45}
\eeq
We have used the shorthand
\beq
X_{12,3} \equiv s_{12} f^{(1)}_{12}( s_{13} f^{(1)}_{13}+s_{23} f^{(1)}_{23})
\label{red46}
\eeq
for the combination of functions (iv) that does not integrate to any divergent propagators $\sim s_{123}^{-1}$.

Once the replacement (\ref{red45}) is coherently applied to the correlator ${\cal I}_{4,1/2}^e$, the kinematic prefactors accompanying any $f^{(2)}_{ij}(s_{ijk})^{-1}$ or $X_{ij,k}(s_{ijk})^{-1}$ allow to factor out compensating Mandelstam variables $s_{ijk}$. These manipulations require no Mandelstam identity other than overall momentum conservation $\sum^4_{i<j} s_{ij}=0$. The double pole treatment in (\ref{red41}) using integration by parts is crucial to build up these compensating numerator factors.

An analogous regulation procedure for divergent propagators $\sim s_{123}^{-1}$ was used for the 4-point 4-loop amplitude of $N=4$ super Yang--Mills (SYM) \cite{Bern:2012uf}. The kinematic numerators of so-called ``snail graphs'' (see fig.\ 3 in section II.D of the reference) are found to be proportional to $k_4^2=2s_{123}$ such as to cancel the vanishing denominators $(k_1+k_2+k_3)^2$ from the external propagators. In the same way as these finite contributions are essential for the 4-loop UV divergence of $N=4$ SYM, our way of minahaning the 4-point 1-loop amplitude for open strings will crucially impact its low-energy limit.

Once any instance of $s_{ijk}$ is cancelled, the correlator takes the form
\begin{align}
{\cal I}_{4,1/2}^e &= -2  F_{1/2}^{(2)}(\gamma)  t_8(1,2,3,4) + \big[ X_{12,3} K_{123|4} + X_{13,2} K_{132|4} + (4\leftrightarrow 3,2,1) \big] \notag \\
& \ \ \ \ \ \ \ \ +
\big[ X_{12} X_{34} K_{12|34} +  L_{12|34}(f^{(2)}_{12} + f^{(2)}_{34}) + \te{cyc}(2,3,4) \big] \ 
\label{red47}
\end{align}
with the somewhat bulky kinematic factors
\begin{align}
K_{12|34} &\equiv \frac{1}{s_{12}} \Big[ s_{12}(e_1\cdot e_4)(e_2\cdot e_3)-s_{12}(e_1\cdot e_3)(e_2\cdot e_4) +(s_{13}-s_{23})(e_1\cdot e_2)(e_3\cdot e_4)  \notag \\
&\ \ \ \ + (e_1\cdot e_2) \big( (k_1\cdot e_3)(k_2\cdot e_4) - (k_2\cdot e_3)(k_1\cdot e_4) \big) \notag \\
&\ \ \ \ + (e_3\cdot e_4) \big( (k_4\cdot e_2)(k_3\cdot e_1) - (k_4\cdot e_1)(k_3\cdot e_2) \big) 
\Big]
  \label{red48} \\
K_{123|4} &\equiv    (e_1\cdot e_3)(e_2\cdot e_4) - \frac{1}{2}(e_1\cdot e_2)(e_3\cdot e_4)-\frac{1}{2}(e_1\cdot e_4)(e_2\cdot e_3) \notag \\
&\ \ \ \ + (e_1\cdot e_2) \Big[ \frac{ (k_2\cdot e_3) - (k_1\cdot e_3) }{2s_{12}}(k_3\cdot e_4) - \frac{ (k_1\cdot e_4) (k_2\cdot e_3)}{s_{23}} \Big]   \notag \\
&\ \ \ \ + (e_2\cdot e_3) \Big[ \frac{ (k_2\cdot e_1) - (k_3\cdot e_1) }{2 s_{23}}(k_1\cdot e_4) - \frac{ (k_2\cdot e_1) (k_3\cdot e_4)}{s_{12}} \Big] \notag \\
&\ \ \ \ + (e_1\cdot e_3) \Big[ \frac{ (k_1\cdot e_2)(k_3\cdot e_4) }{s_{12}} +  \frac{ (k_1\cdot e_4)(k_3\cdot e_2) }{s_{23}}  \Big]\ ,
   \label{red49} \\
 L_{12|34} &\equiv t(1,3)t(2,4) + t(1,4)t(2,3) - t(1,2)t(3,4) - t(1,4,2,3) \ .
 \label{red50}
   \end{align}
  We will reorganize these in more compact and suggestive
  form in the following sections. 
Note that the symmetries $X_{12,3}+X_{23,1}+X_{31,2}=0$ and $K_{123|4}+K_{231|4}+K_{312|4}=0$ are dual to each other and ensure that the two terms $X_{12,3}K_{123|4}+X_{13,2}K_{132|4}$ are permutation invariant in $1,2,3$. The coefficients $L_{ij|kl}$ of the functions $f^{(2)}_{ij}$ have higher mass dimension than the $K_{\ldots|\ldots}$ and by (\ref{red42}) add up to
\beq
 L_{12|34}+ L_{13|24}+ L_{14|23} = 2 t_8(1,2,3,4) \ .
\label{red50a}
\eeq
The low-energy limit associated with the ordering $1,2,3,4$ is obtained by setting
\begin{align}
X_{12} X_{34} &\rightarrow 1  \co X_{41}X_{23} \rightarrow 1  \co X_{13}X_{24} \rightarrow 0 
\label{red51}
\\
X_{12,3}&\rightarrow 1 \co X_{13,2} \rightarrow 0  \ ,
\label{red52}
\end{align}
along with cyclic permutations of (\ref{red52}). We then obtain the 4-point tree amplitude of Yang--Mills (YM) field theories in the overall low-energy limit,
\begin{align}
{\cal I}_{4,1/2}^e &\rightarrow K_{12|34} + K_{41|23} +K_{123|4}+K_{234|1}+K_{341|2}+K_{412|3} \notag \\
&= \frac{ 2}{s_{12}s_{23}} t_8(1,2,3,4) = 2 A^{\te{tree}}(1,2,3,4)  \ ,
\label{red53}
\end{align}
which is obviously consistent with its 3-point analogue (\ref{red37}) under factorization, see appendix \ref{sec:appB1} for a factorization check beyond the low-energy limit.

\subsubsection{Integration by parts}

Apart from the fact that its integral is regular in 4-point kinematics $s_{123}\rightarrow 0$, a key virtue of the function $X_{12,3}$ defined in (\ref{red46}) is its suitability for integration by parts. By discarding a total derivative of $X_{12} \Pi_4$ with respect to $z_3$, we find
\beq
0=\partial_3 X_{12} \Pi_4 =
X_{12} (X_{31}+X_{32}+X_{34}) \Pi_4 \co
X_{12,3} \Pi_4 = X_{12} X_{34} \Pi_4 \ .
\label{nred1}
\eeq
Permutations of (\ref{nred1}) as well as $X_{12,3}+X_{23,1} + X_{31,2}=0$ allow us to express any integral in (\ref{red47}) with two factors of $f^{(1)}_{ij}$ in terms of the two-element basis $\{X_{23,4} , X_{24,3}\}$:
\begin{align}
{\cal I}_{4,1/2}^e &= -2  F_{1/2}^{(2)}(\gamma)  t_8(1,2,3,4)+
\big[  L_{12|34}(f^{(2)}_{12} + f^{(2)}_{34}) + \te{cyc}(2,3,4) \big] \notag \\
 & \ \ \ \ \ \ \ \ \  + 2 \big[ X_{23,4} A^{\te{tree}}(1,2,3,4) + X_{24,3} A^{\te{tree}}(1,2,4,3) \big] \ .
\label{nred2}
\end{align}
Similarly to the maximally supersymmetric case \cite{Mafra:2012kh}, this amounts to eliminating all instances of $z_1$ in the arguments of $f^{(1)}_{ij}$. The tree amplitudes of YM have been identified on the basis of (\ref{red53}), and permutation invariance of the second line of (\ref{nred2}) follows from the photon decoupling identity $A^{\te{tree}}(1,2,3,4)+A^{\te{tree}}(1,3,4,2)+A^{\te{tree}}(1,4,2,3)=0$. All the kinematic constituents $L_{ij|kl},A^{\te{tree}}(i,j,k,l)$ and $ t_8(1,2,3,4)$ in the
representation (\ref{nred2}) of the correlator separately manifest gauge invariance.

\subsubsection{Comparison with \cite{Bianchi:2015vsa}}

Even though our result in (\ref{nred2}) is written in terms of the same basis functions as the four-dimensional expression in
section 5.3 of \cite{Bianchi:2015vsa}, the kinematic coefficients along with $f^{(1)}f^{(1)}$ vanish in the computation of the reference, regardless of the helicity configuration. This causes a discrepancy with the non-zero second line of 
our (\ref{nred2}) which can be traced back to the minahaning procedure. In the present infrared regularitzation scheme, intermediate kinematic factors proportional to $s_{123}$ are kept in the spirit of section \ref{sec:minahan} since they might later on cancel a divergent propagator $(s_{123})^{-1}$ and contribute after integral manipulations such as (\ref{red45}). In \cite{Bianchi:2015vsa}, on the other hand, spinor-helicity variables are introduced at an early stage,
which implicitly drops contributions proportional to $ s_{123}$ irrespective of the accompanying worldsheet functions. It will be interesting to check whether infrared-safe observables in field
theory computed from (\ref{nred2}) and the analogous expression in \cite{Bianchi:2015vsa} might match in
spite of the above differences in the string correlator.

Just like the result in \cite{Bianchi:2015vsa}, the $D$-dimensional expression (\ref{nred2}) obeys the $D=4$ corollary of supersymmetric Ward identities that amplitudes with 3 or 4 particles of alike helicity vanish \cite{Grisaru:1976vm, Grisaru:1977px}. All the kinematic factors $L_{ij|kl},A^{\te{tree}}(i,j,k,l)$ and $ t_8(1,2,3,4)$ have been tested for this property after dimensional reduction to $D=4$ and conversion to spinor-helicity variables. As will be demonstrated in a companion paper \cite{FT}, we are under the impression that the $f^{(1)}f^{(1)}$ contributions in the second line of (\ref{nred2}) are important to identify the onset of UV-divergences of half-maximal SYM amplitudes in $D=4$ dimensions.

\subsection{Half-maximal parity-even amplitudes of higher multiplicity}
\label{sec:fivefive}

Half-maximal amplitudes of higher multiplicity can be evaluated using the same principles. The required spin sums for up to eight external states are available in (\ref{red127a}) to (\ref{red127g}) and can be easily extended using the techniques of \cite{Broedel:2014vla}. In the same way as the final form (\ref{nred2}) of the 4-point correlator augments the simplest function $F_{1/2}^{(2)}(\gamma)$ of the orbifold twist $\gamma=kv$ with the maximally supersymmetric kinematic factor $t_8(1,2,3,4)$, the coefficient of $F_{1/2}^{(2)}(\gamma)$ in higher-multiplicity amplitudes will reproduce the maximally supersymmetric correlators in their dimensional reduction to $D=6$. Starting from six points, new combinations of the $f^{(i)}$ will emerge where the $\gamma$-dependence $F_{1/2}^{(4)},F_{1/2}^{(6)},\ldots$ carries higher modular weight. The hierarchy of various $F_{1/2}^{(k)}(\gamma)$ in the amplitudes captures the model-dependent particle-content running in the loop.

In the sector of $1=F_{1/2}^{(0)}(\gamma)$, the factors of $f^{(1)}_{ij}$ from the spin sums and the contractions between $\partial X^m$ and $e^{k\cdot X}$ give rise to up to $n{-}2$ simultaneous kinematic poles. In the ``maximally overlapping'' configuration of labels $i,j$ in $f^{(1)}_{ij}$ (cf.~$f^{(1)}_{12} f^{(1)}_{13}$ versus $f^{(1)}_{12} f^{(1)}_{34}$), the kinematic poles describe an $(n{-}1)$-particle factorization channel which is plagued by a divergent propagator such as $s_{12\ldots n-1}^{-1}$. Once the complete contribution to this channel is assembled from the correlator, the kinematic numerator is expected to yield compensating Mandelstam invariants (using no other relation than $\sum_{i<j}^n s_{ij}=0$), see section \ref{sec:minahan}. Generalizations of the functions $X_{ij}$ and $X_{ij,k}$ in (\ref{red35}) and (\ref{red46}) which remain smooth after integration over $z_j$ can be found in the context of maximally supersymmetric 1-loop correlators~\cite{Mafra:2012kh}.

\subsection{Quarter-maximal generalizations in the parity-even sector}
\label{sec:fivesix}

In the parity-even sector, the quarter-maximal counterparts of the above correlators ${\cal I}_{3,1/2}^e $ and ${\cal I}_{4,1/2}^e $ can be obtained from a minor modification: according to the discussion of section \ref{sec:fourtwo}, the net difference between the quarter-maximal and half-maximal partition functions in (\ref{IN1}) and (\ref{IN2}) is captured by the straightforward shift (\ref{promote}) in the functions of the orbifold twist. Explicitly, the modified spin sums ${\cal G}_k(\gamma,-\gamma,\ldots) \rightarrow {\cal G}_{k+1}(\gamma_1,\gamma_2,\gamma_3,\ldots)$ in (\ref{red32}) and (\ref{red38}) yield $F_{1/2}^{(l)}(\gamma)\rightarrow F_{1/4}^{(l+1)}(\gamma_j)$ in the final expressions (\ref{red36a}) and (\ref{nred2}) such that
\begin{align}
{\cal I}_{3,1/4}^e &= F_{1/4}^{(1)}(\gamma_j) X_{23} A^{\te{tree}}(1,2,3)
\label{newred36a} \\
{\cal I}_{4,1/4}^e &= -2  F_{1/4}^{(3)}(\gamma_j)  t_8(1,2,3,4)+
F_{1/4}^{(1)}(\gamma_j) \big[  L_{12|34}(f^{(2)}_{12} + f^{(2)}_{34}) + \te{cyc}(2,3,4) \big] \notag \\
 & \ \ \ \ \ \ \ \ \  + 2F_{1/4}^{(1)}(\gamma_j)  \big[ X_{23,4} A^{\te{tree}}(1,2,3,4) + X_{24,3} A^{\te{tree}}(1,2,4,3) \big]  \ .
\label{newnred2}
\end{align}
The same mechanism applies to any higher multiplicity: The dictionary (\ref{promote}) between half-maximal and quarter-maximal spin sums guarantees that the parity-even parts of the integrands are related as
\beq
{\cal I}_{n,1/4}^e = {\cal I}_{n,1/2}^e
\Big|_{
F_{1/2}^{(k)}(\gamma)\rightarrow F_{1/4}^{(k+1)}(\gamma_j)}
\label{from2to4}
\eeq
at any multiplicity $n$.

\subsection{Parity-odd integrands at lowest multiplicity}
\label{sec:fiveseven}

In the parity-odd sector, the zero-mode saturation rule (\ref{zeropsi}) in $D=6$ spacetime dimensions requires at least three external legs in the prescription (\ref{IN1O}). After soaking up the $\psi^m$ from the picture changing operator and vertex operators -- see (\ref{red21}) and (\ref{oddstuff}) -- the 3-point parity-odd integrand can be written as
\beq
{\cal I}^o_{3,D=6} = i \left( \sum_{j=1}^3 f^{(1)}_{0j} k_j^m \right) \epsilon_{m}(e_1, k_2, e_2, k_3, e_3) = 0 \ .
\label{CPodd1}
\eeq
Here and in later equations, we use the shorthand notation
\beq
\epsilon(v_1, v_2, \ldots,v_D) \equiv \epsilon_{mn\ldots p} v_1^m v_2^n \ldots v_D^p
\co
\epsilon_{m}(v_2, v_3, \ldots,v_D) \equiv \epsilon_{mnp\ldots q} v_2^n v_3^p \ldots v_D^q
\label{noind}
\eeq
for vectors $v_1^m,v_2^n,\ldots,v_D^p$, to avoid proliferation of indices. By antisymmetry of the $\epsilon$-tensor, contributions from the sum in (\ref{CPodd1}) with $j=2,3$ drop out immediately, and momentum conservation $k_1=-k_2-k_3$ leads to the same conclusion for the term with $j=1$. Hence, the dependence on the position $z_0$ of the picture changing operator via $f^{(1)}_{0j}$ is spurious,
as expected from general arguments (cf. section \ref{sec:halfmaximalsusy}).

This reasoning can be straightforwardly generalized to arbitrary even dimensions $D$: At the lowest multiplicity $\frac{D}{2}$ with a sufficient number of zero modes of the worldsheet spinors in the integrand, the kinematic argument above still makes the parity-odd correlator vanish,
\begin{align}
{\cal I}^o_{n,D} &= 0 \ , \ \ \ \ \ \ n < \frac{D}{2}
\label{CPodd2} \\
{\cal I}^o_{\frac{D}{2},D} &= i \left( \sum_{j=1}^{D/2} f^{(1)}_{0j} k_j^m \right) \epsilon_{m}( e_1, k_2, e_2, k_3, e_3,  \ldots ,k_{D/2} ,e_{D/2}) = 0 \ .
\label{CPodd3}
\end{align}
As we will see, the first truly non-zero parity-odd correlator ${\cal I}^o_{N,D}$ for open strings occurs at multiplicity $N=\frac{D}{2}+1$. Regardless of their multiplicity, parity-odd open-string amplitudes in $D\geq 4$ dimensions do not exhibit any factorization channel that requires minahaning. In their contribution to closed-string amplitudes, however, parity-odd terms in $D=4$ might introduce spurious divergent propagators.

\subsection{Parity-odd integrands at next-to-lowest multiplicity}
\label{sec:fiveeight}

The simplest non-vanishing parity-odd contribution to half-maximal open-string amplitudes in $D=6$ dimensions occurs at the 4-point level. A subtle chain of integral manipulations and kinematic rearrangements detailed in appendix \ref{sec:oddsc} confirms independence on the position $z_0$ of the picture changing operator, and the only $z_i$-dependence turns out to enter through the non-singular function $f^{(2)}$ in (\ref{red13}),
\begin{align}
{\cal I}^o_{4,D=6} &= \big[ f^{(2)}_{12} E_{12|3,4} + (2\leftrightarrow 3,4) \big] + \big[ f^{(2)}_{23} E_{1|23,4} + (23\leftrightarrow 24,34) \big] \label{CPodd4} \\
E_{12|3,4} &\equiv i \big[ (e_1\cdot k_2) \epsilon(k_2, e_2, k_3, e_3, k_4, e_4) + (1\leftrightarrow 2) \big] - i s_{12} \epsilon( e_1, e_2, k_3, e_3, k_4, e_4) 
\label{CPodd5} \\
E_{1|23,4} &\equiv  i  \big[  s_{23} e_2^m - (e_2\cdot k_3) k_2^m  \big] \epsilon_{m}(e_1, k_3, e_3, k_4, e_4) + (2\leftrightarrow 3) \ .
\label{CPodd6}
\end{align}
Note that (\ref{CPodd4}) and its generalizations to higher multiplicity vanish for external gauge bosons upon dimensional reduction to $D<6$. That is why parity-odd contributions are excluded for amplitudes (\ref{AN3}) in K3$\times T^2$ compactifications.

Just like in the parity-even counterpart (\ref{nred2}), the kinematic coefficients of $f^{(2)}_{12}$ and $f^{(2)}_{34}$ in (\ref{CPodd4}) turn out to agree by
\beq
E_{12|3,4} = E_{1|34,2} \ ,
\eeq
which is a special case of (\ref{appO15}) and (\ref{appO17}) and can be checked using the ``overantisymmetrization'' identity $\eta^{m[n} \epsilon^{pqrstu]}=0$. Moreover, (\ref{CPodd5}) and (\ref{CPodd6}) are unaffected by linearized gauge transformations $e_j^m \rightarrow k_j^m$ for legs $j=2,3,4$, while the first external leg with the vertex operator in the $-1$ superghost picture breaks gauge invariance. For instance, $e_1^m \rightarrow k_1^m$ yields $E_{12|3,4} \rightarrow 2i s_{12} \epsilon(k_2, e_2, k_3, e_3, k_4, e_4)$ and $E_{1|23,4} \rightarrow 2i s_{23} \epsilon(k_2, e_2, k_3, e_3, k_4, e_4)$ and thereby signals a gauge anomaly of the form $F \wedge F \wedge F$. The connection between $\sum_{i<j}^n s_{ij} f^{(2)}_{ij}$ and a boundary term w.r.t.~$\tau_2$ is explained in section 3.3 of \cite{maxsusy}.

The $D=6$ results in (\ref{CPodd4}) to (\ref{CPodd6}) are readily generalized to multiplicity $N\equiv \frac{D}{2}+1$ in arbitrary even dimensions $D$,
\begin{align}
{\cal I}^o_{N,D} &= \big[ f^{(2)}_{12} E_{12|3,\ldots,N} + (2\leftrightarrow 3,\ldots,N) \big] + \big[ f^{(2)}_{23} E_{1|23,4,\ldots,N} + (23\leftrightarrow 24,34,\ldots,(N-1)N) \big] \label{CPodd7} \\
E_{12|3,4,\ldots,N} &\equiv i \big[ (e_1\cdot k_2) \epsilon(k_2, e_2, k_3, e_3, \ldots ,k_N, e_N) + (1\leftrightarrow 2) \big]  - i s_{12} \epsilon( e_1, e_2, k_3, e_3, \ldots ,k_N, e_N)
\label{CPodd8} \\
E_{1|23,4,\ldots ,N} &\equiv i \big[ s_{23} e_2^m -  (e_2\cdot k_3) k_2^m \big] \epsilon_{m} (e_1,k_3, e_3, k_4, e_4,  \ldots ,k_N, e_N) + (2\leftrightarrow 3) \ ,
\label{CPodd9}
\end{align}
where the permutation sum in (\ref{CPodd7}) along with $f^{(2)}_{23}$ includes any pair $i,j$ subject to $2\leq i<j\leq N$. These expressions are derived in appendix \ref{sec:oddsc}, where the ten-dimensional six-point analysis \cite{maxsusy} is carried out in a dimension-agnostic manner.


\section{Berends--Giele organization of open-string amplitudes}
\label{sec:six}

In this section, the kinematic organizing principles of the above open-string results are explored. They rely on bosonic Berends--Giele currents $\mathfrak e^m_{12\ldots p}$ which recursively resum Feynman diagrams with $p$ external on-shell states and an additional off-shell leg. While Berends--Giele currents were
first used in the 1980's to elegantly address gluonic tree amplitudes \cite{Berends:1987me} in YM theories, the value of this concept for superstring theories became apparent in \cite{Mafra:2011kj, Mafra:2011nv, Mafra:2011nw}. In these references, tree-level amplitudes for any number of massless open-superstring states were computed in the pure spinor formalism \cite{Berkovits:2000fe}. The underlying supersymmetric Berends--Giele currents have been generalized and streamlined in \cite{Mafra:2012kh, Mafra:2014oia, Lee:2015upy}, connected with the component currents from the 80's in \cite{Lee:2015upy, Mafra:2015vca} and exploited to compute and compactly represent loop amplitudes of the pure spinor superstring in \cite{Mafra:2012kh, Green:2013bza, Gomez:2013sla, Gomez:2015uha, maxsusy}. 

The Berends--Giele representation of maximally supersymmetric string amplitudes led to a variety
of insights on ten-dimensional SYM amplitudes in pure spinor superspace. In addition to the field-theory limit $\alpha' \rightarrow 0$ of superstring amplitudes, ten-dimensional SYM amplitudes have been obtained from first principles --- locality and BRST invariance. Locality amounts to imposing the Feynman-diagram content in the Berends--Giele constituents of the desired amplitude, and BRST invariance powerfully embodies both maximal supersymmetry and gauge invariance of bosonic components \cite{Berkovits:2000fe}. This program has been successfully applied at tree level \cite{Mafra:2010ir, Mafra:2010jq}, one loop \cite{Mafra:2014gsa, Mafra:2014gja} and two loops \cite{Mafra:2015mja}.

It will now be demonstrated that the Berends--Giele approach to string amplitudes can be extended to half- and quarter maximal supersymmetry. The structure of the above half-maximal 3- and 4-point amplitudes will be clarified using the bosonic components of supersymmetric Berends--Giele currents \cite{Berends:1987me, Lee:2015upy, Mafra:2015vca}. Apart from the conceptual benefit of extending the pure spinor methods, this will pave the way for a compact and enlightening representation of the closed-string computations in section \ref{sec:closedstring}. Moreover, a first-principles approach to half-maximal SYM 1-loop amplitudes obtained in the field-theory limit will be discussed in a companion paper \cite{FT}.

\subsection{Definition of bosonic Berends--Giele currents}

\begin{figure}[h]
\begin{center}
\begin{tikzpicture} [scale=0.8, line width=0.30mm]
\begin{scope}[xshift=-7.5cm]
\draw (-3,0) node{$\efrak^{m}_{12},\ffrak^{mn}_{12} \  \leftrightarrow  $};
\draw (0,0) -- (-1,1) node[left]{$2$};
\draw (0,0) -- (-1,-1) node[left]{$1$};
\draw (0,0) -- (0.8,0);
\draw (0.5,0.3) node{$s_{12}$};
\draw (1.7, 0) node{${\cdots} \  \ , $};
\end{scope}
\draw (-3,0) node{$\efrak^{m}_{123},\ffrak^{mn}_{123} \  \leftrightarrow  $};
\draw (0,0) -- (-1,1) node[left]{$2$};
\draw (0,0) -- (-1,-1) node[left]{$1$};
\draw (0,0) -- (1.8,0);
\draw (0.5,0.3) node{$s_{12}$};
\draw (1,0) -- (1,1) node[right]{$3$};
\draw (1.65,-0.3) node{$s_{123}$};
\draw (2.7, 0) node{${\cdots} \  \ + $};
\scope[xshift=-0.4cm]
\draw (5.5,0) -- (4.5,1) node[left]{$3$};
\draw (5.5,0) -- (4.5,-1) node[left]{$2$};
\draw (5.5,0) -- (7.3,0);
\draw (6,-0.3) node{$s_{23}$};
\draw (6.5,0) -- (6.5,-1) node[right]{$1$};
\draw (7,0.3) node{$s_{123}$};
\draw (7.8, 0) node{${\ldots} $};
\endscope
\end{tikzpicture}
\caption{Cubic-vertex subdiagrams with an off-shell $\cdots$ leg whose kinematic contributions are captured by Berends--Giele currents $\efrak^{m}_{12},\ffrak^{mn}_{12}$ and $\efrak^{m}_{123},\ffrak^{mn}_{123}$, respectively.}
\label{berends}
\end{center}
\end{figure}
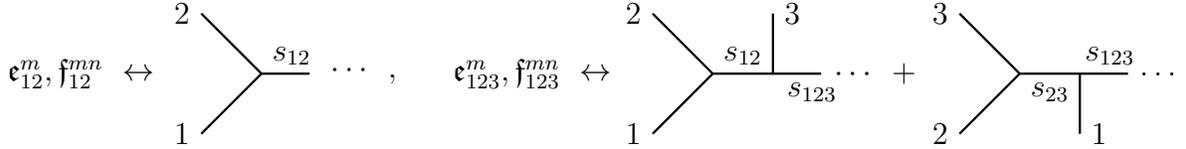

We will only define the minimal set of Berends--Giele currents that appear in half-maximal amplitudes with no more than four external legs\footnote{The all-multiplicity generalizations of $\efrak^{m}_{12\ldots p} $ and $\ffrak^{mn}_{12\ldots p} $ in the present conventions can be found in \cite{Lee:2015upy, Mafra:2015vca}.}. Bosonic currents with a maximum of three on-shell legs are defined recursively via \cite{Berends:1987me, Lee:2015upy, Mafra:2015vca}
\begin{align}
\efrak^{m}_{1}  &\equiv e^m_1
\notag \\
\efrak^{m}_{12}  &\equiv  {1 \over 2 s_{12}} \bigl[ \efrak_{2}^m (k_2\cdot  \efrak_{1})
+ (\efrak_{2})_{n}  \ffrak_1^{mn}
- (1 \leftrightarrow 2)\bigr] 
\label{Giele01} \\
\efrak^{m}_{123}  &\equiv  
 {1 \over 2 s_{123}} \Big\{ \bigl[ \efrak_{3}^m (k_3\cdot  \efrak_{12})
+ (\efrak_{3})_{n}  \ffrak_{12}^{mn}
- (12 \leftrightarrow 3)\bigr] 
 + \bigl[ \efrak_{23}^m (k_{23}\cdot  \efrak_{1})
+ (\efrak_{23})_{n}  \ffrak_{1}^{mn}
- (1 \leftrightarrow 23)\bigr] \Big\}
\notag
\end{align}
along with their non-linear field-strength representatives (in conventions where $2k_1^{[m} \efrak_1^{n]}=k_1^m \efrak_1^n - k_1^n \efrak_1^m$ and $k_{12\ldots p} \equiv k_1 + k_2 + \ldots + k_p$)
\begin{align}
\ffrak^{mn}_1 &\equiv 2k_1^{[m} \efrak_1^{n]}\notag \\
\ffrak^{mn}_{12}&\equiv 2k_{12}^{[m} \efrak_{12}^{n]} 
- 2 \efrak_1^{[m} \efrak_2^{n]} 
\label{Giele02} \\
\ffrak^{mn}_{123}&\equiv 2k_{123}^{[m} \efrak_{123}^{n]} 
-2 ( \efrak_{12}^{[m} \efrak_3^{n]} 
+\efrak_{1}^{[m} \efrak_{23}^{n]} ) \ .
\notag
\end{align}
The cubic diagrams associated with the 2-particle and 3-particle currents $\efrak^{m}_{12},\ffrak^{mn}_{12}$ and $\efrak^{m}_{123},\ffrak^{mn}_{123}$ are depicted in fig.\ \ref{berends}. Appropriate choices of $\efrak^{m}_{\ldots}$ versus $\ffrak^{mn}_{\ldots}$ as suggested by string theory guarantee that quartic Feynman vertices of YM theories are absorbed into these cubic diagrams \cite{Mafra:2015vca}, in line with the BCJ duality between color and kinematics \cite{Bern:2008qj}.

These diagrammatic interpretations allow to derive the Berends--Giele symmetries
\beq
\efrak^{m}_{12} = - \efrak^{m}_{21} \ , \ \ \ \
\ffrak^{mn}_{12} = - \ffrak^{mn}_{21} \co
\efrak^{m}_{123} = \efrak^{m}_{321} = - \efrak^{m}_{231} - \efrak^{m}_{312} \ , \ \ \ \
\ffrak^{mn}_{123} = \ffrak^{mn}_{321} = - \ffrak^{mn}_{231} - \ffrak^{mn}_{312}
\label{Giele03}
\eeq
solely from the antisymmetry of kinematic factors upon flipping a cubic vertex.

\subsection{Scalar building blocks for half-maximal loop amplitudes}

In a multiparticle notation where $A=12\ldots p$ (and similarly $B,C,\ldots$) can contain any number $p$ of on-shell legs, we define the fundamental scalar building block
\beq
M_{A,B} \equiv - \frac{1}{2} \ffrak^{mn}_A  \ffrak^{mn}_B  = M_{B,A}
\label{Giele04}
\eeq
such that for instance $M_{1,2} =  (k_1 \cdot e_2) (k_2 \cdot e_1) - s_{12} (e_1 \cdot e_2) = t(1,2)$. Following the minahaning prescription in section \ref{sec:minahan}, one can straightforwardly check that 
\begin{align}
M_{12,3} &= (e_1\cdot e_2) (k_1\cdot e_3) = s_{12}^{-1} K_{12|3}
\label{Giele05}
\\
M_{123,4} &= K_{123|4}  \co  M_{12,34} = K_{12|34}  
\label{Giele06}
\end{align} 
reproduce the kinematic dependence of the half- and quarter-maximal open-string correlators, see (\ref{red48}) and (\ref{red49}) for the 4-point expressions. Note that (\ref{Giele05}) and (\ref{Giele06}) only hold in massless 3-particle and 4-particle momentum phase space, respectively. It is striking to see the kinematic factors $K_{12|3}, K_{123|4}, K_{12|34}$ decompose into two Berends--Giele currents once the dust of their string-theory origin (including the spin sums in section \ref{sec:fourtwo} and the integral manipulations (\ref{red41}) and (\ref{red45})) has settled. This shows the value of the integral processing in section \ref{sec:fivefour}: it incorporates field-theory insights into the organization of string amplitudes.


The emergence of tree amplitudes $A^{\te{tree}}(\ldots)$ in half-maximal open-string amplitudes yields a representation in terms of the scalar building block (\ref{Giele04}),
\begin{align}
A^{\te{tree}}(1,2,3) &= M_{12,3} + M_{23,1} + M_{31,2}
\label{Giele07} \\
2 A^{\te{tree}}(1,2,3,4) &= M_{123,4} + M_{234,1} + M_{341,2} + M_{412,3} + M_{12,34} + M_{41,23} \ .
\label{Giele08} 
\end{align}
Once the Berends--Giele currents $\ffrak^{mn}_A$ are resummed to yield a solution $F^{mn}$ of the non-linear YM~field equations, the expressions in (\ref{Giele07}) and (\ref{Giele08}) can be generated from the Lagrangian $\sim F^{mn} F_{mn}$, evaluated on this perturbative solution \cite{Mafra:2015gia, Lee:2015upy, Mafra:2015vca}.
Note that the scalar building block in (\ref{Giele04}) is reminiscient of the maximally supersymmetric 1-loop building blocks defined in section 5.2 of \cite{Mafra:2014oia} (see \cite{Mafra:2012kh} for pioneering work) which were later identified as local box numerators in ten-dimensional SYM \cite{Mafra:2014gja}.

\subsection{Vector \& tensor building blocks for half-maximal loop amplitudes}

While the scalar building block in (\ref{Giele04}) completely captures
the kinematic coefficient of $f^{(1)}$ in half-maximal open-string amplitudes at multiplicity $n\leq 4$, the $f^{(2)}$ terms as well as the closed string will require various extensions. We will design vectorial and tensorial building blocks such that parity-even and parity-odd contributions to half-maximal string integrands are unified. For this purpose, the following basic building block for parity-odd kinematics is introduced,
\beq
{\cal E}^m_{A|B,C} \equiv
\frac{i}{4}\epsilon^{m}{}_{npqrs} \efrak_A^n \ffrak_B^{pq} \ffrak_C^{rs} = {\cal E}^m_{A|C,B} \ ,
\label{Giele09}
\eeq 
where the vertical-bar notation $A|B,C$ is a reminder of the special role of the first slot, ${\cal E}^m_{A|B,C} \neq {\cal E}^m_{B|A,C}$, and ${\cal E}^m_{1|2,3}=\epsilon^m(e_1,k_2,e_2,k_3,e_3)$ is recovered in the single-particle case. We define the following frequently occurring composition of parity-even and parity-odd kinematics,
\beq
M^m_{A|B,C} \equiv \efrak^m_A M_{B,C}+\efrak^m_B M_{A,C}+\efrak^m_C M_{A,B} + {\cal E}^m_{A|B,C} = M^m_{A|C,B} \ ,
\label{Giele10}
\eeq
where only the parity-even constituents are permutation invariant in $A,B,C$. This definition is reminiscient of the maximally supersymmetric vector building blocks defined in section 5.4 of \cite{Mafra:2014oia}, see \cite{Green:2013bza, maxsusy} and \cite{Mafra:2014gja} for their role in closed-string amplitudes and pentagon numerators in SYM amplitudes, respectively.

In the same way as the maximally supersymmetric vectors were recursively extended to tensors of arbitrary rank \cite{Mafra:2014gsa}, we define a two-tensor counterpart to the bosonic vector in (\ref{Giele10}):
\beq
M^{mn}_{A|B,C,D} \equiv 2 \big[  \efrak^{(m}_A  \efrak^{n)}_B M_{C,D} +  (AB\leftrightarrow AC,AD,BC,BD,CD)  \big] + 2 \big[ \efrak_B^{(m} {\cal E}^{n)}_{A|C,D}+ (B\leftrightarrow C,D) \big]
\label{Giele11}
\eeq
It will play an essential role for the closed-string 4-point function in section \ref{sec:seventhree} and the loop-momentum dependent part of Feynman-diagram numerators in the field-theory limit \cite{FT}. 

Note that the combination of parity-even and parity-odd parts in (\ref{Giele10}) and (\ref{Giele11}) are tailor-made for half-maximal supersymmetry in $D=6$. By the universality result (\ref{from2to4}), the dimensional reduction of $M_{A,B}, M^m_{A|B,C}$  and $M^{mn}_{A|B,C,D}$ to $D=4$ (suppressing parity-odd contributions $\sim {\cal E}^m_{A|B,C}$) also appears in quarter-maximal amplitudes. However, the parity-odd contributions in quarter-maximal settings follow different patterns as compared to the half-maximal case, see the discussion in sections \ref{sec:fiveseven} and \ref{sec:fiveeight}.

\subsection{Gauge-(pseudo-)invariant kinematic factors}
\label{sec:pseu}

Gauge transformations of the above building blocks yield a rewarding web of relations involving lower-multiplicity counterparts. These gauge variations resemble the BRST variations in pure spinor superspace \cite{Mafra:2014oia, Mafra:2014gsa} and will be thoroughly discussion in the companion paper \cite{FT}. For our present purposes, we simply state the gauge invariant combinations of the scalar, vectorial and tensorial building blocks (\ref{Giele04}), (\ref{Giele10}) and (\ref{Giele11}) which will find prominent appearance in half-maximal amplitudes of the open and closed string.

Since any Berends--Giele current $\efrak^m_A$ and $\ffrak^{mn}_A$ (other than the single-particle $\ffrak^{mn}_1$) is affected by linearized gauge transformations $e^m_i \rightarrow k^m_i$, gauge-invariant quantities usually require several building blocks with different partitions of the external legs. One can check that the scalar combinations
\begin{align}
C_{1|23} &\equiv M_{1,23} + M_{12,3}  - M_{13,2}   
\label{BG17}
\\
C_{1|234} &\equiv M_{1,234} + M_{123,4}  + M_{412,3} + M_{341,2} + M_{12,34} + M_{41,23}  
\label{BG19}
\end{align}
and the vector combinations
\begin{align}
C^m_{1|2,3} &\equiv M^m_{1|2,3} + k_2^m M_{12,3} + k_3^m M_{13,2}
\label{BG18}
\\
C^m_{1|23,4} &\equiv M^m_{1|23,4} + M^m_{12|3,4}  - M^m_{13|2,4}  - k_2^m M_{132,4} + k_3^m M_{123,4} - k_4^m (M_{41,23} + M_{412,3} - M_{413,2}) 
\label{BG20}
\end{align}
are invariant under linearized gauge transformation of any external leg in the appropriate momentum phase space. Note that the expansions in terms of $M_{\ldots}$ and $M^m_{\ldots}$ closely resemble the maximally supersymmetric BRST invariants $C_{1|23,4,5},C_{1|234,5,6},C^m_{1|2,3,4,5}$ and $C_{1|23,4,5,6}^m$ defined in section 5 of \cite{Mafra:2014oia}.

 The situation for tensors is slightly different since their trace carries the fingerprints of the gauge anomaly noticed in section \ref{sec:fiveeight}. The tensorial combination
\begin{align}
C^{mn}_{1|2,3,4} &\equiv M^{mn}_{1|2,3,4} +  2\big[ k_2^{(m} M^{n)}_{12|3,4}+(2\leftrightarrow 3,4) \big] - 2 \big[ k_2^{(m} k_3^{n)} M_{213,4}+ (23\leftrightarrow 24,34)\big]
\label{BG21}
\end{align}
is gauge invariant under $e^m_i \rightarrow k^m_i$ with $i=2,3,4$, but the transformation $e^m_1 \rightarrow k^m_1$ on the first leg yields 
\beq
 C^{mn}_{1|2,3,4} \big|_{e^m_1 \rightarrow k^m_1} =2 i \eta^{mn} 
 \epsilon(k_2,e_2,k_3,e_3,k_4,e_4) \ .
\label{BG21var}
\eeq
Following the terminology of \cite{Mafra:2014gsa}, we will refer to quantities whose gauge variations can be exclusively expressed in terms of  $\epsilon_{mnpqrs} \ffrak_B^{mn}\ffrak_C^{pq}\ffrak_D^{rs}$ as ``pseudo-invariant''. Apart from the tensor (\ref{BG21}), the following scalar is pseudo-invariant,
\beq
P_{1|2|3,4} \equiv (e_2)_m (e^m_1 M_{3,4} + {\cal E}^m_{1|3,4}) + \frac{1}{2} \big[ (e_2 \cdot e_3) M_{1,4} + (3\leftrightarrow 4) \big] + k_2^m M^m_{12|3,4} + s_{23} M_{123,4} + s_{24} M_{124,3} \ ,
\label{BG24}
\eeq 
i.e.~invariant under $e^m_i \rightarrow k^m_i$ with $i=2,3,4$, but subject to the following anomalous gauge variation: 
\beq
P_{1|2|3,4} \big|_{e^m_1 \rightarrow k^m_1}  =2 i  \epsilon(k_2,e_2,k_3,e_3,k_4,e_4) \ .
\label{BG24var}
\eeq
Again, the construction of the 4-point kinematic factors (\ref{BG21}) and (\ref{BG24}) is inspired by six-point counterparts in the maximally supersymmetric case. More specifically, the associated expressions for $C^{mn}_{1|2,3,4,5,6}$ and $P_{1|2|3,4,5,6}$ in pure spinor superspace are given in (3.14) and (5.22) of \cite{Mafra:2014gsa}.

\subsection{Rewriting the open-string correlator}

In terms of the above pseudo-invariants, the parity-even and parity-odd parts of the 3- and 4-point correlators (\ref{red36a}),  (\ref{nred2}) and (\ref{CPodd4}) can be combined to yield 
\begin{align}
{\cal I}_{3,1/2}& = X_{23} C_{1|23}
\label{new3pt}
\\
{\cal I}_{4,1/2}&=  X_{23,4} C_{1|234} + X_{24,3} C_{1|243}  + \big[ s_{12} (f^{(2)}_{12}+f^{(2)}_{34}) P_{1|2|3,4} +(2\leftrightarrow 3,4) \big] -2  F_{1/2}^{(2)}(\gamma)  t_8(1,2,3,4) \ .
\label{new4pt}
\end{align}
%
In other words, the parity-even and parity-odd parts of the pseudo-invariant $s_{12}P_{1|2|3,4} $ in (\ref{BG24}) reproduce the quantities $L_{12|34}$ and $E_{12|3,4}$ defined in (\ref{red50}) and (\ref{CPodd5}), respectively,
\beq
s_{12}P_{1|2|3,4} \, \big|_{\te{parity-even}}=  L_{12|34}
\co
s_{12}P_{1|2|3,4} \, \big|_{\te{parity-odd}}=  E_{12|3,4} \ .
\label{pseud3}
\eeq
Note that the structure of the half-maximal 4-point correlator (\ref{new4pt}) closely resembles the maximally supersymmetric six-point correlator in section 3 of \cite{maxsusy}. Moreover, the expansion of $C_{1|234}$ and $P_{1|2|3,4} $ in terms of Berends--Giele building blocks mirrors their higher-multiplicity counterparts $C_{1|234,5,6}$ and $P_{1|2|3,4,5,6} $ in pure spinor superspace \cite{maxsusy}.

The virtue of organizing the kinematic factors of half-maximal string amplitudes in terms of the building blocks $M_{A,B}$ and their tensorial generalizations will become particularly obvious from the closed-string amplitudes discussed in the following section.

\section{Closed-string scattering amplitudes}
\label{sec:closedstring}

In this section, we evaluate and simplify half-maximal 3-point and 4-point closed-string amplitudes involving gravitons, B-fields and dilatons in $D=6$ dimensions. Similar to the integration-by-parts reduction of the open-string correlators, we will cast the worldsheet functions from the closed-string prescription into an integral basis. The accompanying kinematic factors then manifest gauge invariance or pseudo-invariance, see subsection \ref{sec:pseu}. The simplified expressions for the amplitudes are suitable to extract 4-point couplings in the type II effective action and to appreciate the structural similarity to the maximally supersymmetric 6-point amplitude of \cite{maxsusy}. We will give some parity-even examples
of novel effective couplings and check some known results, but
we will not address field redefinitions, rescalings or frame-changing (see section \ref{sec:ambi}).
Our focus here is the string amplitudes, and we leave a detailed study
of the loop-corrected string effective action to the future.

\subsection{Vertex operators and left-right interactions}
\label{sec:sevenone}

Massless NSNS-excitations of the closed superstring are represented by vertex operators
\begin{align}
{ V}^{(0,0)}(e,\tilde e,k) &\equiv e_m (\partial X^m + (k\cdot \psi) \psi^m) \  \tilde e_n (\bar \partial X^n + (k\cdot \tilde \psi) \tilde \psi^n) \, e^{k\cdot X} 
\label{nred9} \\
{ V}^{(0,-1)}(e,\tilde e,k) &\equiv e_m (\partial X^m + (k\cdot \psi) \psi^m) \  \tilde e_n \bar \psi^n e^{-\tilde \phi} \, e^{k\cdot X} 
\label{nred999} \\
{ V}^{(-1,-1)}(e,\tilde e,k) &\equiv e_m \psi^m  e^{-\phi} \ \tilde e_n \tilde \psi^n e^{-\tilde \phi} \, e^{k\cdot X}  \ .
\label{nred9new}
\end{align}
Apart from the exponentials, they are double-copies of the open-string vertex operators $V^{(0)},V^{(-1)}$, and the tensor product of polarization vectors $e_m \otimes \tilde e_n$ comprises gravitons, B-fields and a dilaton. The left- and right-moving fermions $\psi^m,\tilde \psi^n$ do not interact and yield the holomorphic and antiholomorphic correlation functions, respectively, from the open-string sectors. Accordingly, the spin sums can be carried out separately for left- and right-movers using the techniques of section \ref{sec:fourtwo}. The bosons, on the other hand, entangle left- and right-movers through an additional zero-mode contraction,
\beq
\langle \partial X^m(z)\bar \partial X^n(0) \rangle =  \eta^{mn} \pi \Big( \frac{1 }{\Im( \tau)}  - \delta^{2}(z,\bar z) \Big)\ ,
\label{nred13}
\eeq
where the delta-function on the right-hand side does not contribute in the 
presence of the Koba--Nielsen factor $\Pi_n$ in (\ref{red23}) and will therefore be suppressed\footnote{Note that with our convention
of absorbing a factor $i$ in $ X$ (see footnote \ref{convfootnote} in section \ref{sec:fiveone}) the delta
 function has a negative coefficient, the opposite
 of textbook conventions like \cite{Polchinski:1998rr}.}. Note that the closed-string picture changing operators in (\ref{JN2eo}) and (\ref{JN2oo}) are double-copies of the open-string counterparts in (\ref{oddstuff}), i.e.
 \beq
P^{(0,+1)} \equiv \bar \partial X_m \tilde \psi^m e^{ \tilde \phi} \co
P^{(+1,+1)} \equiv \partial X_m \psi^m e^{\phi}  \ \bar \partial X_n \tilde \psi^n e^{ \tilde \phi} \ .
\label{cpic}
\eeq
Integration by parts relations introduce additional interactions\footnote{We collectively refer to the contributions of (\ref{nred13}) and (\ref{nonholo1}) to closed-string correlators as ``left-right interactions'' since both of them originate from the zero modes common to the fields $\partial X_m$ and $\bar \partial X_m$ from the left- and right-moving sector.} between left- and right-movers since the worldsheet functions defined by (\ref{red11}) are no longer holomorphic at non-zero genus \cite{Broedel:2014vla},
\beq
\bar \partial f^{(n)}(z) \equiv \frac{ \partial f^{(n)}(z)}{\partial \bar z} = -\frac{ \pi }{\Im (\tau)} f^{(n-1)}(z) \ .
\label{nonholo1}
\eeq
By (\ref{nred13}) and (\ref{nonholo1}), $n$-point closed-string correlators receive
additional terms $\sim \left( \frac{ \pi }{\Im \tau} \right)^k$ compared to the square of their open-string counterparts, with $k=0,1,2,\ldots,n{-}2$ in the half-maximal case. We shall illustrate both sources of corrections through 3-point and 4-point examples.

\subsubsection{Zero mode contractions between $\partial X$ and $\bar \partial X$}

The contribution of the zero-mode contractions (\ref{nred13}) to closed-string correlators can be studied independently in the left- and right-moving sector. The key information stems from summing over all possibilities to isolate $k$ zero-modes $\partial X^m$ from open-string quantities $V^{(0)}$ or $P^{(+1)}$. 
The subsequent ``tensorial integrands''
\beq
{\cal I}_{n,1/2}^{m_1\ldots m_k}(\vec{v}_k)  \equiv {\cal I}_{n,1/2}^{m_1\ldots m_k,e}(\vec{v}_k)  + {\cal I}_{n,D=6}^{m_1\ldots m_k,o}
\label{closedcomb}
\eeq 
with parity-even and parity-odd generalizations of the scalar integrands (\ref{IN2}) and (\ref{IN1O})
\begin{align}
 {\cal I}_{n,1/2}^{m_1\ldots m_k,e}(\vec{v}_k) &\equiv  \sum_{\nu=2}^{4}  \frac{(-1)^{\nu} }{\Pi_n}  
\left[ \frac{\theta_{\nu}(0) \theta_{\nu}(kv)  }{ \theta'_1(0) \theta_1(kv)} \right]^2  \, \langle V_1^{(0)} V_2^{(0)} \, \ldots \, V_n^{(0)} \big|_{\partial X_{m_1} \partial X_{m_2} \ldots \partial X_{m_k} } \rangle_\nu
\label{clcomb1} \\
{\cal I}_{n,D}^{m_1\ldots m_k,o} &\equiv
\frac{1}{\Pi_n} \langle P^{(+1)}(z_0) V_1^{(-1)} V_2^{(0)}  \, \ldots \, V_n^{(0)}   \big|_{\partial X_{m_1} \partial X_{m_2} \ldots \partial X_{m_k} }  \rangle_{\nu=1}^{D} 
\label{clcomb2} 
\end{align}
keep track of the combinatorics to peel off zero modes of $\partial X_{m_1} \partial X_{m_2} \ldots \partial X_{m_k} $.

At the 3-point level, only a single zero mode of $\partial X^m$ can be extracted from $V^{(0)}_1 V^{(0)}_2 V^{(0)}_3$ while maintaining non-vanishing sums over parity-even spin structures,
\begin{align}
{\cal I}_{3,1/2}^{m,e}& \equiv e_1^m t(2,3)+e_2^m t(3,1)+e_3^m t(1,2) \label{nred16} \\
& = e_1^m (e_2 \cdot k_3)(e_3 \cdot k_2)
+e_2^m (e_1 \cdot k_3)(e_3 \cdot k_1)+e_3^m (e_2 \cdot k_1)(e_1 \cdot k_2) \ .
\notag
\end{align}
In the parity-odd sector, the only zero-mode extraction of $\partial X^m$ while saturating the zero modes of $\psi^n$ can originate from the picture changing operator of $P^{(+1)} V^{(-1)}_1 V^{(0)}_2 V^{(0)}_3 $, leading to
\beq
{\cal I}_{3,D=6}^{m,o} \equiv i \epsilon^{m}(e_1, k_2, e_2, k_3, e_3) 
\label{nred16odd}
\eeq
in the notation of (\ref{noind}). In the 4-point amplitude, vectorial and tensorial expressions arise after extraction of zero modes $\partial X^m$ and $\partial X^m \partial X^n$, respectively. Their parity-even instances
\begin{align}
{\cal I}_{4,1/2}^{m,e} &\equiv e_1^m \big[ t(2,3) (e_4 \cdot Q_4) + t(3,4)(e_2 \cdot Q_2) + t(4,2)(e_3\cdot Q_3) + t(2,3,4) (f_{23}^{(1)} +f_{34}^{(1)} +f_{42}^{(1)} )  \big]  \notag \\
& \ \ \ \ \ \ + (1\leftrightarrow 2,3,4) \label{nred24} \\
&= f_{12}^{(1)} K^m_{12|3|4} + (12\leftrightarrow 13,14,23,24,34)
\notag
\\
{\cal I}_{4,1/2}^{mn,e} &= 2e_1^{(m} e_2^{n)} t(3,4) + (12\leftrightarrow 13,14,23,24,34)  \ ,
\label{nred25}
\end{align}
can be easily obtained from the spin sums (\ref{red127a}) and (\ref{red127b}), with the shorthand
\begin{align}
K^m_{12|3|4}  &\equiv \big[ e_2^m (e_1 \cdot k_2) - e_1^m(e_2 \cdot k_1) \big] t(3,4) + e_3^m \big[ t(2,4)(e_1 \cdot k_2) - t(1,4) (e_2 \cdot k_1) + t(1,2,4) \big] \notag \\
& \ \ \ \ + e_4^m \big[ t(2,3)(e_1 \cdot k_2) - t(1,3) (e_2 \cdot k_1) + t(1,2,3) \big] \ .
\label{nred26}
\end{align}
Their parity-odd counterparts are given by 
\begin{align}
{\cal I}_{4,D=6}^{m,o} &= \big[ f_{12}^{(1)} s_{12} {\cal E}^m_{12|3,4} + (2\leftrightarrow 3,4) \big] +
\big[ f_{23}^{(1)} s_{23}{\cal E}^m_{1|23,4} + (23\leftrightarrow 24,34)\big]
\label{nred71} \\
\,\,\,\,\,\,\,&  \ \ \ \ \ \ +i\big[( f_{20}^{(1)}-f_{10}^{(1)})k_2^m \epsilon(e_1, e_2, k_3, e_3, k_4, e_4) +(2\leftrightarrow 3,4)\big]
\notag 
\\
{\cal I}_{4,D=6}^{mn,o} &= 2 i e_2^{(m}\epsilon^{n)}(e_1, k_3, e_3, k_4, e_4) + (2\leftrightarrow 3,4)  \ ,
\label{nred72}
\end{align}
where intermediate steps leading to the building blocks ${\cal E}^m_{A|B,C}$ in (\ref{Giele09}) are displayed in appendix \ref{sec:oddve}. The above expressions will later on be rewritten in terms of the building blocks of section \ref{sec:six}. Note that the spurious dependence of ${\cal I}_{4,D=6}^{m,o}$ on the position $z_0$ of  the picture changing operator via $f_{j0}^{(1)}$ will cancel once the contributions from left-right interacting integrations by parts are taken into account.

\subsubsection{Left-right interacting integration by parts}
\label{sec:IBP}

In order to cast the worldsheet integrals into a specified basis, we follow the reduction scheme from the open-string discussion and eliminate any appearance of the first leg in $f^{(1)}_{1j}$ and $\bar f^{(1)}_{1j}$. In a 4-point setting, the additional contributions from $\partial\bar f^{(1)}=\bar\partial f^{(1)} = -\frac{ \pi}{\Im \tau}$ lead to identities such as
\begin{align}
X_{12} \bar X_{13} &= \frac{s_{23} \pi}{\Im \tau} + (X_{23}+X_{24})( \bar X_{32}+ \bar X_{34}) \notag\\
X_{12} \bar X_{12} &= 2\frac{s_{12} \pi}{\Im \tau} + (X_{23} +X_{24})  (\bar X_{23} + \bar X_{24})
\label{nonholo4}\\
X_{12} \bar X_{23} &= -\frac{s_{23} \pi}{\Im \tau} + (X_{23} +X_{24}) \bar X_{23} \ ,
 \notag
\end{align}
see \cite{Richards:2008jg, Green:2013bza} for analogous relations in maximally supersymmetric 5-point amplitudes. For the double-copy of the parity-even integrand ${\cal I}^e_{4,1/2}$ in (\ref{red47}), repeated use of $\partial \bar f^{(1)} = -\frac{ \pi}{\Im \tau}$ yields for instance
\begin{align}
X_{12,3} \bar X_{12,3} &= X_{34,2} \bar X_{34,2} + 4 \Big(\frac{\pi}{\Im \tau} \Big)^2 s_{12}(s_{13}+s_{23})  \label{nonholo2}\\
&+ \frac{2\pi}{\Im \tau} (s_{13}+s_{23})(X_{23}+X_{24})(\bar X_{23}+\bar X_{24}) + \frac{2\pi}{\Im \tau} s_{12} X_{34} \bar X_{34}  \ , 
\notag  
\end{align}
and a more exhaustive list of 4-point integral manipulations can be found in appendix \ref{app:4int}.

\subsection{Low-energy prescriptions}
\label{sec:sevenfour}

To study the implications of closed-string amplitudes for the low-energy effective action, the $\alpha' \rightarrow 0$ behavior of the worldsheet integrals has to be extracted. Since a discussion of the Feynman diagrams in the supergravity limit along the lines of \cite{Green:1982sw} is relegated to the companion paper \cite{FT}, we will follow the procedure of \cite{Green:1999pv, Green:2008uj, Richards:2008jg, Green:2013bza} to truncate the integrals to their analytic momentum-dependence.

The leading low-energy behavior of closed-string integrals is determined by the piece with the highest number of kinematic poles. They originate from a ``diagonal'' pair of worldsheet singularities $f^{(1)}(z) \bar f^{(1)}(\bar z) \sim |z|^{-2}$ where the left-and right moving arguments match, following the general pole prescription
\beq
\int \dd^2 z \ |z|^{s-2} g(z)= \frac{ \pi}{s} g(0) + {\cal O}(s^0) 
\label{nred17}
\eeq
for functions $g(z)$ that are regular at the origin. By repeated use of (\ref{nred17}), 
only diagonal combinations of $X_{ij}$ and $\bar X_{kl} $ affect the low-energy limit, e.g.
\begin{align}
X_{12} \bar X_{12} &\rightarrow s_{12} + {\cal O}(s_{ij}^2) \co \ \ \ \ \ \ \ \ \ \ \ \ \ X_{12} \bar X_{13} \rightarrow  {\cal O}(s_{ij}^2) \label{nred18} \\
X_{12} X_{34} \bar X_{12} \bar X_{34} &\rightarrow s_{12} s_{34} +  {\cal O}(s_{ij}^3) \co   X_{12} X_{34} \bar X_{13} \bar X_{24} \rightarrow  {\cal O}(s_{ij}^3) \ , \notag
\end{align}
where the '$\rightarrow$'-notation is understood to only keep track of the leading order of $\alpha'$ occurring in the amplitude under discussion.

For the nested product $X_{ij,k}$ defined in (\ref{red46}), the analogous rules are determined by
\beq
X_{12,3} \bar X_{12,3} \rightarrow s_{12}(s_{13}+s_{23}) +  {\cal O}(s_{ij}^3) \co X_{12,3} \bar X_{23,1} \rightarrow - s_{12}s_{23}+  {\cal O}(s_{ij}^3)  \ ,
\label{nred19}
\eeq
whereas different triplets of arguments do not yield any low-energy contribution
at leading order, e.g.
\beq
X_{12,3} \bar X_{12,4} \rightarrow {\cal O}(s_{ij}^3)    \ , \ \ \ \ X_{12,3} \bar X_{23,4} \rightarrow {\cal O}(s_{ij}^3)   \ , \ \ \ \ X_{12,3} \bar X_{42,3} \rightarrow {\cal O}(s_{ij}^3) \ , \ \ \ \
 X_{12,3} \bar X_{12} \bar X_{34} \rightarrow {\cal O}(s_{ij}^3)  \ .
\label{nred20}
\eeq
Factors of $\frac{\pi}{\Im \tau}$ from the interactions (\ref{nred13}) or (\ref{nonholo1}) between left- and right-movers are of the same order in the low-energy expansion as a diagonal pair $f^{(1)}  \bar f^{(1)}$, e.g.
\beq
\big( \frac{\pi}{\Im \tau} \Big)^n \rightarrow 1+{\cal O}(s_{ij})  \co \frac{\pi}{\Im \tau} X_{12} \bar X_{12} \rightarrow s_{12} + {\cal O}(s_{ij}^2)  \co  \frac{\pi}{\Im \tau} X_{12} \bar X_{13} \rightarrow {\cal O}(s_{ij}^2)  \ .
\label{nred21}
\eeq
These schematic rules will be used in the following to extract matrix elements of the $R^2$ interaction from the low-energy limit of 3-point and 4-point closed-string amplitudes. Note that integrals involving non-singular worldsheet functions $f^{(2)}_{ij}$ and $F^{(2)}_{1/2}$ on either the left-moving or the right-moving side do not contribute to the 4-point low-energy limit.

Subleading terms in the analytic low-energy expansions exhibit a gap at the mass dimension of $R^3$ such that the first non-vanishing interaction beyond the low-energy limit occurs at the order of $R^4$. This follows from the low-energy behavior of torus integrals over $z_j$ in presence of $f^{(1)} \bar f^{(1)}$ \cite{Richards:2008jg, Green:2013bza} where any tentative contribution at subleading order in $\alpha'$ is found to integrate to zero. The results of \cite{Green:2013bza} for 5- and 6-point integrals in the maximally supersymmetric case directly carry over to the subsequent 3- and 4-point integrals in the half-maximal case.

\subsection{Half-maximal 3-point amplitude}
\label{sec:seventwo}

The treatment of left-right interactions outlined in section \ref{sec:sevenone} is easily applied to the 3-point amplitude. The calculation can be found in the literature (see \cite{Gregori:1997hi} and references therein), and we recalculate it using our methods and the notation of the previous sections to prepare for the 4-point generalization. With the open-string kinematic factors in (\ref{red36}) and (\ref{CPodd1}) as well as the chiral halves (\ref{nred16}) and (\ref{nred16odd}) of left-right contractions, the half-maximal closed-string correlator is given by
\beq
{\cal J}_{3,1/2} \equiv {\cal I}_{3,1/2} \tilde {\cal I}_{3,1/2} 
+
\frac{\pi}{\Im \tau}{\cal I}^m_{3,1/2} \tilde {\cal I}^m_{3,1/2} 
 \ .
\label{cl3a}
\eeq
By comparison with the vector building block in (\ref{Giele10}), parity-even and parity-odd terms combine into
\beq
{\cal I}^m_{3,1/2} = M^m_{1|2,3} \ .
\label{cl3b}
\eeq
The tilde along with $\tilde{ {\cal I}}^{\ldots}_{\ldots}$ in (\ref{cl3a}) is understood to map $e_i^m \rightarrow \tilde e_i^m$ as well as $f_{ij}^{(n)} \rightarrow \bar f_{ij}^{(n)}$. Moreover, the sign of the right-moving parity-odd part e.g.~in $\tilde M^m_{1|2,3}$ differs between type IIB and type IIA due to the different GSO projections in the RR sector, as is clear from the partition function in appendix \ref{opf}. This sign can be implemented by hand in the amplitude by the simple prescription of flipping
the sign of the Levi-Civita tensor, $\epsilon \rightarrow - \epsilon$:
\beq
\tilde M^m_{A|B,C} = \left\{ \begin{array}{cl}
M^m_{A|B,C} \big|_{e_i \rightarrow \tilde e_i} &:\ \te{IIB} \\
M^m_{A|B,C} \big|_{e_i \rightarrow \tilde e_i
\atop{\epsilon \rightarrow - \epsilon}
 } &:\ \te{IIA}
\end{array}\right. \co
\tilde M^{mn}_{A|B,C,D} = \left\{ \begin{array}{cl}
M^{mn}_{A|B,C,D} \big|_{e_i \rightarrow \tilde e_i} &:\ \te{IIB} \\
M^{mn}_{A|B,C,D} \big|_{e_i \rightarrow \tilde e_i
\atop{\epsilon \rightarrow - \epsilon}
} &:\ \te{IIA}
\end{array}\right.
\label{cl3c}
\eeq
At 3 points, any integral of the form $X_{ij}\bar X_{pq}$ is accompanied by regular kinematic factors -- double-copies of $(e_i\cdot e_j) (k_i \cdot e_p)$ -- and proportional to at least one Mandelstam invariant: The entire low-energy expansions of $X_{12} \tilde X_{12}$ and $X_{12} \tilde X_{13}$ is proportional to $s_{12}$ and $s_{12}s_{13}$, respectively, see (\ref{nred18}). Hence, the left-right factorizing part vanishes when we invoke 
momentum conservation of the 3-point function at the end of the calculation which gives $s_{ij}=0$, and we are left with
\beq
{\cal J}_{3,1/2} = \frac{\pi}{\Im \tau}   M^m_{1|2,3}\tilde M^m_{1|2,3} 
= \frac{\pi}{\Im \tau}   C^m_{1|2,3}\tilde C^m_{1|2,3} \ .
\label{cl3d}
\eeq
The last equality involving the vector invariant $C^m_{1|2,3}$ in (\ref{BG18}) follows from 3-particle kinematics such as $k_j^m  M^m_{1|2,3}=0$ or $s_{ij}=0$ and manifests the structural similarity with the maximally supersymmetric 5-point amplitude in section 4.1 of \cite{maxsusy}.

In absence of worldsheet singularities, the Koba--Nielsen factor along with (\ref{cl3d}) can be replaced by its Taylor expansion which trivializes to $\Pi_3 = 1$ by 3-particle kinematics. Hence, the low-energy limit
\beq
{\cal J}_{3,1/2} \rightarrow  {\cal M}^{R^2}(1,2,3)  \equiv M^m_{1|2,3}\tilde M^m_{1|2,3}
 \label{low3}
 \eeq 
obtained from (\ref{nred21}) does not receive any corrections at higher order in $\alpha'$, and its type IIB and IIA components will be discussed further  in section \ref{sec:sevenfive}. We remind the reader that we will not perform any integrals over the worldsheet modulus $\tau$  in this paper. It  is of course important to do so to extract the moduli-dependence of the string effective action, and we would like to return to this issue in the future. 

The slightly abusive notation ${\cal M}^{R^2}(1,2,\ldots,n)$ for the low-energy limit refers to matrix elements involving any combination of $n$ NSNS sector states
at the same order in $\alpha'$ as the gravitational $R^2$ correction. The $n$-graviton component due to the $R^2$ interaction can be straightforwardly extracted by setting $\tilde e_i^m \rightarrow e_i^m$ and $(e_i\cdot e_i)\rightarrow0$.

\subsection{Half-maximal 4-point amplitude}
\label{sec:seventhree}

The 4-point closed-string correlator due to half-maximal orbifold sectors has contributions with zero, one and two left-right contractions,
\begin{align}
{\cal J}_{4,1/2} &\equiv {\cal I}_{4,1/2} \tilde {\cal I}_{4,1/2}+
\frac{\pi}{\Im \tau} {\cal I}^m_{4,1/2} \tilde {\cal I}^m_{4,1/2} 
+\frac{1}{2}\left( \frac{\pi}{\Im \tau} \right)^2{\cal I}^{mn}_{4,1/2} \tilde {\cal I}^{mn}_{4,1/2} \ .
\label{cl3e}
\end{align}
The vector and tensor integrands ${\cal I}_{4,1/2}^{m}$ and ${\cal I}_{4,1/2}^{mn}$ can be reconstructed from (\ref{nred24}), (\ref{nred71}) and (\ref{nred25}), (\ref{nred72}), respectively. After converting the kinematic factors into the building blocks of section \ref{sec:six} via
\begin{align}
M^m_{12|3,4}=  {\cal E}^m_{12|3,4} + \frac{1}{s_{12}} \Big[K^m_{12|3,4} + \frac{1}{2}(k^m_1-k^m_2)(e_1\cdot e_2)M_{3,4}\Big]
\end{align}
and a similar identity for $M^m_{1|23,4}$, we arrive at
\begin{align}
{\cal I}^{mn}_{4,1/2} &= M^{mn}_{1|2,3,4} \notag \\
{\cal I}^m_{4,1/2} &= \big[ X_{12} M^m_{12|3,4} + (2\leftrightarrow 3,4) \big] 
+ \big[ X_{23} M^m_{1|23,4} + (23\leftrightarrow 24,34) \big]  \label{simpV}\\
&\! \! \! \! + \big[ k_2^m (f^{(1)}_{02}\!-\!f^{(1)}_{01}) (e_2\! \cdot \!{\cal E}_{1|3,4}) + (2\leftrightarrow 3,4) \big]  + 
\frac{1}{2}\big[ f^{(1)}_{12} (k_2^m \! -\! k_1^m) (e_1\!\cdot \!e_2) M_{3,4}
+ (12\leftrightarrow 13,14,23,24,34) \big] \ . \notag
\end{align}
Note that the last line of (\ref{simpV}) will conspire with left-right interacting integrations by part and eventually contribute to the first three terms of the pseudo-invariant $P_{1|2|3,4}$ in (\ref{BG24}).

In view of the discussion in section \ref{sec:IBP} and appendix \ref{app:4int}, it is crucial to use the expressions for the left-right factorizing kinematic factors prior to any integrations by parts. More specifically, (\ref{cl3e}) requires the representation (\ref{red38}) for the parity-even part ${\cal I}_{4,1/2}^{e}$ and (\ref{trueodd}) for the parity-odd part ${\cal I}_{4,D=6}^{o}$.
We reduce the integrals in (\ref{cl3e}) to a basis by eliminating any instance of the first leg in $f^{(1)}_{1j}$ and $\bar f^{(1)}_{1j}$ through the integration-by-parts rules of section \ref{sec:IBP} and appendix \ref{app:4int}. In this process, various corrections $\sim  \frac{\pi}{\Im\tau}$ and $\left( \frac{\pi}{\Im\tau} \right)^2$ to the square of the simplified open-string correlator in (\ref{new4pt}) arise. Also, spurious dependences on $z_0$ as seen in (\ref{simpV}) and the derivatives within (\ref{trueodd}) will cancel in this process.

It turns out that the vector invariant $C^m_{1|23,4}$ in (\ref{BG20}) as well as the pseudo-invariants $C^{mn}_{1|2,3,4}$ and $P_{1|2|3,4}$ in (\ref{BG21}) and (\ref{BG24}) are tailor-made to express the closed-string 4-point correlator in a minimal form: They combine all the parity-even and parity-odd open-string constituents and capture the kinematic factors along with the basis integrals: 
\begin{align}
{\cal J}_{4,1/2} &\equiv \Big|  X_{23,4} C_{1|234} + X_{24,3} C_{1|243} + \big[ s_{12} (f^{(2)}_{12}+f^{(2)}_{34}) P_{1|2|3,4} +(2\leftrightarrow 3,4) \big] -2  F_{1/2}^{(2)}(\gamma)  t_8(1,2,3,4)
 \Big|^2 \notag \\
 & \ \ \ \ \ + \frac{\pi}{\Im\tau} ( X_{23} C^m_{1|23,4}+X_{24} C^m_{1|24,3}+X_{34} C^m_{1|34,2}  ) ( \bar X_{23} \tilde C^m_{1|23,4}+\bar X_{24} \tilde C^m_{1|24,3}+\bar X_{34} \tilde C^m_{1|34,2}  )
\label{nred102} \\
& \ \ \ \ \ + \Big( \frac{ \pi }{\Im \tau} \Big)^2 \big( \tfrac{1}{2} C^{mn}_{1|2,3,4}\tilde C^{mn}_{1|2,3,4}  - P_{1|2|3,4}\tilde P_{1|2|3,4} - P_{1|3|2,4}\tilde P_{1|3|2,4} - P_{1|4|2,3} \tilde P_{1|4|2,3}  \big)  \ . \notag
 \end{align}
 By the modular weight $(n,0)$ of the functions $f^{(n)}$ \cite{Broedel:2014vla}, every term in (\ref{nred102}) exhibits uniform modular weight $(2,2)$, where factors of $F^{(2)}_{1/2}$ additionally mix different orbifold sectors $k,k'$ in (\ref{TOR2}). Together with the six-dimensional closed-string measure in (\ref{closedmeas}), the weights of $\dd^2 \tau, \, \tau^{-D/2}$ and $\prod_{j=2}^4 \dd^2 z_j$ are compensated. Hence, (\ref{nred102}) manifests modular invariance of the closed-string amplitude.
 
In the last line of (\ref{nred102}), one can understand
the presence of the ``extra'' $ P_{1|2|3,4}\tilde P_{1|2|3,4} +(2\leftrightarrow 3,4)$  pieces
as follows. They compensate for the anomalous gauge transformation of the tensor contraction $\tfrac{1}{2} C^{mn}_{1|2,3,4}\tilde C^{mn}_{1|2,3,4}$ as can be verified by combining the variations (\ref{BG21var}) and (\ref{BG24var}) with the trace identity
\beq
\eta_{mn} C^{mn}_{1|2,3,4} = 2(P_{1|2|3,4}+ P_{1|3|2,4}+ P_{1|4|2,3} ) \ .
\label{trace}
\eeq
Note that the bilinears in pseudo-invariants seen in (\ref{nred102}) mimic the patterns in the maximally supersymmetric 6-point amplitude, see section 4.2 of \cite{maxsusy}.

The anomalous gauge variations along with factors of $f^{(2)}_{ij}$ in the first two lines of (\ref{nred102}) conspire to total derivatives in $\tau$ and the $z_j$. This follows from the same arguments as given for the maximally supersymmetric 6-point torus amplitude discussed in section 4.4 of \cite{maxsusy}.

The low-energy limit of (\ref{nred102}) can be easily performed by means of the rules in section \ref{sec:sevenfour} and takes a very compact form:
\begin{align}
{\cal J}_{4,1/2} \rightarrow {\cal M}^{R^2}(1,2,3,4) &\equiv \frac{1}{2} C^{mn}_{1|2,3,4} \tilde C^{mn}_{1|2,3,4} + \big[  s_{23}C^m_{1|23,4} \tilde C^m_{1|23,4} - P_{1|2|3,4}\tilde P_{1|2|3,4} + \te{cyc}(2,3,4) \big] \ .
\label{nred29A}
\end{align}
We have discarded the scalar contribution
\beq
s_{23} s_{34} C_{1|234} \tilde C_{1|234} + \te{cyc}(2,3,4) = 0
\eeq
which vanishes by the BCJ relations $s_{12}C_{1|234} = s_{13}C_{1|324} $ of $C_{1|234} = 2 A^{\te{tree}}(1,2,3,4)$ \cite{Bern:2008qj}.
We also note that the parity-odd/odd part of (\ref{nred29A}) can be simplified to yield
\beq
{\cal J}_{4,1/2}^{o,\tilde o} \rightarrow \frac{1}{2} {\cal E}^{mn}_{1|2,3,4}\tilde {\cal E}^{mn}_{1|2,3,4} + \big[
s_{12} {\cal E}^{m}_{12|3,4}\tilde {\cal E}^{m}_{12|3,4}  +s_{23}{\cal E}^{m}_{1|23,4}\tilde {\cal E}^{m}_{1|23,4} - (e_2 \cdot {\cal E}_{1|3,4})(\tilde e_2 \cdot \tilde {\cal E}_{1|3,4})
+\te{cyc}(2,3,4)\big] \ ,
\label{odod}
\eeq
with ${\cal E}^m_{A|B,C}$ in (\ref{Giele09}) and $ {\cal E}^{mn}_{1|2,3,4} \equiv 2 e_2^{(m} {\cal E}^{n)}_{1|3,4} + (2\leftrightarrow 3,4)$, see appendix \ref{sec:clfactor} for its factorization properties.

We pause to contrast the expression above
with the half-maximal closed-string 4-point 
amplitude discussed in \cite{Tourkine:2012vx, Ochirov:2013xba}. That discussion was specialized
early on to the field-theory limit and spinor-helicity expressions. 
After the manipulations performed here, 
we believe the present string amplitude clearly exhibits several 
interesting features that were not manifest in  \cite{Tourkine:2012vx, Ochirov:2013xba}.
Apart from its applicability to arbitrary dimensions $D\leq 6$,
one important aspect is the presence and limitations of double-copy structure in this string amplitude.
More precisely, the $P\tilde{P}$ structure in the last line 
of \eqref{nred102} obstructs the naive expectation to find a pure tensor contraction
$T_{mn} \tilde T^{mn}$ along with $( \frac{ \pi }{\Im \tau} )^2$. We expect this
to be the source of the tension between worldsheet correlators and double copies of
gauge-theory BCJ numerators observed in \cite{Ochirov:2013xba}. We hope to say more about the
implications of \eqref{nred102} for the BCJ-duality between color and kinematics in the future. 

\subsection{The low-energy limit in type IIB and type IIA}
\label{sec:sevenfive}

This section is devoted to the type IIB and IIA components of the low-energy limits ${\cal M}^{R^2}(1,2,\ldots,n)$ in (\ref{low3}) and (\ref{nred29A}). The 3-point case has already been investigated in \cite{Gregori:1997hi} where the parity-even IIB components were found to vanish for any combination of gravitons, B-fields and dilatons. The IIB cancellation relies on the interplay between the even/even and odd/odd spin structures and does not occur for type IIA because of the different GSO projections \cite{Gregori:1997hi}:
\begin{align}
{\cal M}^{R^2}(1,2,3) \, \big|_{\te{even}} &= M^m_{1|2,3}\tilde M^m_{1|2,3} \, \big|_{\te{even}}  = \left\{ \begin{array}{cl}
\!-2
\epsilon^{m}(e_1, k_2, e_2, k_3, e_3)
\epsilon_{m}(\tilde e_1, k_2, \tilde e_2, k_3, \tilde e_3) \!
 &: \, \te{IIA} \\
0 \!&: \, \te{IIB}
\end{array} \right. 
\label{eff01}
\end{align}
The contraction of $\epsilon$ tensors can be converted to the dot products seen in (\ref{nred16}) via Gram determinants,
\beq
\epsilon^{m}(v_1,v_2,\ldots v_5) 
\epsilon_{m}(w_1,w_2,\ldots, w_5) = \det_{i,j=1,2,\ldots,5}( v_i \cdot w_j) \ .
\label{eff02}
\eeq
Note that the parity-even type IIA result in (\ref{eff01}) vanishes for an odd number of B-fields.

In the parity-odd sector, on the other hand, the GSO projections of type IIB and IIA yield \cite{Gregori:1997hi}
\begin{align}
&{\cal M}^{R^2}(1,2,3) \, \big|_{\te{odd}} = M^m_{1|2,3}\tilde M^m_{1|2,3} \, \big|_{\te{odd}} \label{eff03} \\
& \ \ = \left\{ \begin{array}{cl}
i \big[ e_1^m (e_2\cdot k_3)(e_3\cdot k_2) + \te{cyc}(1,2,3) \big] \epsilon_{m}( \tilde e_1, k_2, \tilde e_2, k_3, \tilde e_3) - (e_i \leftrightarrow \tilde e_i) &: \, \te{IIA, odd \# of B-fields} \\
i \big[ e_1^m (e_2\cdot k_3)(e_3\cdot k_2) + \te{cyc}(1,2,3) \big] \epsilon_{m}( \tilde e_1, k_2, \tilde e_2, k_3, \tilde e_3) + (e_i \leftrightarrow \tilde e_i)&: \, \te{IIB, two B-fields} \\
0 &: \, \te{otherwise}
\end{array} \right. \ ,
\notag
\end{align}
signaling type IIA interactions of schematic form $B \wedge R \wedge R$ and $B \wedge \nabla H \wedge \nabla H$, as well as a type IIB coupling $H \wedge H \wedge R$ \cite{Gregori:1997hi}.

\subsubsection{Comparison with the heterotic string}
\label{sec:het}

Matrix elements of the $R^2$ interaction also appear in tree-level amplitudes of the heterotic string \cite{Gross:1986mw} and the bosonic string \cite{Metsaev:1986yb} upon expanding to the linear order in $\alpha'$. This yields a KLT-like double copy of YM amplitudes and $F^3$ matrix elements known from the $(\alpha')^1$-order of the bosonic open string \cite{Broedel:2012rc},
\begin{align}
{\cal M}^{R^2}_{\te{het}}(1,2,3) &= A^{F^3}(1,2,3)  \tilde A^{\te{tree}}(1,2,3)
\label{odd34} \\
{\cal M}^{R^2}_{\te{het}}(1,2,3,4) &=  A^{F^3}(1,2,3,4) s_{12}  \tilde A^{\te{tree}}(1,2,4,3) \ ,
\label{odd35}
\end{align}
which also matches the bosonic-string result. The $F^3$-constituents are given by \cite{inprogress}
\begin{align}
A^{F^3}(1,2,3)  &= (e_1\cdot k_2) (e_2 \cdot k_3) (e_3 \cdot k_1)
\label{odd36} \\
A^{F^3}(1,2,3,4) &= s_{13} \Big\{ 
\frac{ t(1,2)t(3,4) }{ s_{12}^2} +\frac{ t(1,3) t(2,4) }{ s_{13}^2} +\frac{ t(1,4) t(2,3) }{s_{23}^2} - \frac{ g_1 g_2 g_3 g_4
}{s_{12}^2 s_{13}^2 s_{23}^2 }   \Big\} 
\label{odd37} \\
g_i &\equiv (k_{i-1}\cdot e_i) s_{i,i+1} - (k_{i+1} \cdot e_i) s_{i-1,i} \ ,
\end{align}
where the right-hand side of (\ref{odd37})
 manifests gauge invariance at the expense of manifest locality. Note that the structure of $A^{F^3}(1,2,3,4)=s_{13} \times \{ \te{totally symmetric quantity} \}$ guarantees that the BCJ-relations of $A^{\te{tree}}(\ldots)$ \cite{Bern:2008qj} are also obeyed by $A^{F^3}(\ldots)$ \cite{Broedel:2012rc} and that (\ref{odd35}) is permutation invariant. 

This discussion connects to that about field redefinitions in section \ref{sec:ambi}:
in $D=4$, any tensor structure for the  $R^2$ interaction is on-shell equivalent to the  Gauss--Bonnet combination, that is topological if there is no moduli-dependent coefficient, cf.\ \eqref{moduli}. 
The on-shell vanishing of (\ref{odd34}) and (\ref{odd35}) in $D=4$ can be seen from the fact that there is no combination of graviton helicities where both $A^{\te{tree}}(\ldots)$ and $A^{F^3}(\ldots)$ are non-zero \cite{Broedel:2012rc}.

The 3-graviton component agrees between the type IIA 1-loop low-energy limit (\ref{eff01}), (\ref{eff03}) and the heterotic tree-level coupling (\ref{odd34}),
\begin{align}
{\cal M}^{R^2}(1,2,3) \, \big|^{\te{3 gravitons}} &=  \left\{ \begin{array}{cl}
{\cal M}^{R^2}_{\te{het}}(1,2,3) \, \big|^{\te{3 gravitons}} &: \, \te{IIA} \\
0 \!&: \, \te{IIB}
\end{array} \right.  \ .
\label{eff05}
\end{align}
B-fields and dilatons, however, give rise to different component amplitudes. This is expected since the left-right contractions of the form $(e_i\cdot \tilde e_j)$ are absent at tree-level.

\subsubsection{The 4-point low-energy limit}
\label{efftwoB}

While parity-even type IIB couplings vanish for any triplet of NSNS sector states, see (\ref{eff01}), non-zero results appear at the 4-point level: For type IIB gravitons and dilatons with $e^m_j = \tilde e ^m_j$, we have
\beq
{\cal M}^{R^2}(1,2,3,4) \, \big|^{\tilde e_j \rightarrow e_j}_{\te{IIB, even}} =
(s_{12}^2 + s_{13}^2 + s_{23}^2) (e_1\cdot e_1)  (e_2\cdot e_2)  (e_3\cdot e_3)  (e_4\cdot e_4)  \ ,
\label{odd40}
\eeq
which vanishes in presence of gravitons and signals a 4-dilaton interaction $(\partial \phi)^4$ with four derivatives. In presence of B-fields, to be denoted by $1_B, 2_B,\ldots$ in the following, the non-vanishing amplitudes are
\begin{align}
{\cal M}^{R^2}(1_B,2_B,3,4) \, \big|^{\tilde e_{3,4} \rightarrow e_{3,4}}_{\te{IIB, even}} &=
\Big[ H^{1\, pq}_{m}H^{2}_{npq} k_3^{(m} k_4^{n)} -
\frac{1}{6} (k_3\cdot k_4) H^{1}_{mnp}H^{2\,mnp} \Big]
(e_3\cdot e_3)  (e_4\cdot e_4)  
\label{odd40a} \\
{\cal M}^{R^2}(1_B,2_B,3_B,4_B) \, \big|_{\te{IIB, even}} &= \frac{1}{2} \Big[
 H^{1}_{mn(p} H^{2 \, mn}_{q)} H^{3\ p}_{rs} H^{4 \, qrs}
  -\frac{1}{6} (H^1_{mnp} H^{2 \, mnp})( H^3_{qrs} H^{4 \, qrs})
  +\te{cyc}(2,3,4) \Big]  \notag \\
  & \ \ \ \ \ \
- H^{1}_{mnp} H^{2\ m}_{qr}  H_{s}^{3 \, nq} H^{4 \, prs} \ ,
\label{odd40b}
\end{align}
where gauge invariance is manifest from the linearized 3-form field strength
\beq
H^{mnp} \equiv 6 k^{[m} e^n \,  \tilde e^{p]} \ .
\label{odd40c}
\eeq
These results signal interactions of schematic form $H^2 (\partial \phi)^2$ and $H^4$, whose tensor structure is determined by (\ref{odd40a}) and (\ref{odd40b}). Odd numbers of B-fields, on the other hand, yield vanishing low-energy limits
\beq
{\cal M}^{R^2}(1_B,2,3,4) \, \big|^{\tilde e_{2,3,4} \rightarrow e_{2,3,4}}_{\te{IIB, even}} =
{\cal M}^{R^2}(1_B,2_B,3_B,4) \, \big|^{\tilde e_{4} \rightarrow e_{4}}_{\te{IIB, even}}  
= 0 \ .
\label{odd40d}
\eeq
In the parity-odd sector of the type IIB low-energy limit, we have checked the vanishing of the 4-graviton component,
\beq
{\cal M}^{R^2}(1,2,3,4) \, \big|_{\te{IIB, odd}}^{\te{4 gravitons}} = 0 \ ,
\eeq
and expect generalizations of the $H \wedge H \wedge R$ interaction \cite{Gregori:1997hi} seen in (\ref{eff03}).

In the type IIA low-energy limit, we have checked agreement of the 4-graviton component with the $R^2$ coupling (\ref{odd35}) in the heterotic string,
\beq
{\cal M}^{R^2}(1,2,3,4) \, \big|_{\te{IIA}}^{\te{4 gravitons}} =  {\cal M}^{R^2}_{\te{het}}(1,2,3,4)\, \big|^{\te{4 gravitons}} \ .
\label{odd41}
\eeq
Further investigations of the 1-loop low-energy effective action are planned for future work.

\subsection{Quarter-maximal closed-string amplitudes}
\label{sec:sevensix}

The universality results on the parity-even part of quarter- and half-maximal open-string amplitudes in section \ref{sec:fivesix} can be extended to the closed string. The additional left-right contractions (\ref{nred13}) do not alter the key observation (\ref{promote}) about the sum over even spin structures in the left- and right-moving sector: Half-maximal and quarter-maximal cases only differ in the functions $F^{(k)}_{1/2}(\gamma_{k,k'})$ and $F^{(k+1)}_{1/4}(\gamma^j_{k,k'})$ of orbifold twists $\gamma_{k,k'}\equiv (k+k'\tau)v$ and $\gamma^j_{k,k'}\equiv (k+k'\tau)v_j$. Hence, the parity-even/even parts of $n$-point closed-string correlators are related by
\beq
{\cal J}^{e,e}_{n,1/4} = {\cal J}^{e,e}_{n,1/2}  \Big|^{\bar F^{(k)}_{1/2}(\bar \gamma_{k,k'}) \rightarrow \bar F^{(k+1)}_{1/4}(\bar \gamma^j_{k,k'})}_{F^{(k)}_{1/2}(\gamma_{k,k'}) \rightarrow F^{(k+1)}_{1/4}(\gamma^j_{k,k'})} \ .
\label{closed2to4}
\eeq
In presence of parity-odd admixtures from either left- or right-movers, the universality breaks down by the discussion in section \ref{sec:fiveeight}. From (\ref{CPodd7}), for instance, parity-odd/odd contributions to quarter-maximal 3-point amplitudes involve worldsheet functions of the type $f^{(2)}_{ij} \bar f^{(2)}_{pq}, \ \frac{ \pi}{\Im\tau}f^{(1)}_{ij} \bar f^{(1)}_{pq}$ and $\left(\frac{ \pi}{\Im\tau}\right)^2$. This departs from the factors of $F^{(1)}_{1/4}(\gamma^j_{k,k'})  \bar F^{(1)}_{1/4}(\bar\gamma^j_{k,k'})  \frac{ \pi}{\Im\tau}$ in the parity-even quarter-maximal terms (\ref{closed2to4}) as well as their half-maximal counterparts $\sim \frac{ \pi}{\Im\tau}$ in (\ref{cl3d}).


These structural differences in parity-odd contributions to half-maximal and quarter-maximal amplitudes also affect the low-energy behavior. For example, up to $n-1$ left-right contractions are compatible with the four-dimensional version of the $n$-point parity-odd/odd prescription (\ref{JN2oo}), leading to tensorial 3-point kinematic factor $\sim e^{(m}_2 \epsilon^{n)}(e_1,k_3,e_3) + (2\leftrightarrow 3)$. This ties in with the counting of loop momenta in quarter-maximal SYM amplitudes \cite{Johansson:2014zca}.

We see that just as for half-maximal above,  the  parity-even sector  of the low-energy limit
of the closed-string 4-point function on Calabi--Yau orbifolds has the mass dimension of $R^2$, so it does not produce
a loop correction to the  Einstein--Hilbert action, 
as expected from general arguments, see section \ref{sec:oneloop}. Only the parity odd/odd part of Calabi--Yau amplitudes has the right mass dimension to produce
a loop correction to the  Einstein--Hilbert action. However, this is delicate to see since it might require further minahaning; 
in the calculations above, we used strict momentum conservation in the odd/odd sector.


\section{Conclusions and outlook}

We made progress on calculating 1-loop string amplitudes
with reduced supersymmetry through three key methods:
\begin{itemize}
\item modular functions $f^{(n)}$ that let us generalize spin sums from the maximally supersymmetric case
\item the minahaning procedure of relaxing momentum conservation as an infrared regularization
\item building blocks of Berends--Giele type to capture gauge (pseudo-)invariant kinematic factors
\end{itemize}
A companion paper \cite{FT} on the field-theory limit will elaborate on the value of the Berends--Giele organization of kinematic factors for 1-loop amplitudes of half-maximal SYM in six and lower dimensions.


Another domain of application of the current results that we have not pursued in  detail is the string effective action.
We have set the stage for a systematic $\alpha'$-expansion by expressing integrands in terms of useful modular objects,
but we did not discuss their integration over $\tau$ here.\footnote{For the $\alpha'$-expansion of open-string amplitudes, the framework of elliptic MZVs is suitable for integrations over the vertex operator insertions $z_i$ at a given order, see \cite{Broedel:2014vla} for its application in the maximally supersymmetric case. In the closed-string sector, torus integrals over $z_i$ can be universally addressed using the techniques of \cite{Green:2008uj, D'Hoker:2015foa, D'Hoker:2015qmf, D'Hoker:2016jac}, see in particular \cite{Richards:2008jg, Green:2013bza} for connections between amplitudes of different multiplicities.}

One issue with this is that we have not been too specific about string-theory models; for compact open-string models, we should include orientifolds for tadpole cancellation.  As in for example \cite{Bianchi:2006nf},
we believe that this can be done straightforwardly from our results.

We have not touched on RR fields at all in this paper.
One interesting class of calculations concerns the completion of the dilaton and the NSNS field strength $H_3$ to 
the NSNS+RR axio-dilaton and self-dual field strength $G_3$. 
As an example, the action at order $\alpha'^3$ contains for example $|G_3|^2R^3$ 
(see e.g.\ \cite{Kehagias:1997cq,Conlon:2005ki,Policastro:2008hg} as well as \cite{Green:2013bza} for
S-duality properties of higher-derivative corrections). 

As emphasized earlier, it is important to remember that
these calculations are performed at the orbifold point, and generalizations
to smooth Calabi--Yau manifolds (including smooth K3) with the same amount of supersymmetry
may vary from straightforward to highly nontrivial \cite{Lin:2015dsa,Lin:2015wcg}. 
Whether or not these results are representative of generic points in moduli space,
experience shows that explicit results at specific points
will provide useful and highly needed guidance for 
generalizations.

It would be very interesting to revisit our amplitudes in a manifestly supersymmetric formalism -- either by using the hybrid formalism \cite{Berkovits:1994wr, Berkovits:1999im} or by deforming the pure spinor formalism \cite{Berkovits:2000fe}  to preserve half-maximal supersymmetry in $D=6$ dimensions.

\section*{Acknowledgement}

We are grateful to Massimo Bianchi, Martin Cederwall, Dario Consoli, Louise Dolan, Henrik Johansson, Joe Minahan
and Yang Sun for enlightening discussions and to Michael Haack, Piotr Tourkine and Dimitrios Tsimpis  for valuable comments on a draft of the manuscript. OS is grateful to Carlos Mafra for collaboration on related topics and to Karlstad University for financial support and kind hospitality during several stages of this project. Also, OS thanks the Institute of Advanced Studies in Princeton for warm hospitality during final stages of this work.


\appendix 

\section{Orbifold Partition Functions}
\label{opf}
In this appendix, we give further details on the vacuum amplitudes associated with the prescriptions in section \ref{sec:prescr}. 
In compactifications of type I to $D$ dimensions  on orbifold limits of Calabi--Yau threefolds or K3,  the cylinder vacuum amplitude (partition function) for open strings stretching between D9-branes can be written as\footnote{In this paper we consider only D9-branes with no background fluxes. }
\cite{Aldazabal:1998mr} 
\be
\mathcal{C} = 
\frac{V_D}{8 N}
\int_0^{\infty}\frac{\dd \tau_2}{\tau_2\,\,(8\pi^2\alpha' \tau_2)^{\frac{D}{2}}}\,\,\sum_{k=0}^{N-1}\mathcal{Z}^k_{\scriptscriptstyle \mathcal{C}} \ .
\ee 
Analogously, in orbifold compactifications of type IIA and type IIB, the torus vacuum  amplitude (partition function) reads \cite{Font:2005td} 
\be
\mathcal{T}= 
\frac{V_D}{8 N}
\int_{\mathcal{F}}\frac{\dd^2\tau}{ \,\,\tau_2 \,(4\pi^2 \alpha' \tau_2)^{\frac{D}{2}}} 
\sum_{k,k'=0}^{N-1}\mathcal{Z}^{k,k'}_{\scriptscriptstyle \mathcal{T}}\,.
\ee
In the main text we discuss gauge boson and graviton amplitudes for various orbifold 
compactifications, namely for
$\mathbb R^{1,5} \times{T^4}/{\mathbb{Z}_N }$ , $\mathbb R^{1,3}\times {T^6}/{\mathbb{Z}_N }$ and $\mathbb R^{1,3}\times T^2\times {T^4}/{\mathbb{Z}_N }$. To write general expressions that cover all these cases and to account for the possible presence of half-maximal sub-sectors in $D=4$, which depends on the rank $N$, we introduce the following slightly
non-standard notation:
by $d_k$ we denote the number of internal dimensions where for the given $k$ the orbifold has a fixed direction, and we set $D_k=D + d_k$.
For orbifold compactifications preserving some supersymmetry, which we always assume, the open-string partition function integrands can be expressed as
\footnote{In the literature, orbifold partition functions are often expressed in terms of $\theta$ functions with characteristics. These can easily be related to the above expressions using the basic definitions 
\cite{Mum1, Polchinski:1998rq} and the supersymmetry constraint $\sum_i v_i=0$. 
}
\be\label{ZC0}
\mathcal{Z}^0_{\mathcal{C}}={\Gamma^{(6)}_{\mathcal{C}}}\,\sum_{\nu=1}^4 (-1)^{\nu-1}
\left[\frac{{\theta_{\nu}(0,\tau)}}{{\theta'_1(0, \tau)}} \right]^4 \,(\text{tr}\,\gamma_0)^2
\ee 

\be\label{ZCk}
\mathcal{Z}^{k }_{\mathcal{C}}={ \Gamma^{(d_k)}_{\mathcal{C}}}\,\hat{\chi}_k \,
 \sum_{\nu=1}^{4} (-1)^{\nu-1} \left[ \frac{{\theta_\nu(0,\tau)}}{{\theta'_1(0, \tau)}} \right]^{\frac{D_k-2}{2}}\,
\prod_{i=1}^{\scriptscriptstyle 5- \frac{D_k}{2} } \,\frac{{\theta_{\nu}(kv_i, \tau)}}{{\theta_{1}(kv_i, \tau)}} \,\,
(\text{tr}\,\gamma_k)^2\,,
\ee
where $\hat{\chi}_k=\prod_{i=1}^{\scriptscriptstyle 5-\frac{D_k}{2}} [(-2\sin \pi k v_i)/(2\pi)]$, the set $\{\gamma_k\}_{k=0}^{N-1}$ spans a matrix representation of the orbifold group acting in the adjoint on the  $SO(32)$ Chan--Paton Lie algebra\footnote{This is schematic, but standard \cite{Gimon:1996ay, Aldazabal:1998mr}.
Explicit expressions for the matrices $\gamma_k$ are also given in the companion paper \cite{FT}.
}, the $\Gamma^{(2n)}_{ \mathcal{C}}$ represent sums over open-string momenta on the $T^{2n}$  tori with trivial orbifold action and $\Gamma^{(0)}_{ \mathcal{C}}\equiv1$ .

The closed-string integrands read
\be\label{ZT0}
\mathcal{Z}^{0,0}_{\mathcal{T}}= {\Gamma^{(6)}_{\mathcal{T}}}\,
\sum_{\nu,\tilde{\nu}=1}^4 (-1)^{\nu+\tilde{\nu} + \mu \delta_{\tilde{\nu},1}}
\left[\frac{{\theta_{\nu}(0,\tau)}}{{\theta'_1(0, \tau)}} 
\frac{{\bar{\theta}_{\tilde{\nu}}(0,\bar{\tau})}}{{\bar{\theta}'_1(0, \bar{\tau})}}
\right]^4
\ee
\begin{align}\label{ZTk}
\mathcal{Z}^{k ,k'}_{\scriptscriptstyle\mathcal{T}}=
\,{\Gamma^{(d_k)}_{\mathcal{T}}}\,
 {\hat{\chi}_{k,k'}}     
 \sum_{\nu,\tilde{\nu}=1}^4 (-1)^{\nu + \tilde{\nu} + \mu \delta_{\tilde{\nu},1}}
\left[\frac{{\theta_{\nu}(0,\tau)}}{{\theta'_1(0, \tau)}}
\frac{{\bar{\theta}_{\tilde{\nu}}(0,\bar{\tau})}}{{\bar{\theta}'_1(0, \bar{\tau})}} \right]^{\frac{D_k-2}{2}}\times\\
\nonumber{}
\prod_{i=1}^{\scriptscriptstyle 5-\frac{D_k}{2}} 
\frac{{\theta_{\nu}((k+k'\tau)v_i,\tau)}}{{\theta_{1}((k+k'\tau)v_i,\tau)}}
\frac{{\bar{\theta}_{\tilde{\nu}}((k+k'\bar{\tau})v_i,\bar{\tau})}}{{\bar{\theta}_{1}((k+k'\bar{\tau})v_i,\bar{\tau})}} \ ,
\end{align}
where $\mu$  takes the value 0 or 1 in type IIB or IIA, respectively.
 Eq. \eqref{ZTk} applies to all supersymmetric orbifolds of the kind $\mathbb R^{1,5}\times{T^4}/{\mathbb{Z}_N }$ and $\mathbb R^{1,3}\times T^2\times {T^4}/{\mathbb{Z}_N }$, but for Calabi--Yau limits
 it is only valid for $\mathbb R^{1,3}\times {T^6}/{\mathbb{Z}_N }$ with no fixed direction, i.e.\ for $N$ prime,
and requires a slight generalization if not.

We have introduced coefficients ${\hat{\chi}_{k,k'}}={{\chi}_{k,k'}}/(2\pi)^{10-D_k}$, where ${\chi}_{k,k'}$ denotes the number of simultaneous fixed points under the $\Theta^k$ and $\Theta^{k'}$ orbifold actions.
The textbook way to generate ${\chi}_{k,k'}$ 
 \cite{Blumenhagen:2013fgp,Font:2005td,Ibanez:2012zz} is  by starting with $k'=0$ and
acting with modular transformations, for example the T transformation takes
$k' \rightarrow k+k'$. Individual orbifold sectors
mix under modular transformations,
but the full amplitude is of course invariant by construction. See also the comment below (\ref{nred102}).

Finally, the  $\Gamma^{(2n)}_{ \mathcal{T}}$ represent the sum over closed string momentum states and winding states on tori where the orbifold projection is trivial.  For factorized spacetime tori $
T^6=(T^2)^3$, as we assume, explicit examples of  lattice sums are
\be
\Gamma^{(2n)}_{\cal C} =\prod_{i=1}^n \Bigg\{  \frac{ T_2^i}{\alpha'\tau_2} \sum_{n_1, n_2\in \mathbb{Z}} \text{exp}\left({-\frac{\pi T_2^i}{\alpha'\tau_2} \frac{|n_1+n_2U^i|^2}{U_2^i}}\right)\Bigg\}
\ee
and
\be
\Gamma^{(2n)}_{\cal T} = \prod_{i=1}^n \Bigg\{ \frac{2 T_2^i}{\alpha' \tau_2} \sum_{A\in {GL}(2, \mathbb{Z})} \text{exp}\left(
{-\frac{4\pi i\, T^i}{\alpha'}\, \text{det}{A} \,- \frac{2\pi T_2^i}{\alpha'\tau_2U_2^i}\Big|(1, U^i)A 
\Big(\!\begin{array}{c}
\tau\\
1
\end{array}\!\Big)
\Big|^2} \right) \Bigg\} \; ,
\ee
where $T^i$ and $U^i$ are K\"ahler and complex structure moduli of
the $i^{\te{th}}$ spacetime torus, see e.g.\ \cite{Gregori:1997hi,Kiritsis:2007zza}.


\section{Explicit examples of factorization}
\label{sec:factorization}

\subsection{Open string}
\label{sec:appB1}

In section, we verify that the representation of the 4-point open string correlator in (\ref{red47}) factorizes correctly upon integration. We have to show that the residue of the kinematic pole in $s_{12}$ can be written in terms of the 3-point integrand (\ref{red36}) with a cubic vertex of SYM attached. This cubic vertex can be represented using the two-particle polarization vector
\beq
e_{12}^m \equiv e_2^m (e_1 \cdot k_2) - e_1^m (e_2 \cdot k_1) + \frac{1}{2}(k_1^m-k_2^m)(e_1\cdot e_2)\ ,
\label{nred3}
\eeq
subject to $(k_{12}\cdot e_{12}) = 0$ which follows from peeling off $e_3^m$ from $A^{\te{tree}}(1,2,3)$ and coincides with the Berends--Giele current $s_{12} \efrak^m_{12}$ in (\ref{Giele01}). Factorization of the 4-point K3 amplitude on the $s$-channel allows for a 3-point K3 amplitude involving either $(e_{12},e_3,e_4)$ or $(e_1,e_2,e_{34})$ both of which are only defined up to $s_{12}=s_{34}$. The statement to prove is
\beq
\te{Res}_{s_{12}=0}{\cal A}_{1/2}(1,2,3,4) = {\cal A}_{1/2}(0,3,4) \, \big|^{k_0 \rightarrow k_{12}}_{e_0 \rightarrow e_{12}}  + {\cal A}_{1/2}(1,2,0) \, \big|^{k_0 \rightarrow k_{34}}_{e_0 \rightarrow e_{34}}   \ ,
\label{nred4}
\eeq
where the right-hand side is defined modulo $s_{12}$ by modifications of the integrand (\ref{red36}) such as
\beq
{\cal I}_{3,1/2}(12,3,4) \equiv X_{1+2,3} (e_{12} \cdot e_3) (e_4 \cdot k_{12}) + X_{3,4}(e_3\cdot e_4) (e_{12} \cdot k_3) + X_{4,1+2} (e_4 \cdot e_{12})(e_3\cdot k_4) \ ,
\label{nred5}
\eeq
using the vanishing of parity-odd contributions (\ref{CPodd1}). The notation $(12,3,4)$ instructs us to evaluate the 3-point integrand at polarizations $e_{12},e_3,e_4$, momenta $k_{12},k_3,k_4$ and coinciding positions $z_1{=}z_2$, e.g. 
\beq
X_{1+2,3} \equiv X_{13}+ X_{23} \big|_{z_1=z_2} = (s_{13}+s_{23}) f^{(1)}_{23}  \big|_{z_1=z_2} \ .
\label{nred5a}
\eeq
The residue of the 4-point amplitude in $s_{12}=s_{34}$ required by (\ref{nred4}) is unaffected by the parity-odd part in (\ref{CPodd4}) since the integrals involving $f^{(2)}_{ij}$ are local. The only kinematic poles in the parity-even integrand (\ref{red47}) stem from $K_{123|4},K_{124|3},K_{341|2}$, $K_{342|1}$ and $K_{12|34}$, see (\ref{red48}) and (\ref{red49}). The accompanying worldsheet functions must be mapped to their $s_{12}\rightarrow 0$ regime,
\beq
X_{12} = \delta(z_1-z_2) + {\cal O}(s_{12}) \co X_{34} = \delta(z_3-z_4) + {\cal O}(s_{34})  \ ,
\label{nred6}
\eeq
which only holds after integration against $\Pi_4$. In this limit,
\begin{align}
&\te{Res}_{s_{12}=0}{\cal I}_{4,1/2}  =\te{Res}_{s_{12}=0} \Big\{X_{12,3} K_{123|4} + X_{12,4} K_{124|3}+ X_{34,1} K_{341|2} + X_{34,2} K_{342|1} + X_{12} X_{34} K_{12|34} \Big\}
\notag \\
& \ \ = \delta(z_1-z_2) \big\{X_{1+2,3} (k_3\cdot e_4)\big[ (e_1\cdot e_3) (k_1\cdot e_2) - (e_2 \cdot e_3) (k_2\cdot e_1) + \tfrac{1}{2}(e_1\cdot e_2) (k^m_2 - k^m_1) e_m^3 \big] \notag \\
&\ \ \ \ \ \ \ \ + X_{1+2,4}(k_4\cdot e_3)\big[ (e_1\cdot e_4) (k_1\cdot e_2) - (e_2 \cdot e_4) (k_2\cdot e_1) + \tfrac{1}{2}(e_1\cdot e_2) (k^m_2 - k^m_1) e_m^4 \big] \notag \\
&\ \ \ \ \ \ \ \ + X_{34}(e_3 \cdot e_4) \big[ (k_3\cdot e_2)(k_2\cdot e_1)-(k_3\cdot e_1)(k_1\cdot e_2) + \tfrac{1}{2}(e_1\cdot e_2)(s_{13}-s_{23})\big] \big\}
\label{nred7} \\
&\ \ \ \ \ + \delta(z_3-z_4) \big\{ X_{3+4,1}  (k_1\cdot e_2)\big[ (e_1\cdot e_3) (k_3\cdot e_4) - (e_4 \cdot e_1) (k_4\cdot e_3) + \tfrac{1}{2}(e_3\cdot e_4) (k^m_4 - k^m_3) e_m^1 \big] \notag \\
&\ \ \ \ \ \ \ \ + X_{3+4,2}  (k_2\cdot e_1)\big[ (e_2\cdot e_3) (k_3\cdot e_4) - (e_4 \cdot e_2) (k_4\cdot e_3) + \tfrac{1}{2}(e_3\cdot e_4) (k^m_4 - k^m_3) e_m^2 \big]  \notag \\
&\ \ \ \ \ \ \ \ + X_{12}(e_1 \cdot e_2) \big[ (k_1\cdot e_4)(k_4\cdot e_3)-(k_1\cdot e_3)(k_3\cdot e_4) + \tfrac{1}{2}(e_3\cdot e_4)(s_{13}-s_{23})\big] \big\} \ ,
\notag
\end{align}
where the $[\ldots ]$ on the right-hand side can be identified with dot products of the two-particle polarization vector (\ref{nred3}). Hence, we recover the modified 3-point correlators in (\ref{nred5}),
\beq
\te{Res}_{s_{12}=0}{\cal I}_{4,1/2}  = \delta(z_1-z_2) {\cal I}_{3,1/2} (0,3,4) \, \big|^{k_0 \rightarrow k_{12}}_{e_0 \rightarrow e_{12}}  + \delta(z_3-z_4) {\cal I}_{3,1/2}(1,2,0) \, \big|^{k_0 \rightarrow k_{34}}_{e_0 \rightarrow e_{34}} \ .
\label{nred8}
\eeq
Upon integration over vertex operator positions, (\ref{nred8}) implies the desired factorization of the 4-point amplitude in (\ref{nred4}).

\subsection{Closed string}
\label{sec:clfactor}

As a sample of factorization of closed-string amplitudes, we consider the parity-odd/odd contribution to the 4-point low-energy limit in (\ref{odod}). In the pole channel of $s_{12}=s_{34}$, the 4-point expression has to reproduce the low-energy expression 
\beq
{\cal J}^{o,\tilde o}_{3,1/2}(0,3,4) \, \big|^{k_0 \rightarrow k_{12}}_{e_0 \rightarrow e_{12}}  \rightarrow 
-
\epsilon^{m}(e_{12}, k_3, e_3, k_4, e_4)
\epsilon_{m}( \tilde e_{12}, k_3, \tilde e_3, k_4, \tilde e_4)
\label{nred501}
\eeq
involving double-copies of the two-particle polarizations $e^m_{12}$ and $e^m_{34}$. Given the poles in (\ref{odod}) from 
\begin{align}
{\cal E}^m_{12|3,4} &= \frac{i\epsilon^{m}( e_{12}, k_3, e_3, k_4, e_4)}{ s_{12} }\co
{\cal E}^m_{1|2,34} = \frac{i\epsilon^{m}(e_{1},k_2, e_2, k_{34}, e_{34}) + {\cal O}(s_{34})}{ s_{34} } \ ,
\end{align} 
where the non-linearity of $\ffrak^{mn}_{34} \rightarrow - 2e^{[m}_3 e^{n]}_4$ in ${\cal E}^m_{1|2,34} $ is suppressed, we have
\begin{align}
&\te{Res}_{s_{12}=0}{\cal J}^{o,\tilde o}_{4,1/2}  \rightarrow  \te{Res}_{s_{12}=0} \Big\{ s_{12} {\cal E}^m_{12|3,4} \tilde {\cal E}^m_{12|3,4}  + s_{34} {\cal E}^m_{1|2,34} \tilde {\cal E}^m_{1|2,34}  \Big\} \notag \\
&\ \ \ = - \epsilon^{m}(e_{12}, k_3, e_3, k_4, e_4)  \epsilon_{m}( \tilde e_{12},k_3, \tilde e_3, k_4, \tilde e_4) - 
\epsilon^{m}( e_{1}, k_2, e_2, k_{34}, e_{34})  \epsilon_{m}( \tilde e_{1}, k_2, \tilde e_2, k_{34}, \tilde e_{34}) \ , \\
& \ \ \ \sim  {\cal J}^{o,\tilde o}_{3,1/2}(0,3,4) \, \big|^{k_0 \rightarrow k_{12}}_{e_0 \rightarrow e_{12}}   + {\cal J}^{o,\tilde o}_{3,1/2} (1,2,0) \, \big|^{k_0 \rightarrow k_{34}}_{e_0 \rightarrow e_{34}}   \ .\notag
\end{align}
Note that this check is again valid for any combination of gravitons, B-fields and dilatons.



\section{Kinematics of massless 3-point functions}
\label{kinematics}
In this appendix we give a few reminders about basic on-shell kinematics and connect the discussion with an interpretation of the minahaning procedure in section \ref{sec:minahan}.

\subsection{Scalar 3-particle special kinematics}
Massless 3-point functions of scalars vanish on-shell by momentum conservation. Here is a quick reminder why this is the case. Momentum conservation with all momenta ingoing is
\be
k_1+k_2+k_3 =0  \; .
\label{momcons}
\ee
Take the scalar product of this with $k_1$ and use on-shell masslessness $k_1^2=0$ to obtain $k_1\cdot k_2=-k_1\cdot k_3$. 
But this leads to
\be \label{momconssq}
0=(k_1+k_2+k_3 )^2=\underbrace{2k_1\cdot k_2+2k_1\cdot k_3}_{=0} + 2k_2 \cdot k_3=2k_2\cdot k_3
\ee
so $k_2\cdot k_3=0$, and similarly for the remaining two Mandelstam variables. 
We see that all Lorentz scalars $k_i^2=k_i\cdot k_j=0$, using momentum conservation 
and on-shell-ness. 

\subsection{Vector 3-particle special kinematics}
``Vectors'' here mainly refer to non-Abelian gauge bosons. 
With vector polarizations $e_i$ we can make nonzero Lorentz scalars. 
A priori there are 6 independent $e_i \cdot k_j$ for each $i\neq j$, but
\be
e_1 \cdot (k_1+k_2+k_3) =0 
\ee
so by  transversality $e_i\cdot k_i=0$ (no sum), we have $e_1\cdot (k_2+ k_3)=0$ and cyclic. In other words,
the only nonzero scalars are polarizations contracted with momentum {\it differences} 
$k_i-k_j$,
which leaves three:
\be  \label{nonzerovector}
e_1 \cdot (k_2-k_3) \; , \quad
e_2 \cdot (k_3-k_1)\; , \quad
e_3 \cdot (k_1-k_2)
\ee
This is enough to write the tree-level 3-point amplitude. 
However, at least in $D=4$, even these three Lorentz scalars vanish due to 3-particle special kinematics. One way to think about this is that the momenta need to be collinear,
so one can always reduce any $e_i \cdot k_j$ to $e_i\cdot k_i=0$.

\subsection{Interpretation of the minahaning procedure}
In quantum field theory, the fact that 3-point amplitudes of massless particles
vanish on-shell is no problem: just
go off-shell, $k_i^2\neq 0$. 
In (first-quantized) string theory
there is no obvious self-consistent way to go off-shell.
In the amplitude literature \cite{Elvang:2010kc}, one routinely uses 3-point functions as building blocks,
but with complex momenta.  
As detailed in section \ref{sec:minahan}, we use the minahaning procedure: we keep real momenta but relax momentum conservation, and maintain on-shell conditions $k_i^2=0$. Then
we have nonzero Lorentz scalars in the 3-point function, at least as an intermediate step.
%
The basic idea is that 
the physical state conditions are not violated by relaxing momentum conservation.

\begin{figure}[h]
\begin{center}
\includegraphics[width=0.5\textwidth]{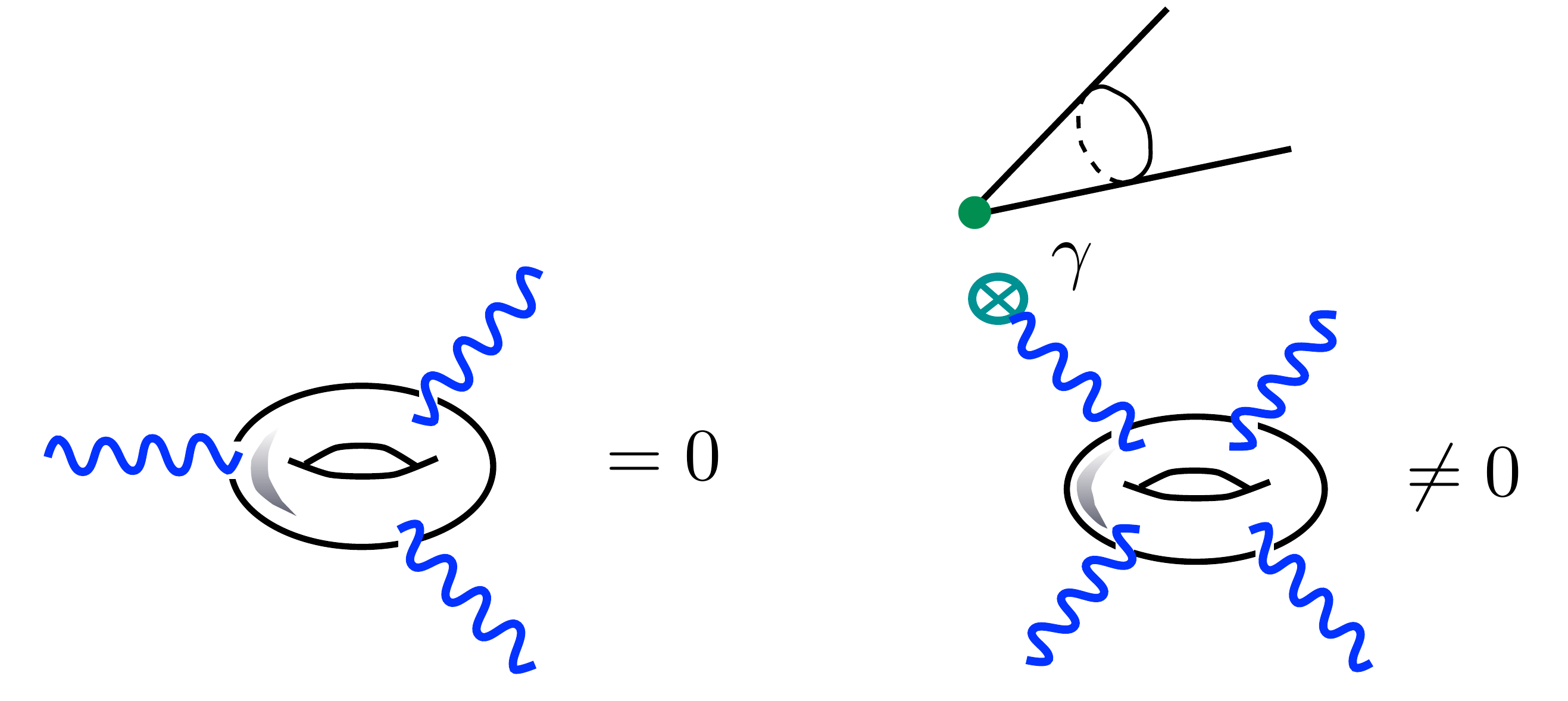}
\vspace{-5mm}
\caption{A schematic picture of minahaning as background field insertion.}
\label{fig:background}
\end{center}
\end{figure}

But what does it mean to relax momentum conservation? One operational way to think of it is that the 3-point function is ``embedded'' in
the 4-point function (so the 4$^{\rm th}$ momentum supplies the deformation), and the 4-point in the 5-point (as embodied in the notation $s_{123}$ for the deformation), and so on. This sounds surprising: why would we need to regularize
the 4-point function, where there is no issue with ``special kinematics'' as above? 
A more  physical way to relax momentum conservation
is to use an external background field, for example a gravitational background, such as AdS
or a sphere \cite{Kiritsis:1994ta}.\footnote{Of course, spheres may not be suitable as regulators if they break supersymmetry  \cite{Kiritsis:1997em}.}
In an orbifold, 
there is delta-function curvature at the fixed points,
so the orbifold twist $\gamma$ insertion mimics a background gravitation field insertion, as in fig.\ \ref{fig:background}.
(However, we emphasize that this is different from the insertion of an ordinary vertex operator,
since the ``position'' of this insertion is the twist $\gamma$, which is not integrated over.) 
Considering a background field may make it clearer
why the 4-point function is affected by the infrared regularization: with a background field, there is potentially a background insertion in {\it every} $n$-point function. 

For completeness, we also mention that D$p$-branes for $p<9$
provide another setting
where  momentum conservation in the naive sense is ``naturally relaxed'':  momentum is not conserved transverse to the D-brane, since the D-brane is very massive in perturbation theory
(see for example \cite{Hashimoto:1996bf}).

\section{Parity-odd contributions}
\label{app:dim}

\subsection{Parity-odd scalar correlator in arbitrary dimensions}
\label{sec:oddsc}

As argued in sections \ref{sec:fiveseven} and \ref{sec:fiveeight}, parity-odd contributions to open-string 1-loop amplitudes in $D$ spacetime dimensions kick in at multiplicity $N\equiv \frac{D}{2}+1$. Following the ten-dimensional six-point analysis in appendix B.2 of \cite{maxsusy}, we shall sketch intermediate steps towards the dimension-agnostic expression for ${\cal I}^o_{N,D}$ in (\ref{CPodd7}).

Using momentum conservation as well as overantisymmetrizations over $D+1$ indices such as
\begin{align}
e^m_1 &\epsilon(k_2,e_2,k_3,e_3,k_4,e_4,\ldots,k_N,e_N) - \big[ e_2^m \epsilon(k_2,e_1,k_3,e_3,k_4,e_4,\ldots ,k_N,e_N) + (2\leftrightarrow 3,4,\ldots,N) \big] \notag \\
&= k_2^m \epsilon(e_1,e_2,k_3,e_3,k_4,e_4,\ldots,k_N,e_N) + (2\leftrightarrow 3,4,\ldots,N)
\label{appO12}
\end{align}
the parity-odd $(N=\frac{D}{2}+1)$-point correlator (\ref{IN1O}) can be shown to yield 
\begin{align}
{\cal I}^o_{N,D}  &= \Big\{ E_{1|23,4,\ldots,N} \big[ \eta_{023} - \eta_{012} - \eta_{013} - (f^{(1)}_{01})^2 \big] + (23\leftrightarrow 24,34,\ldots,(N{-}1)N) \Big\}  \label{appO14}\\
&+\Big\{ \big[ \partial_2 f^{(1)}_{02}  + (f^{(1)}_{02} - f^{(1)}_{01}) \sum_{j\neq 2}^N s_{2j} f^{(1)}_{2j} \big] \epsilon(e_2,e_1,k_3,e_3,k_4,e_4,\ldots,k_N,e_N) + (2\leftrightarrow 3,4,\ldots,N) \Big\}
\ , \notag 
\end{align}
see (\ref{CPodd9}) for the kinematic factor $E_{1|23,4,\ldots,N}$. The shorthand $\eta_{ijk}$ represents the non-singular combination
\beq
\eta_{ijk} \equiv f^{(1)}_{ij} f^{(1)}_{ik} + f^{(1)}_{ji} f^{(1)}_{jk} + f^{(1)}_{ik} f^{(1)}_{jk} = f^{(2)}_{ij}+ f^{(2)}_{ik}+ f^{(2)}_{jk}
\label{appO13}
\eeq
which can be rewritten in terms of $f^{(2)}$ via Fay identities \cite{Broedel:2014vla}. The worldsheet functions along with $E_{1|23,4,\ldots,N}$ then simplify to $f^{(2)}_{23}-f^{(2)}_{12}-f^{(2)}_{13} - (f^{(1)}_{01})^2 - 2 f^{(2)}_{01}$, where the $z_0$-dependent parts drop out by virtue of the corollary
\beq
E_{1|23,4,\ldots,N}  + (23\leftrightarrow 24,34,\ldots,(N-1)N) = 0
\label{appO15}
\eeq
of (\ref{appO12}). The worldsheet functions in the second line of (\ref{appO14}) are total derivatives of $\Pi_N (f^{(1)}_{02} - f^{(1)}_{01})$ w.r.t.\ $z_2$ which do not contribute to open-string amplitudes but play a crucial role for the closed string to confirm the position $z_0$ of the picture changing operator to drop out. To keep track of parity-odd contributions to the closed-string amplitude in $D=6$, we spell out the total derivatives for this case,
\begin{align}
{\cal I}^o_{4,D=6} &= \Big\{ (f^{(2)}_{23}-f^{(2)}_{12}-f^{(2)}_{13} )E_{1|23,4}  + (23\leftrightarrow 24,34) \Big\} \label{trueodd} \\
& \ \ \ \ +\Big\{ \big[ \partial f^{(1)}_{02}  - (f^{(1)}_{02} - f^{(1)}_{01}) ( X_{21}+X_{23}+X_{24}) \big] \epsilon(e_2,e_1,k_3,e_3,k_4,e_4) + (2\leftrightarrow 3,4) \Big\} \ .
\notag
\end{align}
After dropping the total derivatives in (\ref{appO14}) and rearranging the remaining $f^{(n)}_{ij}$, we obtain
\beq
{\cal I}^o_{N,D}  = (f^{(2)}_{23}-f^{(2)}_{12}-f^{(2)}_{13} )E_{1|23,4,\ldots,N}  + (23\leftrightarrow 24,34,\ldots,(N-1)N)  \ .
\label{appO16}
\eeq
By another instance of (\ref{appO12}), the overall coefficient of $f^{(2)}_{12}$ conspires to the expression (\ref{CPodd8}) for~$E_{12|3,4,\ldots,N}$,
\beq
 E_{12|3,4,\ldots,N} = - E_{1|23,4,\ldots,N} - E_{1|24,3,5,\ldots,N}- \ldots - E_{1|2N,3,4,\ldots,N-1} \ .
\label{appO17}
\eeq
In view of (\ref{appO17}), the expressions for the parity-odd correlator in (\ref{appO16}) and (\ref{CPodd7}) are identical.

\subsection{The parity-odd 4-point vector correlator}
\label{sec:oddve}

In this appendix, we display intermediate expressions leading to the compact result (\ref{nred71}) for the parity-odd 4-point vector correlator ${\cal I}_{4,D=6}^{m,o}$. After peeling of a zero mode of $\partial X^m$, one OPE among the conformal fields is compatible with the parity-odd zero-mode saturation (\ref{zeropsi}). Contractions of the picture changing operator at $z_0$ (as indicated by $\big|_{f^{(1)}_{0j}}$) yields spurious poles
\begin{align}
{\cal I}_{4,D=6}^{m,o} \, \big|_{f^{(1)}_{0j}} &= e_1^m f^{(1)}_{01} i\epsilon(k_2,e_2,k_3,e_3,k_4,e_4) + i \big[  e_2^m ( f^{(1)}_{02}- f^{(1)}_{01}) \epsilon(k_2,e_1,k_3,e_3,k_4,e_4) \notag \\
& \ \ \  \ \ \  \ \ \ + f_{02}^{(1)} (k_2^m e_2^p - e_2^m k_2^p)  \epsilon_p(e_1,k_3,e_3,k_4,e_4) + (2\leftrightarrow 3,4) \big] 
\label{appO1} \\
&= i \big[( f_{20}^{(1)}-f_{10}^{(1)})k_2^m \epsilon(e_1, e_2, k_3, e_3, k_4, e_4) +(2\leftrightarrow 3,4)\big] \notag
\end{align}
which will later on conspire with left-right interacting integrations by parts. Terms of the form $e^m_2 f^{(1)}_{02}$ cancel on the spot and the overall coefficient of $f^{(1)}_{01}$ has been rearranged via
\beq
e^m_1 \epsilon(k_2,e_2,k_3,e_3,k_4,e_4) - \big[ e_2^m \epsilon(k_2,e_1,k_3,e_3,k_4,e_4) + (2\leftrightarrow 3,4) \big] = k_2^m \epsilon(e_1,e_2,k_3,e_3,k_4,e_4) + (2\leftrightarrow 3,4)
\label{appO2}
\eeq
which follows from antisymmetrizing in seven vector indices, see (\ref{appO12}) for a generalization.

The contractions among conformal fields in the vertex operators can be regrouped into 
\beq
{\cal I}_{4,D=6}^{m,o}  - \big( {\cal I}_{4,D=6}^{m,o} \, \big|_{f^{(1)}_{0j}} \big)
=
\big[ f_{12}^{(1)} E^m_{12|3,4} + (2\leftrightarrow 3,4) \big] +
\big[ f_{23}^{(1)} E^m_{1|23,4} + (23\leftrightarrow 24,34)\big] \ ,
\label{appO3}
\eeq
where the associated vector building blocks are given by
\begin{align}
E^m_{12|3,4}  &= i \big[  (e_1 \cdot k_2)\epsilon^m(e_2, k_3, e_3, k_4, e_4)  
                                     - (e_2 \cdot k_1)\epsilon^m(e_1, k_3, e_3, k_4, e_4) 
                                     - (e_1 \cdot e_2)\epsilon^m(k_2, k_3, e_3, k_4, e_4) 
                              \big]  \, ,
\notag \\
E^m_{1|23,4}  &= i \big[  (e_2 \cdot k_3)\epsilon^m(e_1, k_{23}, e_3, k_4 ,e_4)  
                                     - (e_3 \cdot k_2)\epsilon^m(e_1, k_{23}, e_2, k_4 ,e_4)
\label{nred74}\\
\,\,\,\,\,\,\, \ \ \ \ &    \ \ \ \ \ \ \              \ \ \ \                
                            - s_{23}  \epsilon^m(e_1, e_2, e_3, k_4, e_4) 
                                    -(e_2 \cdot e_3) \epsilon^m(e_1, k_2, k_3, k_4, e_4)\big]        \ .                
\notag
\end{align}
They can be identified as
\beq
E^m_{12|3,4} = s_{12}{\cal E}^m_{12|3,4}\,, \hspace{2cm}E^m_{1|23,4} = s_{23}{\cal E}^m_{1|23,4}\,,
\label{appOZ}
\eeq
after inserting the rank-two expressions (\ref{Giele01}) and (\ref{Giele02}) for $\efrak^m_{12}$ and  $\ffrak^{mn}_{12}$ into the definition (\ref{Giele09}) of ${\cal E}^m_{A|B,C}$. Combining (\ref{appO1}) with (\ref{appO3}) and (\ref{appOZ}) yields the desired expression for ${\cal I}_{4,D=6}^{m,o}$ in (\ref{nred71}). 

\section{Integral reduction in the 4-point closed-string amplitude} 
\label{app:4int}

In this appendix, we augment the general discussion in section \ref{sec:IBP} with further samples of corrections $\sim \frac{\pi }{\Im(\tau)}$ when reducing the closed-string integrals to a basis without any appearance of $f^{(1)}_{1j}$ and $f^{(n)}_{0j}$. When both left- and right-movers contribute with two factors of $f^{(1)}$ as in (\ref{nonholo2}), further representative examples include
\begin{align}
X_{12,3} \bar X_{13,2} &= X_{34,2} \bar X_{24,3} + \Big(\frac{\pi}{\Im \tau} \Big)^2 \big[ 4 s_{12} s_{13} + 2 (s_{12}+s_{13})s_{23} \big] \label{nonholo3} \\
&+ \frac{\pi}{\Im \tau}  \big[ (2s_{12}+s_{23}) X_{34} (\bar X_{32}+\bar X_{34}) + (2s_{13}+s_{23}) (X_{23}+X_{24}) \bar X_{24} \notag \\
& \ \ \ \ - s_{23} (X_{23}+X_{24})(\bar X_{32}+\bar X_{34}) + s_{23} X_{34} \bar X_{24} \big]  \notag \\
X_{12,3} \bar X_{12} \bar X_{34} &= X_{34,2}  \bar X_{34,2} -2 \Big(\frac{\pi}{\Im \tau} \Big)^2 s_{12}s_{34} + \frac{2\pi}{\Im \tau} s_{12} X_{34}\bar X_{34} \label{nhextra} \\
& \ \ \ \ -  \frac{\pi}{\Im \tau} s_{34} (X_{23}+X_{24})(\bar X_{23}+\bar X_{24})
\notag \\
X_{12} X_{34} \bar X_{12} \bar X_{34} &= X_{34,2}  \bar X_{34,2}  +  \frac{2\pi}{\Im \tau}  s_{12} X_{34} \bar X_{34}
\label{nonholo3a} \\
X_{12} X_{34} \bar X_{13} \bar X_{24}  &= X_{34,2} \bar X_{24,3}  + \Big(\frac{\pi}{\Im \tau} \Big)^2  s_{24}  s_{34} +   \frac{\pi}{\Im \tau} s_{23} X_{34} \bar X_{24} \notag \\
& \ \ \ \ -   \frac{\pi}{\Im \tau} \big(  s_{34}   (X_{23}+X_{24})  \bar X_{24} +  s_{24} X_{34} (\bar X_{32}+ \bar X_{34})  \big)  
\label{nonholo3b}
\end{align}
due to $\bar \partial f^{(n)}_{ij} = -\frac{ \pi }{\Im(\tau)} f^{(n-1)}_{ij}$. Similarly, in presence of $\bar f^{(2)}_{ij}$ on the right-moving side, left-moving integration by parts introduces corrections such as
\beq
X_{12} \bar f^{(2)}_{23} = -\frac{ \pi}{\Im \tau} \bar f^{(1)}_{23} + (X_{23} +X_{24}) \bar f^{(2)}_{23} \ .
\label{nonholo5}
\eeq
Cases of the form $ f^{(2)}_{ij}\bar f^{(2)}_{pq}$ do not admit any reduction via integration by parts and can be taken as basis elements regardless on $i,j,p$ and $q$. For ease of notation, we have suppressed the Koba--Nielsen factor $\Pi_n$ in (\ref{nonholo3}) to (\ref{nonholo5}), i.e.\ relations of this type are understood to hold upon integration over the $z_j$.

Moreover, the integration-by-parts removal of spurious double poles in the 4-point function, see section \ref{sec:double}, introduces extra contributions such as
\begin{align}
\big[ \partial f^{(1)}_{12} + s_{12} (f^{(1)}_{12})^2 \big] \bar X_{12} \bar X_{34} &= \frac{1}{2} f^{(1)}_{12} (X_{23}\!+\!X_{24}\!-\!X_{13}\!-\! X_{14}) \bar X_{12} \bar X_{34}+ \frac{ \pi}{\Im(\tau)} X_{12}\bar X_{34} \label{double1} \\
\big[ \partial f^{(1)}_{12} + s_{12} (f^{(1)}_{12})^2 \big] \big[\bar  \partial \bar f^{(1)}_{12} + s_{12} (\bar f^{(1)}_{12})^2 \big] &= \frac{1}{2} f^{(1)}_{12} (X_{23}\!+\!X_{24}\!-\!X_{13}\!-\! X_{14})  \big[\bar  \partial \bar f^{(1)}_{12} + s_{12} (\bar f^{(1)}_{12})^2 \big]  +  \frac{2 \pi X_{12} \bar f^{(1)}_{12}}{\Im(\tau)} \ .
\notag
\end{align}
Finally, in presence of parity-odd contributions, integration by parts as seen in (\ref{trueodd}) is required to remove the spurious dependence on the position $z_0$ of the picture changing operator,
\begin{align}
\partial f^{(1)}_{02} \bar X_{12} \bar X_{34} &= f^{(1)}_{02} (X_{21}+X_{23}+X_{24})  \bar X_{12} \bar X_{34} +\frac{ \pi}{\Im(\tau)} f^{(1)}_{02} s_{12} \bar X_{34} \label{double110} \\
\partial f^{(1)}_{02} \bar X_{23,4} &= f^{(1)}_{02} (X_{21}+X_{23}+X_{24}) \bar X_{23,4} - \frac{ \pi}{\Im(\tau)} f^{(1)}_{02}  \big[ s_{24} \bar X_{23} + s_{23}(\bar X_{24}+ \bar X_{34})\big] \label{double111} \\
\partial f^{(1)}_{02} \bar f^{(2)}_{2j} &= f^{(1)}_{02} (X_{21}+X_{23}+X_{24}) \bar f^{(2)}_{2j} - \frac{ \pi}{\Im(\tau)} f^{(1)}_{02}  \bar f^{(1)}_{2j}  \label{double112} \\
\partial f^{(1)}_{02} \bar \partial \bar f^{(1)}_{0j} &= f^{(1)}_{02} (X_{21}+X_{23}+X_{24}) \bar \partial \bar f^{(1)}_{0j}  \ ,\label{double113}
\end{align}
where the derivatives are understood as $\partial f^{(1)}_{02} \equiv \partial_0 f^{(1)}_{02}$ and $\bar \partial \bar f^{(1)}_{0j}\equiv \bar \partial_0 \bar f^{(1)}_{0j}$.
Iterating manipulations of the above type yields the final result (\ref{nred102}) for the 4-point closed-string correlator in a basis of integrals.

\linespread{1.1}


\end{document}